\documentclass[11pt,a4paper]{article}
\pdfoutput=1
\usepackage{jheppub}

\usepackage{multirow, graphicx,amssymb,url,mathrsfs,amsmath}
\usepackage{wrapfig,boxedminipage,setspace,subfigure,epsfig}
\usepackage{amsxtra,amstext,latexsym,dsfont,amsfonts}
\usepackage{color,eucal}
\usepackage[dvipsnames]{xcolor}
\usepackage{float}
\usepackage{slashed,comment}
\usepackage{kotex}
\usepackage{contour}

\usepackage{tikz}
\usetikzlibrary{calc,patterns,angles,quotes}
\usetikzlibrary{decorations.pathreplacing,decorations.markings,snakes}

\usepackage{tabularx, array}
\newcolumntype{L}[1]{>{\raggedright\arraybackslash}p{#1}}
\newcolumntype{C}[1]{>{\centering\arraybackslash}p{#1}}
\newcolumntype{R}[1]{>{\raggedleft\arraybackslash}p{#1}}

\renewcommand{\l}{\lambda}

\newcommand{\dd}{\mathrm{d}}

\newcommand{\grad}{\nabla}

\usepackage{xparse}

\ExplSyntaxOn
\NewDocumentCommand{\HS}{m}
 {
  \seq_set_split:Nnn \l_tmpa_seq { ~ } { #1 }
  \seq_map_inline:Nn \l_tmpa_seq { \contour{green}{##1} ~ } \unskip
 }
\ExplSyntaxOff

\title{Deep learning-based holography for $T$-linear resistivity}

\author[a]{Byoungjoon Ahn,}
\author[b]{Hyun-Sik Jeong,}
\author[a]{Chang-Woo Ji,}
\author[a,c]{Keun-Young Kim}
\author[a]{and Kwan Yun}

\emailAdd{bjahn123@gist.ac.kr}
\emailAdd{hyunsik.jeong@csic.es}
\emailAdd{physianoji@gm.gist.ac.kr}
\emailAdd{fortoe@gist.ac.kr}
\emailAdd{ludibriphy70@gm.gist.ac.kr}

\preprint{\texttt{IFT-UAM/CSIC-25-2}}

\affiliation[a]{Department of Physics and Photon Science, Gwangju Institute of Science and Technology, \\
123 Cheomdan-gwagiro, Gwangju 61005, Korea}
\affiliation[b]{Instituto de F\'isica Te\'orica UAM/CSIC, Calle Nicol\'as Cabrera 13-15, 28049 Madrid, Spain}
\affiliation[c]{Research Center for Photon Science Technology, Gwangju Institute of Science and Technology, \\
123 Cheomdan-gwagiro, Gwangju 61005, Korea}

\abstract{
We employ deep learning within holographic duality to investigate $T$-linear resistivity, a hallmark of strange metals. Utilizing Physics-Informed Neural Networks, we incorporate boundary data for $T$-linear resistivity and bulk differential equations into a loss function. This approach allows us to derive dilaton potentials in Einstein-Maxwell-Dilaton-Axion theories, capturing essential features of strange metals, such as $T$-linear resistivity and linear specific heat scaling. We also explore the impact of the resistivity slope on dilaton potentials. Regardless of slope, dilaton potentials exhibit {\it universal} exponential growth at low temperatures, driving $T$-linear resistivity and matching infrared geometric analyses. At a specific slope, our method rediscovers the Gubser-Rocha model, a well-known holographic model of strange metals. Additionally, the robustness of $T$-linear resistivity at higher temperatures correlates with the asymptotic AdS behavior of the dilaton coupling to the Maxwell term. Our findings suggest that deep learning could help uncover mechanisms in holographic condensed matter systems and advance our understanding of strange metals.
}

\begin{document}
\maketitle

%
\section{Introduction}\label{}
Strange metals, found in high-$T_c$ superconductors and related systems, represent one of the enduring puzzles in modern physics~\cite{Coleman_2005,Zaanen:2010yk,Liu_2012,Faulkner:2010da,Phillips:2022nxs}. These materials are characterized by strongly interacting quantum critical states that challenge conventional theoretical frameworks.\footnote{Despite numerous theoretical proposals~\cite{Lohneysen:2007aa}, a compelling explanation for the properties of strange metals remains elusive. Models such as the Sachdev-Ye-Kitaev (SYK) model~\cite{Sachdev:1993aa,Guo:2020aa,Chowdhury:2022aa}, the Yukawa-SYK model~\cite{Guo:2022aa,Patel_2023,Li:2024aa,Sachdev:2024aa,Wang:2024utm,Wang:2025oiz}, and the marginal Fermi liquid~\cite{Varma:1989aa} have provided partial insights, but a fully satisfactory understanding is still lacking.} The holographic duality, also known as AdS/CFT correspondence or gauge-gravity duality, originally discovered in string theory, offers a potential framework to describe these states~\cite{Zaanen:2015oix,Hartnoll:2016apf,Baggioli:2019rrs,Zaanen:2021llz}.

Despite the progress made using this approach, a comprehensive explanation for the experimentally observed (iconic) linear resistivity remains elusive~\cite{Zaanen:2010yk,Varma:1989aa,Hussey__2004}. A variety of strange metals, including cuprates, pnictides, and heavy fermion systems, exhibit a universal linear relationship between resistivity ($\rho$) and temperature ($T$), expressed as
\begin{align}
\begin{split}
\rho \approx T \,,
\end{split}
\end{align}
which contrasts sharply with the $T^2$ scaling of resistivity found in ordinary metals described by Fermi liquid theory. This universality underscores the need for a deeper understanding of the underlying mechanisms driving this behavior.\footnote{As other examples of universal properties, there are the Hall angle~\cite{Kim:2010zq,Blake:2014yla,Zhou:2015dha,Kim:2015wba,Chen:2017gsl,Blauvelt:2017koq,Ahn:2023ciq} at finite magnetic field and Home’s law in superconductors~\cite{Homes:2004wv,Zaanen:2004aa,Erdmenger:2015qqa,Kim:2015dna,Kim:2016hzi,Kim:2016jjk,Jeong:2021wiu}. For a detailed discussion of transport anomalies unique to strange metals, refer to~\cite{Phillips:2022nxs}.}

\paragraph{Infrared geometry analysis of the resistivity.}
In most holographic approaches, the $T$-linear resistivity is attributed to properties of the infrared (IR) geometry described by Einstein-Maxwell-Dilaton with Axion (EMDA) theories~\cite{Charmousis:2010zz, Davison:2013txa, Gouteraux:2014hca, Kim:2015wba, Zhou:2015dha, Ge:2016lyn, Cremonini:2016avj, Chen:2017gsl, Blauvelt:2017koq, Ahn:2017kvc, Jeong:2018tua, Ahn:2019lrh}, \textit{e.g.,} Eq. \eqref{action}, recognized as promising effective holographic models for condensed matter systems.\footnote{Holographic axion theories~\cite{Andrade:2013gsa,Vegh:2013sk,Baggioli:2021xuv} have proven to be a powerful framework for investigating strongly coupled condensed matter systems~\cite{Zaanen:2015oix,Hartnoll:2016apf,Baggioli:2019rrs,Natsuume:2014sfa,Zaanen:2021llz,Faulkner:2010da}. These theories have been extensively applied in studies of conductivity~\cite{Davison:2013txa,Gouteraux:2014hca,Blauvelt:2017koq,Jeong:2018tua,PhysRevLett.120.171602,Ammon:2019wci,Ahn:2019lrh,Jeong:2021wiu,Baggioli:2022pyb,Balm:2022bju,Ahn:2023ciq}, transport coefficients~\cite{Davison:2014lua,Blake:2016jnn,Amoretti:2016cad,Blake:2017qgd,Baggioli:2017ojd,Ahn:2017kvc,Giataganas:2017koz,Davison:2018ofp,Blake:2018leo,Jeong:2019zab,Arean:2020eus,Liu:2021qmt,Jeong:2021zhz,Wu:2021mkk,Jeong:2021zsv,Huh:2021ppg,Jeong:2022luo,Baggioli:2022uqb,Jeong:2023ynk,Ahn:2024aiw,Zhao:2023qms}, and the collective dynamics of strongly coupled phases~\cite{Baggioli:2014roa,Alberte:2015isw,Amoretti:2016bxs,Alberte:2017oqx,Amoretti:2017frz,Amoretti:2017axe,Alberte:2017cch,Andrade:2017cnc,Amoretti:2018tzw,Amoretti:2019cef,Ammon:2019wci,Baggioli:2019abx,Amoretti:2019kuf,Ammon:2019apj,Baggioli:2020edn,Amoretti:2021fch,Amoretti:2021lll,Wang:2021jfu,Zhong:2022mok,Bajec:2024jez}. Furthermore, these theories have been explored in the context of quantum information~\cite{RezaMohammadiMozaffar:2016lbo,Yekta:2020wup,Li:2019rpp,Zhou:2019xzc,Huang:2019zph,Jeong:2022zea,HosseiniMansoori:2022hok} and innovative approaches such as the AdS/Deep Learning correspondence~\cite{Ahn:2024gjf,Ahn:2024jkk}.} These methods classify scaling IR geometries using critical exponents such as the dynamical critical exponent ($z$), hyperscaling violation exponent ($\theta$), and charge anomalous parameter ($\zeta$). The geometries are supported by specific matter fields and couplings.

It has been demonstrated~\cite{Donos:2014cya, Kim:2014bza, Kim:2015sma, Kim:2015wba} that resistivity can be calculated solely from horizon data, involving the metric and matter field values at the horizon $r_h$. This leads to a schematic expression for resistivity as
\begin{align}
\begin{split}
\rho \approx r_h^{\,f_{\rho}(z,\theta,\zeta)} \,,
\end{split}
\end{align}
where $f_{\rho}$ is a function of the critical exponents $z$, $\theta$, and $\zeta$. At low temperatures, the horizon radius $r_h$ can be related to temperature through
\begin{align}
\begin{split}
r_h \approx T^{\,f_{h}(z,\theta,\zeta)} \,,
\end{split}
\end{align}
where $f_{h}$ is a model-dependent function. Substituting this relation, the resistivity becomes
\begin{align}\label{IRLOWT}
\begin{split}
\rho \approx T^{\, f_{\rho}(z,\theta,\zeta)\, f_{h}(z,\theta,\zeta)} \,.
\end{split}
\end{align}
The system parameters, such as the chemical potential ($\mu$), momentum relaxation rate ($\beta$), and couplings, influence the critical exponents and proportionality constants. Consequently, the resistivity is governed by the critical exponents that characterize the IR fixed points of the system.

While this framework provides valuable insights, many of these results are highly sensitive to the specific microscopic details of the IR fixed point under consideration. It remains unclear whether these details represent generic features applicable to real electronic systems or are artifacts of the highly symmetric quantum field theories used in holographic models.

\paragraph{Holographic models of strange metal phenomenology: Gubser-Rocha models.}
The Gubser-Rocha model~\cite{Gubser:2009qt,Davison:2013txa,Jeong:2018tua}, widely recognized as one of the most prominent holographic frameworks exhibiting $T$-linear resistivity, belongs to one class of EMDA theories: Eq. \eqref{GRaction}. In this specific context, the linear resistivity originates directly from the properties of an IR fixed point. This fixed point is categorized as a semilocal quantum liquid~\cite{Iqbal2012}, characterized by an IR geometry described by:
\begin{align} \label{CONADS2R2}
\begin{split}
\dd s^2 = \xi^{-1} \left(\frac{- \dd t^2 + \mathcal{L}^2 \dd \xi^2}{\xi^2} + \sum_{i=1}^{2}\dd x_{i}^{2}\right)  \,,\qquad 
\mathcal{L}^2 := \frac{8}{2 - \beta^2} \,.
\end{split}
\end{align}
This geometry, conformal to AdS$_2 \times \mathbb{R}^2$, emerges in EMDA theories under specific limits~\cite{Goldstein2010,Charmousis2010,Hartnoll2012,Gouteraux:2014hca}
\begin{align}
\begin{split}
z \rightarrow \infty \,, \qquad \theta \rightarrow -\infty \,, \qquad {\theta}/{z} \rightarrow -1 \,,
\end{split}
\end{align}
with scalar couplings in the IR approximated by
\begin{align} \label{IRCONFORMAL}
\begin{split}
V_{\text{IR}}(\phi) \,\approx\, Z_{\text{IR}}(\phi) \,\approx\, e^{\frac{\phi}{\sqrt{3}}} \,,
\end{split}
\end{align}
and the dilaton field $\phi$ exhibiting logarithmic running behavior.

The concept of local quantum criticality, which implies emergent temporal scale invariance accompanied by short-ranged spatial correlations, is closely tied to the marginal Fermi liquid phenomenology developed to describe experimental results in cuprates~\cite{Varma:1989aa}. The Gubser-Rocha model reflects these ideas through its conformal to AdS$_2 \times \mathbb{R}^2$ geometry, which covariantly transforms~\cite{Iqbal2012,Davison:2013txa} (as $\dd s^2 \rightarrow \lambda^{-1} \dd s^2$) under the following scaling symmetry:
\begin{align} \label{}
\begin{split}
t \rightarrow \lambda t \,, \quad x \rightarrow  x \,, \quad y \rightarrow  y \,, \quad \xi \rightarrow  \lambda \xi \,.
\end{split}
\end{align}

The Gubser-Rocha model presents further noteworthy features. First, it allows for (UV completed) analytic solutions, a rare advantage among holographic models at finite temperature. In the IR analysis of resistivity, as discussed in studies like~\cite{Davison:2013txa, Gouteraux:2014hca, Zhou:2015dha, Ge:2016lyn, Cremonini:2016avj, Chen:2017gsl, Blauvelt:2017koq, Ahn:2017kvc,Liu:2024gxr}, solutions are typically valid only at low temperatures. Investigating resistivity at arbitrary temperatures often requires introducing potential terms that yield an asymptotically UV AdS geometry~\cite{Kiritsis:2015oxa, Ling:2016yxy, Bhattacharya:2014dea}. For most holographic models, this necessitates numerical methods, as finite-temperature analytic solutions are generally unavailable. The Gubser-Rocha model is a notable exception, with analytic solutions provided in~\cite{Gubser:2009qt, Davison:2013txa, Jeong:2018tua}.

Furthermore, the Gubser-Rocha model has a robust foundation in string theory. In $(4+1)$ dimensions, it arises from the ten-dimensional type IIB string theory as the near-horizon limit of D3-branes~\cite{Gubser:2009qt, Cvetic:1999xp}. Similarly, in $(3+1)$ dimensions, it emerges from a consistent truncation of eleven-dimensional supergravity compactified on AdS$_4 \times S^7$~\cite{Gubser:2009qt}.

This model also provides a compelling holographic realization of a `general' mechanism for $T$-linear resistivity, as outlined in~\cite{Davison:2013txa}. The mechanism is based on three conditions: (i) \textit{weak} momentum relaxation, which links resistivity to shear viscosity ($\eta$), $\rho \approx \eta$, (ii) the Kovtun-Son-Starinets bound, which relates shear viscosity to entropy density, $\eta \approx s$, and (iii) the entropy density scaling linearly with temperature, $s \approx T$, as observed in the strange metal phase of cuprates, \textit{i.e.}, Sommerfeld heat capacity.\footnote{It is worth noting that the dilaton field $\phi$ is originally introduced in ensuring vanishing entropy at zero temperature.} However, this mechanism is realized in the Gubser-Rocha model only at \textit{low} temperatures.\footnote{More broadly, $T$-linear resistivity is often associated with the characteristics of a specific spacetime geometry known as AdS$_2$~\cite{Faulkner:2009wj}. This geometry plays a significant role in condensed matter physics, as it is closely connected to models for strange metals, including the Sachdev-Ye-Kitaev model~\cite{Chowdhury:2021qpy}.}

Nevertheless, experimentally, $T$-linear resistivity is observed over a wide temperature range, including room temperature ($\sim$ 300 K). Extending this behavior to higher temperatures within the Gubser-Rocha framework requires \textit{strong} momentum relaxation (\textit{e.g.,} $\beta/\mu \sim 10$), as demonstrated in~\cite{Jeong:2018tua}. This study showed that $T$-linear resistivity could persist at high temperatures, even beyond the critical temperature of superconductors as in experiments, provided momentum relaxation is sufficiently strong. Similar findings have been verified in other EMDA models, \textit{e.g.,} ~\cite{Ahn:2019lrh}.

The importance of strong momentum relaxation was first emphasized in~\cite{Hartnoll:2014lpa}, where it was argued that rapid (or strong) momentum relaxation allows transport to be governed by the universal, intrinsic diffusion of energy and charge. As a result, the universality of $T$-linear resistivity emerges in the incoherent regime, characterized by strong momentum relaxation.\footnote{In this regime, in addition to $T$-linear resistivity, the Gubser-Rocha model has also been shown to account for other universal phenomena relevant to high-$T_c$ superconductivity, such as Homes' law~\cite{Jeong:2021wiu, Wang:2023rca}. Additional studies have explored the model’s phase diagram using fermionic spectral functions~\cite{Jeong:2019zab,Lu:2024qxj}, its conductivity properties~\cite{Ling:2013nxa,Zhao:2023qms}, plasmons in a layered strange metal~\cite{Eede:2023rrv}, universal bounds of energy/charge diffusion constant~\cite{Kim:2017dgz,Liu:2021qmt,Jeong:2023ynk}, holographic Schwinger-Keldysh effective field theories for diffusion~\cite{Liu:2024tqe}, and black hole interiors, including spacetime singularities, complexity measures, and thermal $a$-functions~\cite{Arean:2024pzo}.}

\paragraph{Machine learning and holographic condensed matter.}
The traditional holographic approach using EMDA theories provides a bottom-up framework for studying strongly coupled systems. In this method, gravitational bulk physics is constructed to model realistic dual boundary condensed matter systems: the Gubser-Rocha model serves as a paradigmatic example.

In this paper, we utilize deep learning (DL)~\cite{Hinton_2006,Bengio2007ScalingLA,LeCun_2015}, a prominent area of machine learning that leverages deep neural networks and computational science, to develop a data-driven approach for holographic gravity modeling of strongly coupled condensed matter systems.\footnote{Machine learning has emerged as a transformative tool across theoretical and experimental physics, demonstrating remarkable utility in data-rich domains~\cite{Carleo:2019ptp}. Its applications span a broad range of disciplines, including string theory~\cite{RUEHLE20201}, quantum many-body problems~\cite{Carleo:2017aa}, condensed matter physics~\cite{Bedolla_2020}, and nuclear physics~\cite{Boehnlein:2022aa}.} Specifically, we address the inverse problem: reconstructing the geometric bulk physics using input data from the dual boundary perspective.

As a case study, we focus on $T$-linear resistivity as the boundary input data to determine the corresponding bulk dilaton potential $V(\phi)$ and coupling $Z(\phi) $ in EMDA theories. For simplicity, we collectively refer to $V(\phi) $ and $Z(\phi)$ as dilaton potentials. Our results demonstrate that machine learning serves as a powerful tool for uncovering the underlying bulk theory in such inverse problems.

This process, known as bulk reconstruction, tackles foundational questions in holographic duality. These include identifying the gravity theory dual to a given boundary theory and exploring the extent to which holographic duality is valid. Furthermore, the algorithms developed for bulk reconstruction have the potential to be implemented in future tabletop material experiments, serving as an essential tool for studying emergent spacetime.

Through this study, we highlight how machine learning advances our understanding of $T$-linear resistivity within the context of holographic dilaton potentials. Furthermore, this approach may offer valuable insights for exploring quantum gravity in a controlled experimental setting, paving the way for applications in tabletop quantum gravity experiments.

The first concrete implementation of the AdS/CFT correspondence using deep neural networks was pioneered by Hashimoto, Sugishita, Tanaka, and Tomiya~\cite{Hashimoto:2018ftp}: see also the detailed treatment in the textbook~\cite{Tanaka_2021}. Their work demonstrated how a bulk geometry could emerge from given experimental data, highlighting the potential of DL to uncover the nature of emergent geometries in the holographic framework. Once the neural network is trained, the bulk metric can be reconstructed, an approach referred to as the AdS/DL correspondence. For an accessible introduction to this concept, \cite{Song:2020agw} provides an example in the context of a classical mechanics problem.

The connection between the AdS/CFT correspondence and DL has also been discussed in earlier works~\cite{You:2017guh,Gan:2017nyt,Lee:2017skk}. This relationship can be further explored through tensor networks and the AdS/MERA correspondence~\cite{Swingle:2009bg}, offering additional perspectives on the similarities between holography and DL frameworks.

In the context of holographic duality, DL techniques have been effectively utilized with various boundary data from QCD and condensed matter systems. Notable examples include modeling the magnetization curve of strongly correlated materials~\cite{Hashimoto:2018ftp}, extracting lattice QCD data on the chiral condensate~\cite{Hashimoto:2018bnb, Hashimoto:2020jug}, hadron spectra~\cite{Akutagawa:2020yeo}, meson spectrum data~\cite{Hashimoto:2021ihd, Hashimoto:2022eij, Mansouri:2024uwc, Luo:2024iwf}, shear viscosity in strongly coupled systems~\cite{Yan:2020wcd, Gu:2024lrz}, and the lattice QCD equation of state~\cite{Chen:2024ckb, Bea:2024xgv}.
Additionally, DL has contributed to exploring the QCD phase diagram~\cite{Cai:2024eqa}, analyzing optical conductivity~\cite{Li:2022zjc, Ahn:2024gjf}, and studying the high-$T_c$ superconducting phase diagram~\cite{Kim:2024car}. Other applications include investigating specific condensed matter systems on a ring~\cite{Hashimoto:2024yev} and exploring entanglement entropy~\cite{Park:2022fqy, Park:2023slm, Ahn:2024jkk}. These examples highlight the versatility and power of DL in advancing holographic modeling across diverse physical systems.

It is noteworthy that \cite{Ahn:2024gjf} employs machine learning to reconstruct an emergent spacetime using `AC' conductivity data from $\text{UPd}_2\text{Al}_3$, a prototypical example of heavy fermion metals in strongly correlated electron systems. In this work, we extend this approach and, to the best of our knowledge, provide the first demonstration of machine learning applied to reconstruct bulk dilaton potentials alongside the corresponding emergent spacetime using `DC' conductivity data, specifically focusing on the iconic experimental universality observed in strange metals: $T$-linear resistivity.

\paragraph{Deep learning for $T$-linear resistivity.}
Last but not least, we discuss the boundary $T$-linear resistivity data that forms the basis of our machine learning methodology. In various cuprates, a superb $T$-linear resistivity as $T\rightarrow0$ has been experimentally observed~\cite{Legros_2018}. Example include electron-doped cuprates such as Pr$_{2-x}$Ce$_{x}$CuO$_{4\pm\delta}$ (PCCO)~\cite{Fournier:1998aa,Dagan:2004aa,Tafti:2014aa} and La$_{2-x}$Ce$_{x}$CuO$_{4}$ (LCCO)~\cite{Jin_2011,Sarkar:2017aa}, as well as hole-doped cuprates like Bi$_{2}$Sr$_{2}$CuO$_{6+\delta}$ (Bi-2201)~\cite{Martin:1990aa}, La$_{2-x}$Sr$_{x}$CuO$_{4}$ (LSCO)~\cite{Cooper_2009}, and La$_{1.6-x}$Nd$_{0.4}$Sr$_{x}$CuO$_{4}$ (Nd-LSCO)~\cite{Daou_2008,Collignon:2017aa,Doiron_Leyraud_2017}. The corresponding resistivity is expressed as:
\begin{align}\label{EXPRES}
\begin{split}
\rho \approx \rho_0 + A \, T \,,
\end{split}
\end{align}
where $\rho_0$ is the residual resistivity, and $A$ represents the slope of $T$-linear resistivity. In experimental analysis, $\rho_0$ is often subtracted, implicitly distinguishing elastic from inelastic scattering.\footnote{This subtraction assumes that elastic and inelastic scattering are additive and attributes the temperature-independent term to disorder. While this assumption is not guaranteed in unconventional metals without quasiparticles, controlled disordering of cuprates has shown that Matthiessen's rule holds in weakly disordered $T$-linear regimes~\cite{Hartnoll:2021ydi}.}

It is noteworthy that the Planckian relaxation timescale,
\begin{align} \label{PLRT}
\begin{split}
\tau_{\text{Pl}} = \frac{\hbar}{k_{B} T} \,,
\end{split}
\end{align}
is widely recognized as a key factor governing electronic dynamics in cuprate stange metals. For a detailed discussion of the Planckian timescale in conventional and unconventional metals, see~\cite{Hartnoll:2021ydi}.\footnote{Eq. \eqref{PLRT} is the characteristic timescale of quantum critical systems at energy scales where scale invariance emerges, supported by explicit computations in models such as quantum critical magnets and superfluids~\cite{sachdev2011}.} Notably, in several metals, the strength of $T$-linear resistivity corresponds to a scattering rate near the universal value $1/\tau_{\text{Pl}}$~\cite{Bruin_2013}, and has been tested in cuprates. 

The $T$-linear resistivity regime is expected to emerge whenever the scattering rate $1/\tau$ approaches its Planckian limit \eqref{PLRT}, regardless of the specific inelastic scattering mechanism involved~\cite{Zaanen:2004aa}. The Drude formula further supports this relationship:
\begin{align} \label{EXPRES2}
\begin{split}
\rho = \frac{m_{*}}{n e^2} \frac{1}{\tau} = \frac{m_{*}}{n e^2} \frac{k_B}{\hbar} T =: A \, T \,,
\end{split}
\end{align}
where $n$ is the carrier density, $m_{*}$ is the effective mass, and $\tau$ reaches the Planckian limit in the second equality. Experimentally, using aforementioned electron/hole-doped cuprates~\cite{Legros_2018}, $T$-linear resistivity has been observed to coincide with this expectation, $\tau\approx\tau_{\text{Pl}}$, irrespective of the specific inelastic scattering processes involved.
Furthermore, the slope $A$ of $T$-linear resistivity per CuO$_2$ plane increases when the doping is lowered and is significantly larger in hole-doped cuprates than in electron-doped ones, reflecting the observation with the effective mass $A\approx m_{*}$.

Building on these experimental findings, our work will apply machine learning to explore the role of $T$-linear resistivity in holography. Specifically, we investigate how changes in the slope $A$ and the residual resistivity $\rho_0$ influence the reconstructed dilaton potentials. We will also demonstrate that machine learning can identify strong momentum relaxation mechanisms that robustly produce $T$-linear resistivity, consistent with IR scaling analyses.

This paper is organized as follows.
In section \ref{sec2}, we introduce the machine learning methodology: Physics-informed neural networks. In section \ref{sec31}, we review EMDA theories and an outline of the machine learning approach discussed in \ref{sec2} to derive dilaton potentials based on $T$-linear resistivity. Then, in section \ref{sec32}, we present the data-driven dilaton potentials derived from $T$-linear resistivity data. Specifically we provide a rediscovery of the Gubser-Rocha model, and analysis of how the slope ($A$) and residual resistivity ($\rho_0$) influence the results. Section \ref{sec4} is devoted to conclusions.

%
\section{Physics-informed neural networks: a methodology}\label{sec2}
In this section, we introduce our machine learning methodology: Physics-informed neural networks (PINNs)~\cite{RAISSI2019686}, which will be applied in our holographic computations in the next section. 

Since their introduction in 2017~\cite{Raissi:2017aa,Raissi:2017ab} and the consolidated publication in 2019~\cite{RAISSI2019686}, PINNs have garnered significant attention in the computational science and engineering community. With over 12,000 citations of~\cite{RAISSI2019686}, these networks have inspired advancements across a range of scientific domains. Researchers have refined the methodology, expanded its applications, and addressed limitations in the original formulation. Comprehensive reviews, for instance~\cite{Karniadakis_2021,Toscano:2024aa,Cuomo_2022,Farea_2024,Ganga:2024aa,Raissi:2024aa,Zhao_2024,Cai2021,HBWJ2023,Lawal_2022,Raissi_2020}, highlight improvements in network architecture, optimization strategies, uncertainty quantification, and theoretical insights, as well as diverse applications in fields like biomedicine, fluid mechanics, geophysics, and chemical engineering.

PINNs represent a novel approach at the interface of machine learning and physics. Unlike standard neural networks that rely solely on data, PINNs incorporate physical laws, such as ordinary or partial differential equations, directly into the training process. This integration enables the construction of robust models, even with limited or noisy data, and enhances their capacity for extrapolation beyond the training domain.

At the core, the training of a PINN is guided by a loss function comprising two components:  
(I) Data Loss: Ensures the model fits the available data points;
(II) Physics Loss: Encodes the governing physical equations, acting as a regularization term that aligns the model output with the underlying laws.  

In conventional machine learning, regularization prevents overfitting and enhances generalization by limiting the solution space. In PINNs, physical equations themselves serve as the regularization mechanism, guiding the model to solutions that are both physically consistent and compliant with boundary conditions. Through iterative optimization, the network training parameters are adjusted to minimize the combined loss function, balancing data fidelity and physical consistency. This innovative methodology makes PINNs a compelling tool for solving differential equations and modeling complex physical systems.

\paragraph{Schematic structure of PINNs.}
To illustrate the methodology of PINNs, consider a simple ordinary differential equation for a field $\mathcal{F}(r)$:
\begin{equation}\label{PINNODEEQ}
\mathcal{G}[r,\, \mathcal{F},\, \partial_r\mathcal{F};\, \Theta] = 0 \,,
\end{equation}
where $\Theta(r)$ represents a general unknown function involved in the equation.

\begin{figure}[]
  \centering
     {\includegraphics[width=14cm]{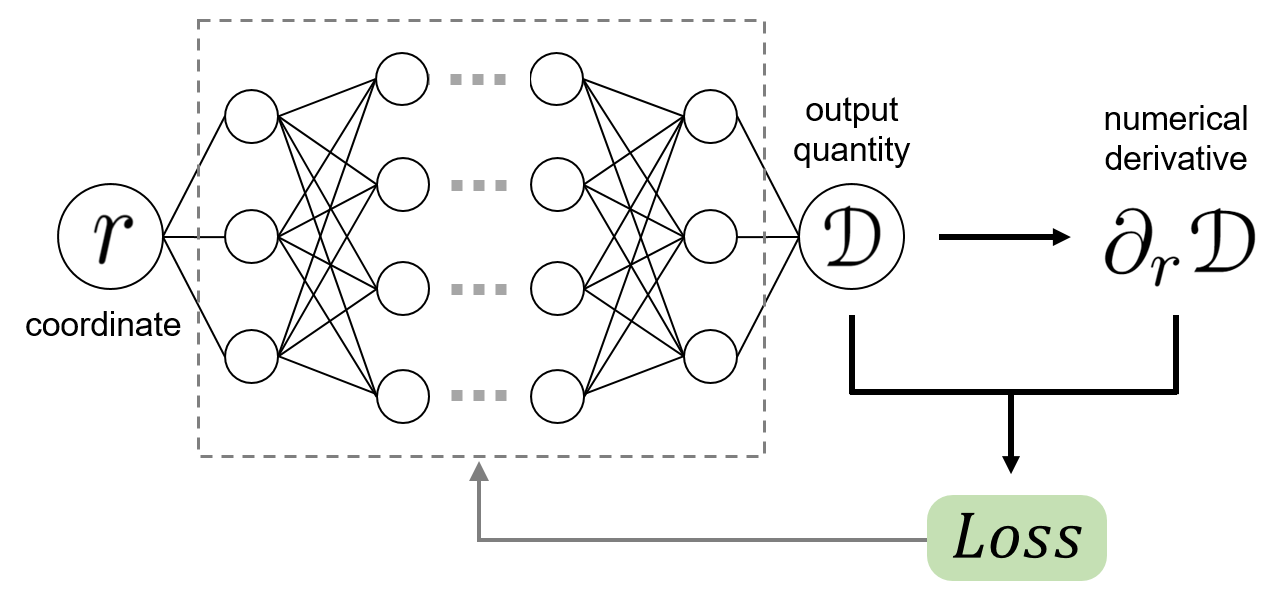} \label{}}
 \caption{A schematic representation of the deep neural network $\mathcal{D}$ in Eq.~\eqref{DNN}. Each layer's nodes are interconnected through weights and biases, with activation functions applied to produce non-linear transformations. In the context of PINNs, the total loss function comprises two components: the physics loss, derived from the derivatives of the deep neural network (such as $\partial_r \mathcal{D}$) to satisfy the equations of motion Eq.~\eqref{DNNEQ}, and the data loss, which incorporates empirical data.}\label{SKETCHFIG}
\end{figure}

In a PINN framework, $\mathcal{F}(r)$ is approximated using a deep neural network $\mathcal{D}(r)$, which allows us to rewrite the equation as:
\begin{equation}\label{DNNEQ}
\bar{\mathcal{G}} := \mathcal{G}[r,\, \mathcal{D},\, \partial_r\mathcal{D};\, \Theta]  \,.
\end{equation}
A deep neural network, a universal approximator~\cite{HORNIK1989359}, is defined as
\begin{equation}\label{DNN}
    \mathcal{D}(r) = W_M\cdot \Phi(\cdots\Phi(W_2\cdot\Phi(W_1 r+b_1)+b_2)\cdots)+b_M \,,
\end{equation}
where $W_M$ and $b_M$ are weights and biases, and $\Phi$ represents the activation function. 
The unknown function $\Theta(r)$ can also be replaced by the deep neural network.
In this manuscript, we employ the $\texttt{Tanh}$ activation function, a commonly used non-linear function in neural network applications. We refer to Fig. \ref{SKETCHFIG} for an illustrative depiction of a deep neural network utilized within the PINNs framework.

The next step involves constructing a loss function comprising the data loss $L_{\text{data}}$ and the physics loss $L_{\text{eom}}$
\begin{equation}
    L :=  L_{\text{data}} + L_{\text{eom}} \,,
\end{equation}
where
\begin{align}\label{PINNLossEom}
    L_{\text{eom}} := \frac{1}{N_r}\sum_r \log \left[ \cosh \big| \bar{\mathcal{G}} \big| \right]  \,,
\end{align}
where $N_{r}$ denotes the number of data points of $r$.
Through iterative optimization, $\mathcal{D}(r)$ is trained to approximate the solution $\mathcal{F}(r)$. The choice of the loss function depends on the problem specifics. In this paper, we use the \texttt{Log-Cosh Loss} as in \eqref{PINNLossEom}, which balances the advantages of \texttt{Mean Absolute Error} (L1 loss) and \texttt{Mean Squared Error} (L2 loss) by handling both small and large errors adaptively.\footnote{\texttt{Log-Cosh Loss} behaves quadratically for small errors (similar to L2 loss) and linearly for large errors (similar to L1 loss). This adaptive nature of \texttt{Log-Cosh Loss} allows our model to handle both small and large prediction errors effectively.} This makes it computationally efficient and suitable for large-scale data.

In our implementation, we utilize the \texttt{PyTorch} framework for  machine learning capabilities, including the \texttt{autograd} engine for efficient numerical differentiation through automatic differentiations. For stochastic optimization, we also employ the \texttt{Adam} algorithm~\cite{Kingma:2014vow}. 

\paragraph{Toy example: projectile motion.}
As a simple example, consider the motion of a projectile under gravity and drag: for the related simulation and code, see~\cite{towardsdatasciencePhysicsInformed}. The acceleration ${d^2 \bar{\mathcal{F}}}/{d t^2}$ is given by
\begin{equation}\label{TEX}
\frac{d^2 \bar{\mathcal{F}}}{d t^2} = - \bar{\Theta} \bigg|\frac{d \bar{\mathcal{F}}}{d t}\bigg| \frac{d \bar{\mathcal{F}}}{d t} - g \,,
\end{equation}
where $\bar{\mathcal{F}}(t)$ is the position of the projectile, $\bar{\Theta}$ is the drag coefficient (unknown), and $g$ is the gravity vector (known). The acceleration of a projectile at a given time $t$ is determined by two factors: (I) air resistance (drag), which acts in the opposite direction of the projectile's motion and is proportional to its current velocity, and (II) gravitational pull, the constant downward force that affects the projectile throughout its trajectory.  The combined effect of these forces determines the projectile's acceleration vector, continuously modifying its speed and direction as it moves.

A PINN minimizes both the data loss, based on the available training data (\textit{e.g.,} for $\bar{\mathcal{F}}(0\leq t \leq5)$), and the physics loss, derived from the governing equations Eq. \eqref{TEX} over the specified domain (\textit{e.g.,} $0\leq t \leq10$). Through the inclusion of the physics loss, the deep neural network is able to infer the function $\bar{\mathcal{F}}(t)$ even for times outside the data range (\textit{e.g.,} $5\leq t \leq10$). Additionally, unknown parameters, such as $\bar{\Theta}$, can be treated as trainable variables, allowing the model to learn their values during the training process. This flexibility illustrates how PINNs effectively combine physical principles with data-driven learning to solve complex problems.

%
\section{Deep learning for $T$-linear resistivity in holography}\label{}
In this section, we present an overview of the Einstein-Maxwell-Dilaton-Axion (EMDA) model and describe how PINNs can be utilized to derive the dilaton potentials associated with $T$-linear resistivity.

%
\subsection{Setup}\label{sec31}

%
\subsubsection{Einstein-Maxwell-Dilaton with Axion model}\label{}

\paragraph{Action and equations of motion.}
We consider the Einstein-Maxwell-Dilaton theory coupled to Axion fields~\cite{Charmousis:2010zz, Davison:2013txa, Gouteraux:2014hca, Kim:2015wba, Zhou:2015dha, Ge:2016lyn, Cremonini:2016avj, Chen:2017gsl, Blauvelt:2017koq, Ahn:2017kvc, Jeong:2018tua, Ahn:2019lrh}, described by the following action
\begin{align}\label{action}
\begin{split}
S = \int \dd^{4}x  \sqrt{-g}\left( R + \mathcal{L}_m \right)   \,, \quad
\mathcal{L}_m = \displaystyle- \frac{1}{2}(\partial \phi)^2 + V(\phi) - \frac{Z(\phi)}{4}F^{2} - \frac{1}{2}\sum_{i=1}^{2} (\partial \chi_i)^2 \,,
\end{split}
\end{align}
where $V(\phi)$ and $Z(\phi)$ are the dilaton potentials, the $U(1)$ gauge field has field strength $F = \dd A$, and the axion fields $\chi_i$ break the translational symmetry in the boundary theory, enabling finite resistivity. The action gives rise to the following equations of motion
\begin{align} \label{eommaster}
\begin{split}
&R_{\mu\nu} = \frac{1}{2}\partial_{\mu}\phi\partial_{\nu}\phi
+\frac{1}{2}\sum_{i=1}^{2}\partial_{\mu}\chi_{i}\partial_{\nu}\chi_{i}+\frac{Z(\phi)}{2}F_{\mu}{^\rho}F_{\nu\rho} -\frac{Z(\phi)F^2}{8}g_{\mu\nu}-\frac{V(\phi)}{2}g_{\mu\nu} \,, \\
&\grad_{\mu}(Z(\phi)F^{\mu\nu}) = 0 \,, \\& \square\phi+V'(\phi)-\frac{1}{4}Z'(\phi)F^2 =0
  \,, \\
&\grad_{\mu}\grad^{\mu}\chi_{i} =0 \,.
\end{split}
\end{align}

To solve these equation, we adopt a homogeneous ansatz for the fields
\begin{align} \label{Ourmodelmetric}
\begin{split}
        \dd s^2 &= \frac{1}{r^2}\left(-f(r) \dd t^2+\frac{1}{f(r)}\dd r^2+h(r)\sum_{i=1}^{2}\dd x_{i}^{2}\right) \;, \\
        \phi &= \phi(r) \,, \qquad A=A_t(r) \dd t \,, \qquad \chi_{i} = \beta x_{i} \,,       
\end{split}
\end{align}
where $r$ represents the radial coordinate, and the functions $f(r)$ and $h(r)$ encode the spacetime geometry. In this manuscript, we place the AdS boundary at $r=0$ and the event horizon at $r=1$, which does not affect the generality of our results.

Within our ansatz, the equations of motion \eqref{eommaster} can be recast in four independent equations
\begin{equation}\label{Ourmodeleq}
    \begin{aligned}
    & 0 = 2 r^2 f(r) h'(r) \phi'(r) + 2 h(r) V'(\phi(r)) \\ 
    & \qquad + r h(r) \Big(r^3 A_t'(r)^2 Z'( \phi(r) )+2 \left(r f'(r)-2 f(r)\right) \phi '(r) + 2 r f(r) \phi''(r)\Big) \;, \\
    & 0 = A_t''(r)+\frac{A_t'(r) h'(r)}{h(r)}+\frac{A_t'(r) \phi'(r) Z'(\phi (r))}{Z(\phi (r))} \;, \\
    & 0 = \frac{2 h''(r)}{h(r)}-\frac{h'(r)^2}{h(r)^2}+\phi'(r)^2 \;, \\ 
    & 0 =  h(r)^2 \Big(r^4 A_t'(r)^2 Z(\phi (r)) -4 r f'(r)+f(r) \left(12-r^2 \phi '(r)^2\right) - 2 V(\phi(r)) \Big) \\
    & \qquad + 2 r h(r) \Big(\left(r f'(r)-4 f(r)\right) h'(r)+\beta ^2 r \Big) + r^2 f(r) h'(r)^2  \,.
    \end{aligned}
\end{equation}
The axion field equations are automatically satisfied.

\paragraph{UV-completion of dilaton potentials.}
To ensure a holographic duality framework within AdS spacetime, \textit{i.e.,} a UV-completed theory, the following boundary conditions must be satisfied~\cite{Charmousis:2010zz}: 
\begin{align}\label{ASYMPO}
V(\phi) = 6 - \frac{1}{2}m^2 \phi^2 + \cdots \,,  \qquad
Z(\phi) = 1 + \cdots  \,,
\end{align} 
near the AdS boundary with $\phi \rightarrow 0$. Here the leading term in $V(\phi)$ corresponds to the cosmological constant of AdS, while $m^2$ presents the mass-squared of the dilaton field. Near the boundary, the dilaton also exhibits the asymptotic behavior
\begin{equation}\label{}
\phi(r) \,\approx\, \phi_{-} \, r^{\Delta_{-}} + \phi_{+} \, r^{\Delta_{+}} \,, \qquad \Delta_{\pm} := \frac{3}{2} \pm \sqrt{\frac{9}{4}+m^2} \,,
\end{equation}
where $\Delta_{\pm}$ are the conformal dimensions of the dual operator.

\paragraph{Thermodynamics and DC resistivity.}
Various thermodynamic quantities of EMDA theories \eqref{action} can be achieved: the Hawking temperature $T$, thermal entropy density $s$, chemical potential $\mu$, and charge density $q$ are determined as follows
\begin{equation}\label{TDC}
T = \frac{-f'(1)}{4\pi}  \,,
\quad 
s = 4\pi \, h(1)  \,,
\quad
\mu = A_t(0)  \,, 
\quad
q =    Z_{H} \, h(1) \, A_t'(1)  \,, 
\end{equation}
where $Z_{H}$ denotes the dilaton potential evaluated at the horizon. The DC electric resistivity $\rho$ of the EMDA, relevant for systems with broken translational invariance, can be expressed as~\cite{Gouteraux:2014hca, Donos:2014cya}:
\begin{align}\label{confor}
\rho = \frac{1}{\sigma_{DC}} \,,\qquad \sigma_{DC} = Z_H  + \frac{q^2}{\beta^2 \, h(1)} \,,
\end{align}
where $\beta$, the axion charge, explicitly breaks translational symmetry, ensuring finite conductivity.

\paragraph{Holographic Gubser-Rocha model.}
Next, we briefly review the holographic Gubser-Rocha model, a framework for studying strange metals~\cite{Gubser:2009qt, Davison:2013txa, Jeong:2018tua}. The model is given by specific choices for dilaton potentials:
\begin{align}\label{GRaction}
\begin{split}
V(\phi) = 6 \cosh \frac{\phi}{\sqrt{3}} \,, \qquad Z(\phi) = e^{\frac{\phi}{\sqrt{3}}} \,,
\end{split}
\end{align}
which admit the following analytic solutions
\begin{align}\label{GRSOL}
\begin{split}
f(r) &=  (1-r) U(r) \,,\qquad h(r) = (1+ Q \, r)^{3/2} \,, \\
A_t(r) &= (1-r) a(r) \,, \qquad\, \phi(r) = \frac{\sqrt{3}}{2} \log (1+ r \, Q) \,,
\end{split}
\end{align}
where $U(r)$ and $a(r)$ are given by
\begin{align} \label{}
\begin{split}
U (r) = \frac{1+(1+3Q) r + \left(1+ 3Q(1+Q)-\frac{1}{2}\beta^2\right)r^2}{(1+ Q r)^{3/2}} \,, \quad
a(r) = \frac{\sqrt{3Q(1+Q)\left( 1- \frac{\beta^2}{2(1+Q)^2} \right)}}{1+Q r} \,.
\end{split}
\end{align}
By using the thermodynamic expressions in \eqref{TDC} and the solutions in \eqref{GRSOL}, it can be shown that the parameter $Q$, which controls the dilaton field, is determined by the physical ratios $T/\mu$ and $\beta/\mu$: see \cite{Jeong:2018tua} for details. Moreover, by comparing \eqref{ASYMPO} and \eqref{GRaction}, the dilaton mass-squared in the Gubser-Rocha model is found to be
\begin{align}\label{GRm2}
\begin{split}
m^2 = -2  \,,
\end{split}
\end{align}
indicating $\phi(r) \,\approx\, \phi_{-} \, r + \phi_{+} \, r^2 $ in the asymptotic AdS region.

\paragraph{$T$-linear resistivity and linear specific heat.}
Using the analytic solutions \eqref{GRSOL} with \eqref{TDC} and \eqref{confor}, the Gubser-Rocha model captures key features of strange metals, including $T$-linear resistivity and linear specific heat (Sommerfeld heat capacity). In the low $T/\mu$ limit, these quantities behave as
\begin{align}\label{linearrho}
\begin{split}
\rho \,\approx\, \frac{2\pi (\beta/\mu)^2 \sqrt{4+6 (\beta/\mu)^2}}{\sqrt{3} \left( 1+(\beta/\mu)^2 \right)^2} \, \frac{T}{\mu} \,=:\, A_{\text{GR}} \, \frac{T}{\mu} \,, \qquad 
\frac{s}{\mu^2} \,\approx\, \gamma_{s} \, \frac{T}{\mu}  \,,
\end{split}
\end{align}
where $\gamma_{s}$, Sommerfeld coefficient, is a complicated function of $\beta/\mu$.

As shown in \cite{Jeong:2018tua}, the $T$-linear resistivity described by \eqref{linearrho} remains robust even at higher temperatures in the regime of strong momentum relaxation (\textit{e.g.,} up to $T/\mu \approx 1$ when $\beta/\mu=10$). In contrast, for the coherent regime, the $T$-linear resistivity holds within a very narrow temperature range (\textit{e.g.,} up to $T/\mu \approx 0.03$ when $\beta/\mu=0.1$). 

\begin{figure}[]
  \centering
     {\includegraphics[width=6.4cm]{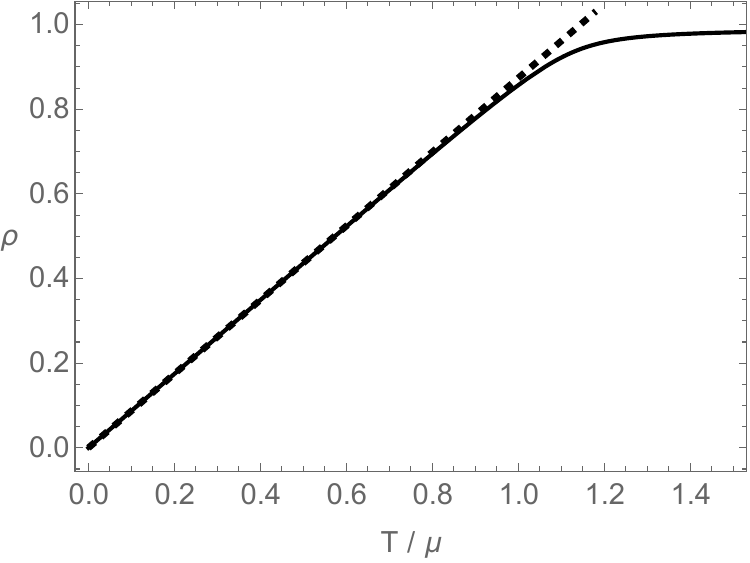} \label{}}
\quad
     {\includegraphics[width=6.8cm]{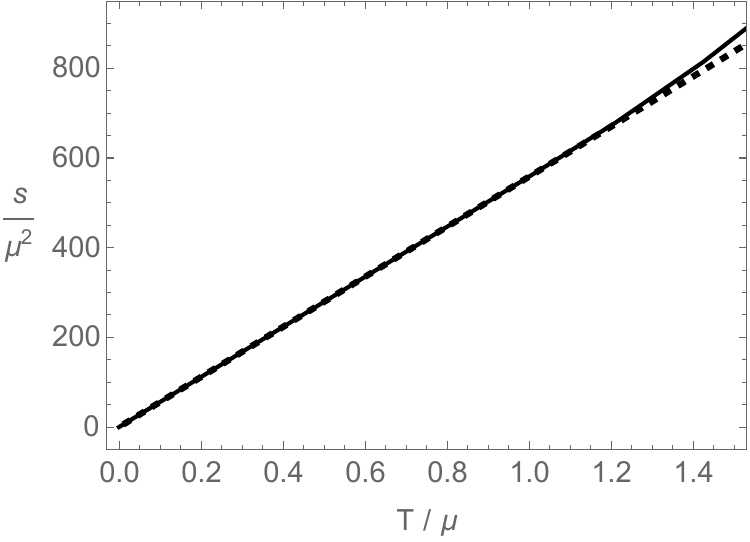} \label{}}
 \caption{The DC resistivity ($\rho$) and thermal entropy density ($s$) of the Gubser-Rocha model when $\beta/\mu = 10$. The solid lines represent results obtained from the exact expressions in \eqref{confor}, while the dashed lines correspond to the approximate linear expression in \eqref{linearrho}. The slopes of the linear fits are numerically determined to be $0.87$ for resistivity and $558$ for entropy density, respectively.}\label{sGRfigure2}
\end{figure}

While $s\approx T$ at low temperatures is also common for ordinary metals, strange metals typically exhibit much larger coefficients, suggesting unusual electronic properties. Interestingly, the Gubser-Rocha model demonstrates that $\gamma_{s}$ grows with $\beta/\mu$. This feature may support the applicability of the Gubser-Rocha model for describing the incoherent regime of strange metals.
For an illustration of $T$-linear resistivity and linear specific heat in the Gubser-Rocha model under strong momentum relaxation, see Fig. \ref{sGRfigure2}.
%

%
\subsubsection{Deep learning dilaton potentials}\label{}
We proceed by outlining the implementation of our machine learning methodology, PINNs, within our holographic framework. To incorporate PINNs, we begin by preparing the relevant equations of motion, as expressed in Eq.~\eqref{PINNODEEQ}. Next, we represent all fields appearing in the equations of motion, $\mathcal{F}(r)$, using a deep neural network as defined in Eq.~\eqref{DNN}. The equations are then reformulated into the form of Eq.~\eqref{DNNEQ} to construct the physics loss, $L_{\text{eom}}$ as in Eq.~\eqref{PINNLossEom}. This physics loss is combined with the data loss, $L_{\text{data}}$, to define the total loss function $L$. Finally, we train the deep neural network by minimizing this total loss function.

\paragraph{Algorithm for implementing PINNs.}
Applying the above algorithm to our problem would be  
\begin{enumerate}
\item{\textbf{Prepare the equations of motion:} Begin by formulating the equations of motion, Eq.~ \eqref{Ourmodeleq}, that will be incorporated into the loss function.}
\item{\textbf{Replace fields and potentials with neural networks:} Replace all fields and potentials $\{f(r),\, h(r),\, A_t(r),\, \phi(r),\, V(\phi),\, Z(\phi)\}$ with corresponding deep neural networks, as defined in Eq.~\eqref{DNN}.\footnote{Note that $\{f(r),\, h(r),\, A_t(r),\, \phi(r)\}$ correspond to $\mathcal{F}$ and $\{V(\phi),\, Z(\phi)\}$ to $\Theta$ in Eq. \eqref{PINNODEEQ}, respectively.}}
\item{\textbf{Construct the physics loss:} Combine the reformulated fields and potentials with the equations of motion (EOM) to define the physics loss $L_{\text{eom}}$ as 
\begin{align}\label{LOSS1}
\begin{split}
L_{\text{eom}} = \frac{1}{N_{(r,T)}}\sum_{(r,T),i}\log\left(\cosh\left(|\text{EOM}_i(r,T)|\right)\right) \,,
\end{split}
\end{align}
where \texttt{Log-Cosh Loss} is used. Here, $i \in [1,4]$ corresponds to the four equations of motion, and $N_{(r,T)}=100$, reflecting the $100$ sampled values of $T/\mu$, with $r\in[0,\,1]$ randomly assigned for each $T/\mu$ in each training iteration.\footnote{To avoid the overflow error in the numerical computation, it is also convenient to rescale $|\text{EOM}_i|$ in Eq. \eqref{LOSS1} as $\mathcal{N} \tanh{\frac{|\text{EOM}_i|}{\mathcal{N}}}$ with some numerical factor $\mathcal{N}$, which does not change any physics.}
}
\item{\textbf{Define the data loss:} Construct the data loss, $L_{\text{data}}$, to encode the $T$-linear resistivity as 
\begin{align}\label{LOSS2}
\begin{split}
L_{\text{data}} = \frac{1}{N_{T}}\sum_{T} |\rho(T/\mu) - \rho_{\text{data}}(T/\mu)| \,,
\end{split}
\end{align}
where the L1 loss is employed. Here, $\rho$ is defined in Eq.~\eqref{confor},  $\rho_{\text{data}}$ represents the given boundary data, and $N_{T}=100$ indicates the number of temperature samples.}
\item{\textbf{Train the neural network:} Minimize total loss, $L=L_{\text{data}}+L_{\text{eom}}$, to train the neural network.\footnote{In our numerical computations, we adopt the common rescaling weight in the loss function as $L= w_{1}\, L_{\text{data}} + w_{2} \, L_{\text{eom}}$ where $w_{2}/w_{1}$ is varied from $10^{-5}$ to $10^{3}$. We also use the cyclical learning rates~\cite{Smith:2015aa,Smith:2017aa}.} Additionally, include $\beta/\mu$ as a training parameter $\Theta$, as outlined in Eq.~ \eqref{DNNEQ}. As such, the bulk dilaton potentials, represented using deep neural networks, are determined to correspond to the specified $T$-linear resistivity.}
\end{enumerate}
This step-by-step process ensures the integration of both physics-based and data-driven constraints into the PINNs framework for effective training and results.

\paragraph{Ansatz with the deep neural networks.}
We now provide additional details regarding the deep neural networks utilized in step 2. For the fields, we adopt the following ansatz
\begin{align}\label{Ourmodelfields}
\begin{split}
f(r) &:= (1-r) - r(1-r)(1 - 4\pi T) + r(1-r)^2 \mathcal{D}_f(r) \,, \\
h(r) &:= r^2 \frac{s}{4\pi} + (1-r)\big(1+r+rf'(0)\big) + r^2 (1-r) \mathcal{D}_h(r) \,, \\
A_t(r) &:= (1-r)\mu + r(1-r) \mathcal{D}_{A_t}(r) \,, \\
\phi(r) &:= r \mathcal{D}_{\phi}(r) \,,
\end{split}
\end{align}
where $\mathcal{D}$'s are distinct deep neural networks, each comprising three layers with 20 nodes per layer. This ansatz, as defined in Eq.~\eqref{Ourmodelfields}, is constructed to naturally satisfy the AdS boundary conditions
\begin{align}\label{Ourmodelcond}
\begin{split}
f(0) = h(0) = 1 \,, \quad A_t(0) = \mu \,, \quad \phi(0) = 0 \,,
\end{split}
\end{align}
as well as horizon conditions
\begin{align}\label{Ourmodelcond2}
\begin{split}
f(1) = 0 \,, \quad A_t(1) = 0 \,,\quad T = \frac{-f'(1)}{4\pi} \,, \quad s = 4\pi \, h(1) \,,
\end{split}
\end{align}
where $f(1) = 0$ ensures the formation of a black hole, $A_t(1) = 0$ enforces regularity at the horizon, and the remaining two conditions are Eq.~\eqref{TDC}. One can also check that the equations of motion \eqref{Ourmodeleq}, combined with the UV completion specified in Eq. \eqref{ASYMPO}, ensure that the condition $h'(0)=f'(0)$ is satisfied, as imposed in our ansatz \eqref{Ourmodelfields}.

For the dilaton potentials, we adopt the following ansatz
\begin{align}\label{Ourmodelpotentials}
\begin{split}
        V(\phi) := 6 - \frac{1}{2} m^2 \phi^2 + \phi^3 \mathcal{D}_V(\phi) \,, \qquad
        Z(\phi) := 1 + \phi \mathcal{D}_Z(\phi)  \,,
\end{split}
\end{align}
where $\mathcal{D}_{V}$ and $\mathcal{D}_{Z}$ are deep neural networks with three layers of 10 nodes each. This construction ensures compatibility with the UV completion given in Eq.~\eqref{ASYMPO}. For simplicity, we set $m^2 = -2$ in this work.

\paragraph{Training data set.}
For the training data, $L_{\text{data}}$, we primarily use the $T$-linear resistivity data $\{T/\mu,\, \rho\}$, as well as the linear specific heat data $\{T/\mu,\, s/\mu^2\}$. To incorporate this in our computations, we set $\mu=1$ and select a temperature range of $T/\mu \in (0.3,\, 0.85)$, which captures both the linear resistivity and specific heat behaviors.\footnote{The $T$-linear resistivity data, $\rho_{\text{data}}$, does not extend below $T/\mu = 0.286 \approx 0.3$ due to numerical instability.}

Specifically, as outlined in the introduction, the training data for the resistivity is modeled as 
\begin{align}\label{RESISDATA}
\begin{split}
\rho_{\text{data}} = \rho_0 + A_{\text{data}} \, \frac{T}{\mu} \,, \qquad  A_{\text{data}} :=  \bar{A} \times A_{\text{GR}} \,,
\end{split}
\end{align}
where $A_{\text{GR}}$ is the slope of the linear resistivity for the Gubser-Rocha model, given in Eq.~\eqref{linearrho}. We examine the effect of the slope $\bar{A}$ (when it reduces to those of Gubser-Rocha model when $\bar{A}=1$) and the residual resistivity $\rho_0$ on the dilaton potentials. To focus on the $T$-linear resistivity, we fix the specific heat data based on the Gubser-Rocha model, as shown in the right panel of Fig. \ref{sGRfigure2}. The effect of the slope in the linear specific heat will be considered in future work.

%
\subsection{Taming $T$-linear resistivity}\label{sec32}

%
\subsubsection{Re-discovering the Gubser-Rocha model}\label{}
Utilizing the machine learning framework outlined in Sec \ref{sec31}, we proceed with the training process. Initially, the $T$-linear resistivity data \eqref{RESISDATA} with $\rho_0=0$ is employed to investigate the dependence of the slope $\bar{A}$ on the dilaton potentials.

\begin{figure}[]
  \centering
     {\includegraphics[width=7cm]{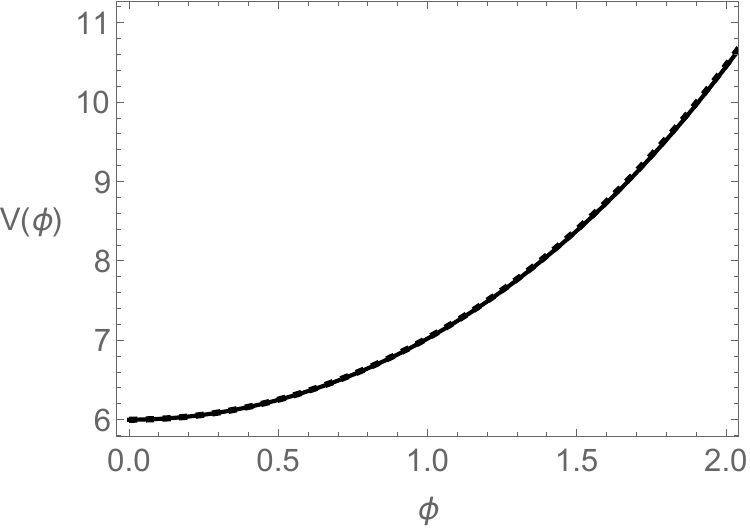} \label{}}
\quad
     {\includegraphics[width=7cm]{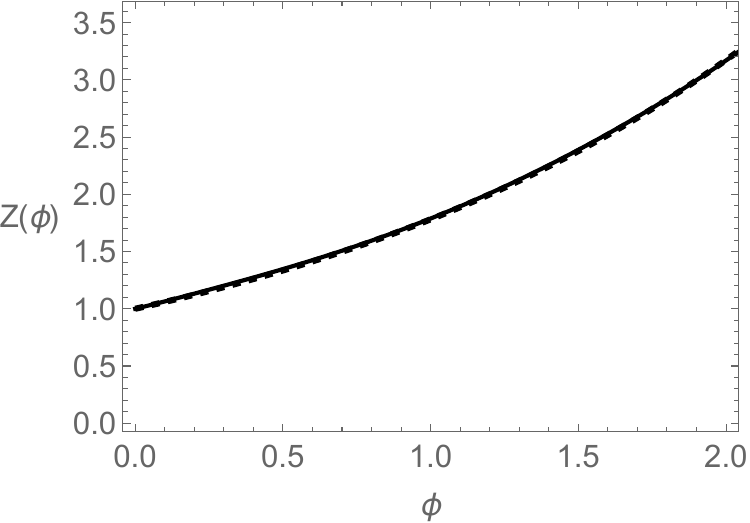} \label{}}
 \caption{Trained dilaton potentials $\{V(\phi),\, Z(\phi)\}$ derived from the $T$-linear resistivity data \eqref{RESISDATA} with $\rho_0=0$ and $\bar{A}=1$. Solid lines indicate the PINNs-trained potentials, while dashed lines correspond to the Gubser-Rocha model \eqref{GRaction}.}\label{MLFIG1}
\end{figure}

For $\bar{A}=1$, corresponding to the $T$-linear resistivity of the Gubser-Rocha model shown in Fig. \ref{sGRfigure2}, our PINNs implementation successfully reconstruct the dilaton potentials of the Gubser-Rocha model $\{V(\phi),\, Z(\phi)\}$ given in \eqref{GRaction}. The results are presented in Fig. \ref{MLFIG1}, where the solid lines represent the trained dilaton potentials obtained from PINNs, and the dashed lines denote the known dilaton potentials of the Gubser-Rocha model.

\begin{figure}[]
 \centering
     {\includegraphics[width=15cm]{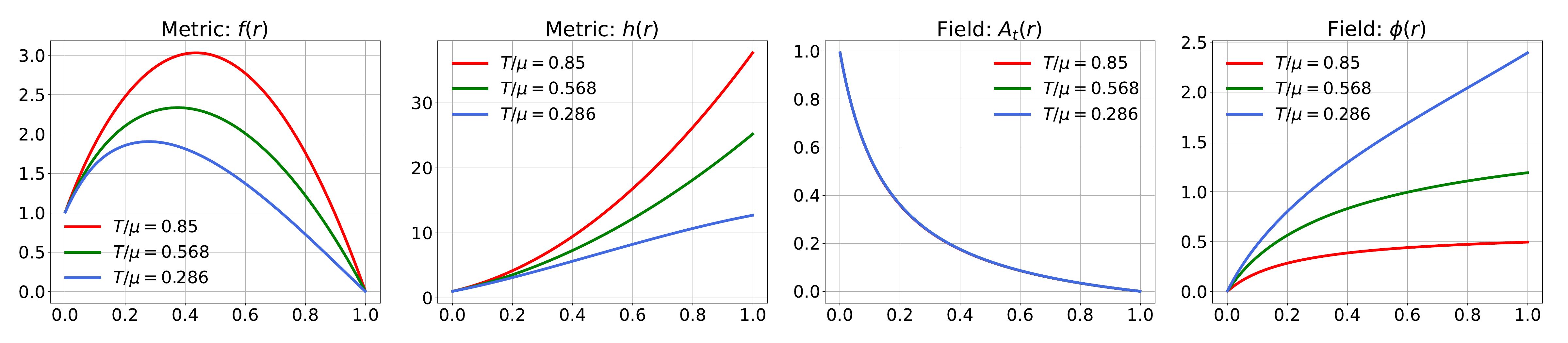} \label{}}
 \vspace{-0.5cm}
 \caption{Trained fields $\{f(r),\, h(r),\, A_t(r),\, \phi(r)\}$ obtained from PINNs at various temperatures with $\rho_0=0$ and $\bar{A}=1$. While the $A_{t}$ profiles appear nearly identical for different temperatures, they differ numerically with a maximum deviation on the order of $0.0015$. These fields agree with the Gubser-Rocha solution \eqref{GRSOL} in the coordinate system where $\mu=1$.}\label{MLFIG2}
\end{figure}

Additionally, the corresponding fields $\{f(r),\, h(r),\, A_t(r),\, \phi(r)\}$ derived from PINNs are shown in Fig. \ref{MLFIG2}. These fields represent the emergent holographic spacetime geometry reconstructed from the $T$-linear resistivity data at the boundary.
Notably, the reconstructed fields agree with the Gubser-Rocha solution \eqref{GRSOL} within the coordinate system set by $\mu=1$~\cite{Jeong:2018tua,Ahn:2019lrh,Jeong:2021wiu,Ahn:2023ciq}. Furthermore, the qualitative behavior of these fields remains similar for other parameter choices of $\left(\rho_0,\, \bar{A}\right)$: for instance, see Fig. \ref{MLFIG5} and \ref{MLFIG9}, where the deviations in $A_t$'s are more pronounced compared to those, Fig. \ref{MLFIG2}, observed in the Gubser-Rocha model.

\begin{figure}[]
  \centering
     {\includegraphics[width=7cm]{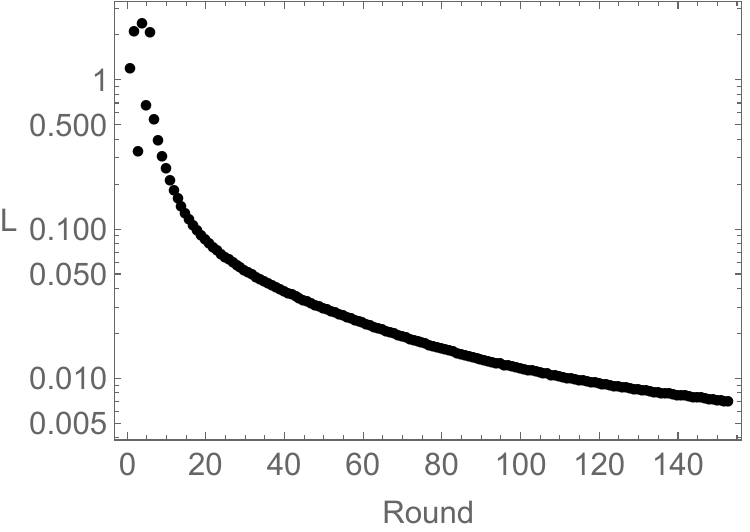} \label{}}
\quad
     {\includegraphics[width=7cm]{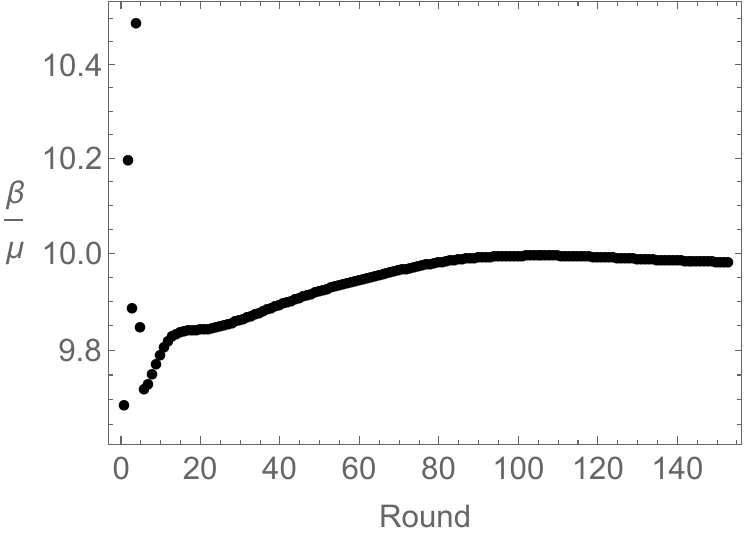} \label{}}
 \caption{The left panel depicts the total loss function for $\rho_0=0$ and $\bar{A}=1$. Here, \texttt{Round} corresponds to $20000$ epochs (or $10000$ epochs for the first six data points). The loss after training, $\texttt{Round}=150$, is $L=0.00698$. The right panel shows the trained value of $\beta/\mu$, which converges to $\beta/\mu = 9.982$, closely matching the value of $10$ used in Fig. \ref{sGRfigure2}.}\label{MLFIG3}
\end{figure}

We also evaluate the loss function $L$, constructed using \eqref{LOSS1} and \eqref{LOSS2}, as determined by our machine learning approach. Across all datasets considered in this manuscript, the value of $L$ consistently stabilizes around $10^{-3}$ after training. For instance, in the case of the Gubser-Rocha model, the left panel of Fig. \ref{MLFIG3} illustrates the total loss function behavior. The right panel of Fig. \ref{MLFIG3} shows the trained $\beta/\mu = 9.982$, which closely approximates the value of $10$ used in Fig. \ref{sGRfigure2}.
%

%
\subsubsection{Deriving dilaton potentials from $T$-linear resistivity}\label{}
\begin{figure}[]
  \centering
     {\includegraphics[width=7cm]{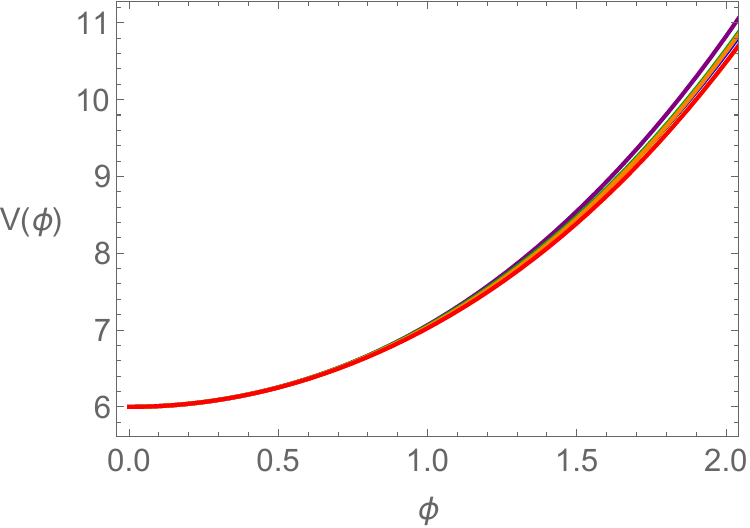} \label{}}
\quad
     {\includegraphics[width=6.8cm]{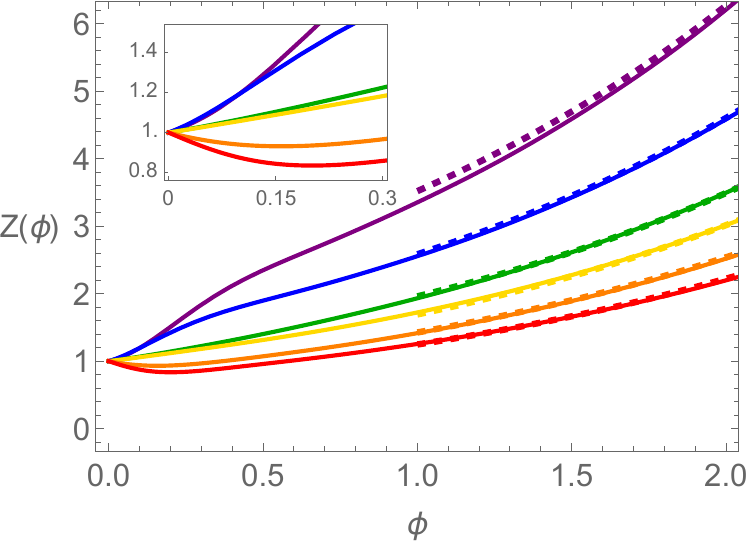} \label{}}
 \caption{Trained dilaton potentials $\{V(\phi),\, Z(\phi)\}$ derived from $T$-linear resistivity data \eqref{RESISDATA} with $\rho_0=0$ and $\bar{A}=(1.5,\, 1.3,\, 1.1,\, 0.9,\, 0.7,\, 0.5)$ (red, orange, yellow, green, blue, purple). The dashed lines in the right panel correspond to the fitting curves \eqref{EXPOFAC}.}\label{MLFIG4}
\end{figure}

\begin{figure}[]
 \centering
     {\includegraphics[width=15cm]{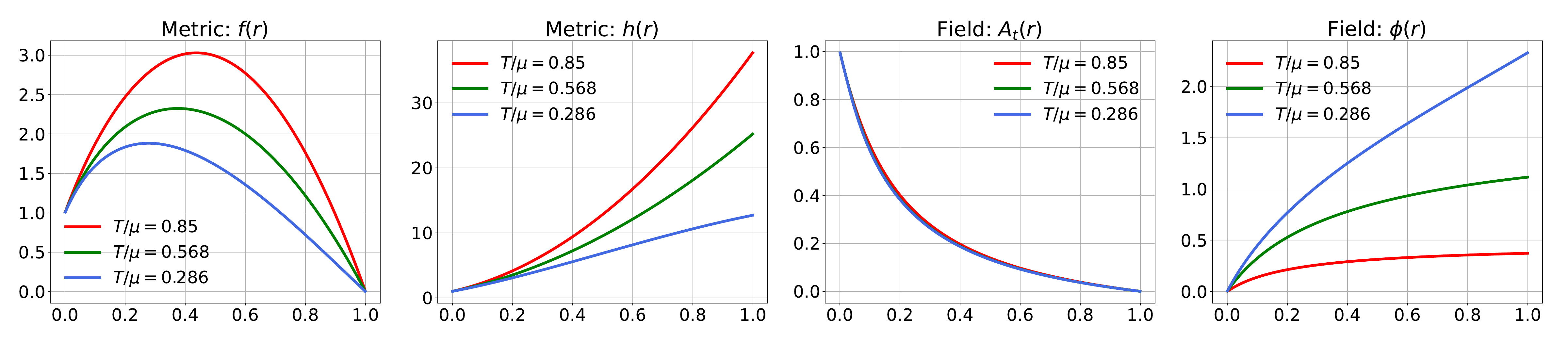} \label{}}
 \vspace{-0.5cm}
 \caption{Trained fields $\{f(r),\, h(r),\, A_t(r),\, \phi(r)\}$ obtained from PINNs at various temperatures with $\rho_0=0$ and $\bar{A}=1.5$.}\label{MLFIG5}
\end{figure}

We now discuss the slope dependence for the case where $\bar{A} \neq 1$. The trained dilaton potentials obtained using PINNs are shown in Fig. \ref{MLFIG4}, along with representative trained fields for $\bar{A} = 1.5$ in Fig. \ref{MLFIG5}. The total loss and the trained value of $\beta/\mu$ for each $\bar{A}$ is summarized in Table \ref{TABLE1}.
\begin{table}[h]
\centering
\begin{tabular}{cccccccc}
    $\bar{A}$      & 1.5 & 1.3 & 1.1 & 0.9 & 0.7  & 0.5    \\
\hline
\hline
Loss $L$ & 0.0085 & 0.0068 & 0.0099 & 0.0090 & 0.0087 & 0.0060  \\
$\beta/\mu$ & 9.86 & 9.92 & 9.83 & 10.08 & 9.99  & 10.17   
\end{tabular}
\caption{The total loss $L$ and $\beta/\mu$ after training with $\rho_0=0$.}\label{TABLE1}
\end{table}

\begin{figure}[]
  \centering
     {\includegraphics[width=6.5cm]{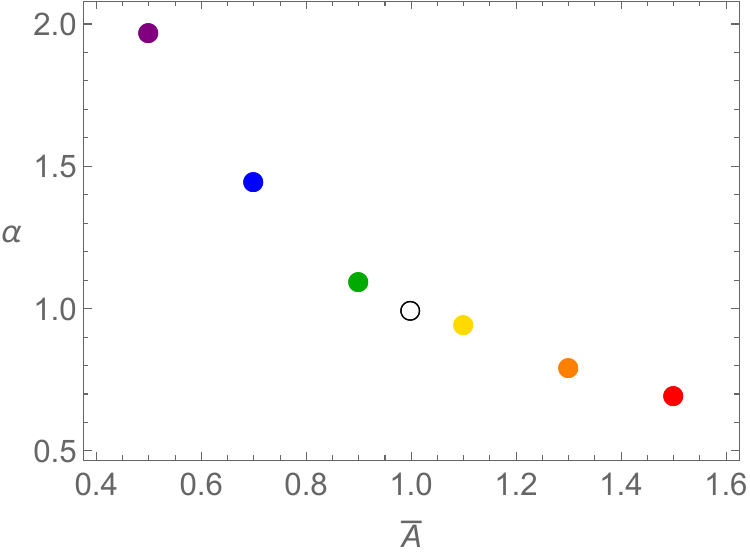} \label{}}
\quad
     {\includegraphics[width=7.1cm]{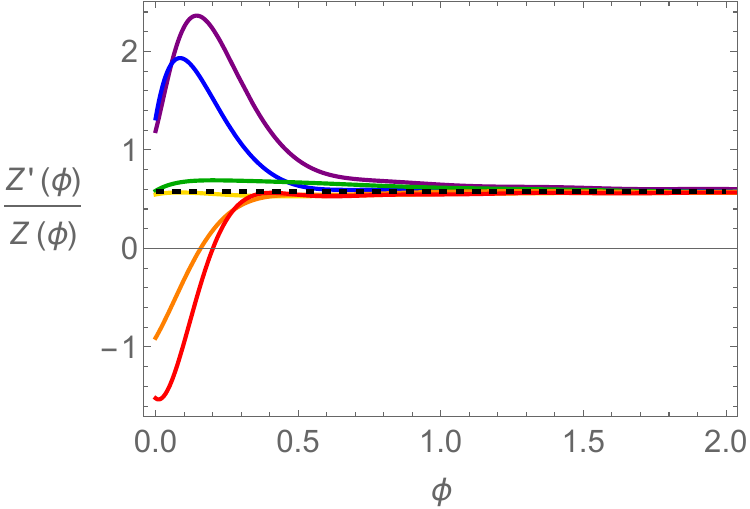} \label{}}
 \caption{The left panel shows $\alpha$ from \eqref{EXPOFAC} with the black empty dot representing the Gubser-Rocha model. The right panel displays the exponential factor $\gamma$ in \eqref{EXPOFAC} at large $\phi$ regime for $\bar{A}=(1.5,\, 1.3,\, 1.1,\, 0.9,\, 0.7,\, 0.5)$ (red, orange, yellow, green, blue, purple). The black dashed line corresponds to the Gubser-Rocha model value $1/\sqrt{3}\approx0.57735$. }\label{MLFIG6}
\end{figure}

\begin{figure}[]
  \centering
     {\includegraphics[width=6.6cm]{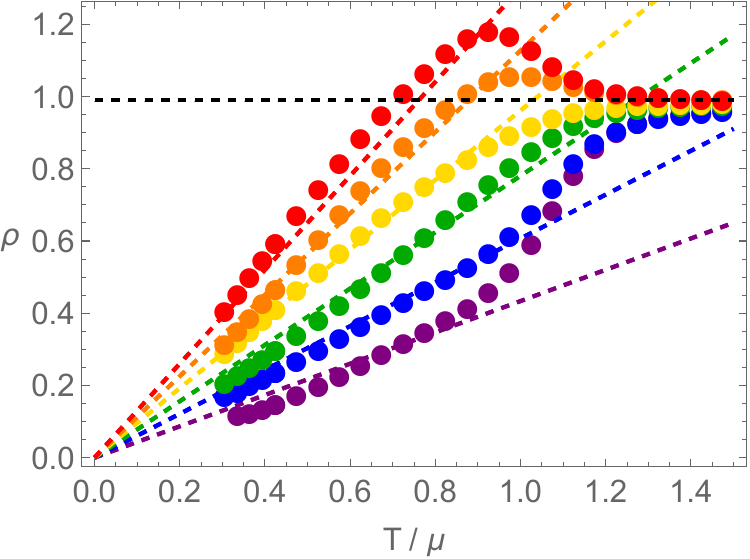} \label{}}
\quad
     {\includegraphics[width=7cm]{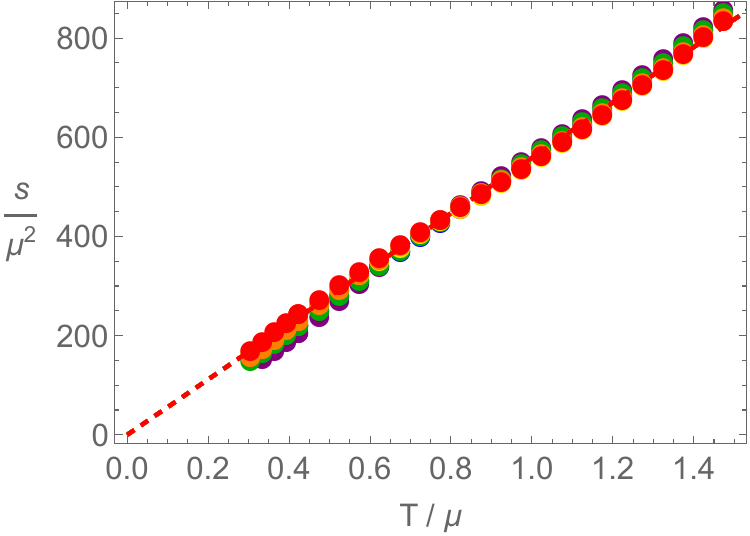} \label{}}
 \caption{Resistivity $\rho$ and thermal entropy $s$ with $\rho_0=0$ and $\bar{A}=(1.5\,,1.3\,,1.1\,,0.9\,,0.7\,,0.5)$ (red-purple). Colored dashed lines represent $T$-linear resistivity \eqref{RESISDATA}, while the black dashed line corresponds to \eqref{LAMRE} with $\beta/\mu$ from Table. \ref{TABLE1}. The slopes of thermal entropy $s/\mu^2$ are almost identical as required by machine learning process.}\label{MLFIG7}
\end{figure}

From Fig. \ref{MLFIG4}, four observations about the $\bar{A}$ dependence can be made.
\begin{enumerate}
\item{\textbf{Similarity in $V(\phi)$:} The dilaton potential $V(\phi)$ remains nearly identical across different values of $\bar{A}$.}
\item{\textbf{Resistivity and $Z(\phi)$:} The behaviour of $Z(\phi)$ in Fig. \ref{MLFIG4} can be understood from the resistivity in Fig. 
 \ref{MLFIG7} by noting that (I) $\rho \approx 1/Z_H$ from \eqref{confor} at large $\beta$, and (II) $\phi_H:=\phi(1)$ decreases as $T$ increases from Fig. \ref{MLFIG5}. For example, for $\bar{A} = 1.5$ (red line), we see that the resistivity increases until the curve reaches the minimum value of $Z(\phi)$, with $T$ increasing from right to left on the curve in Fig. \ref{MLFIG4}. After the minimum (turning around point), $Z(\phi)$ approaches $1^{-}$, and equivalently, the resistivity decreases and saturates to $1^{+}$. This interpretation is valid for every value of $\bar{A}$.}
\item{\textbf{$Z(\phi)$ in the `small' $\phi$ limit:} The dilaton potential $Z(\phi) $ near the AdS boundary (small $\phi$) can be approximated as
\begin{align}\label{Z1FIT}
Z(\phi) \approx 1 + Z^{(1)} \phi + \cdots \,.
\end{align}
By using the same logic as item 2, we may interpret $Z^{(1)}$ is related with the slope near $T/\mu \sim 1.2$ in the left panel of Fig. \ref{MLFIG7}.}
\item{\textbf{Exponential growth of $Z(\phi)$ in the `large' $\phi$ limit:} In the infrared (large $\phi$ regime), $Z(\phi)$ grows exponentially as 
\begin{align}\label{EXPOFAC}
Z(\phi) \approx  \alpha \, e^{\gamma \, \phi}  \,,
\end{align}
with $\gamma$ being independent of $\bar{A}$ and equal to $1/\sqrt{3} \approx 0.57735$. This value can be determined from $Z'(\phi)/Z(\phi)$, as shown in the right panel of Fig. \ref{MLFIG6}.
On the other hand, we find $\alpha$ depends on $\bar{A}$: see the left panel of Fig. \ref{MLFIG6}.}
\end{enumerate}


These findings suggest that the linear $T$-dependence of resistivity (the power of $T$) originates from the IR geometry of the Gubser-Rocha model \eqref{IRCONFORMAL}. Meanwhile, the parameter $\bar{A}$ (the slope of $T$) primarily modifies the flow from the same IR fixed point (conformal to AdS$_2$) to the UV fixed point (AdS$_4$), affecting $Z^{(1)}$ and $\alpha$.

One notable advantage of PINNs is their ability to generalize ``beyond" the training data range, such as $T/\mu \in (0.3, 0.85)$. Using the trained dilaton potentials, we solved the equations with ODE solvers to predict the resistivity $\rho$ over an ``extended" range, $T/\mu \in (0.3, 1.5)$. Fig. \ref{MLFIG7} demonstrates that the trained potentials produce the resistivity behavior effectively.

The $T$-linear resistivity is consistent with the training range $T/\mu \in (0.3, 0.85)$ but transitions to a constant resistivity at high temperatures ($T/\mu \approx 1.5$), as described by 
\begin{align}\label{LAMRE}
\text{(High-temperature limit):} \qquad \rho = \frac{1}{\sigma_{DC}} \,,\qquad \sigma_{DC} = 1  + \frac{\mu^2}{\beta^2} \,,
\end{align}
shown by the black dashed line in the left panel of Fig. \ref{MLFIG7}. This behavior arises because the dilaton field $\phi$ becomes small at higher temperatures (see Fig. \ref{MLFIG5}), reducing the EMDA theory \eqref{action} to linear-axion model \cite{Vegh:2013sk,Andrade:2013gsa,Baggioli:2021xuv} giving a temperature-independent resistivity \eqref{LAMRE}.

In the intermediate range $T/\mu \in (0.85, 1.5)$, an intriguing overshooting behavior in resistivity is observed for larger $\bar{A}$ values (red-orange curves). This behavior appears related to another turnaround of $Z(\phi)$ in the small $\phi$ regime, as seen in the inset of Fig. \ref{MLFIG4}.

Finally, it is worth noting that the saturated resistivity at high temperatures is unrelated to the Mott-Ioffe-Regel limit. Instead, dimensional analysis shows that $\rho \approx T^{3-d}$ for $d$ spacetime dimensions.\footnote{The dimension of the electric field is $[E]=2$ and the dimension of a current density in ($d$) spacetime dimensions is $[J]=d-1$.} In our case, with $d = 3$, this implies a temperature-independent resistivity at high temperatures, consistent with Eq.~\eqref{LAMRE}.

%
\subsubsection{More on data-driven potentials: residual resistivity}\label{}
We now examine the influence of the residual resistivity ($\rho_0$) in \eqref{RESISDATA} on the dilaton potentials. For this analysis, we fix the slope parameter to $\bar{A}=0.5$ and observe that the qualitative behavior remains consistent for other values of $\bar{A}$.

\begin{figure}[]
  \centering
     {\includegraphics[width=7cm]{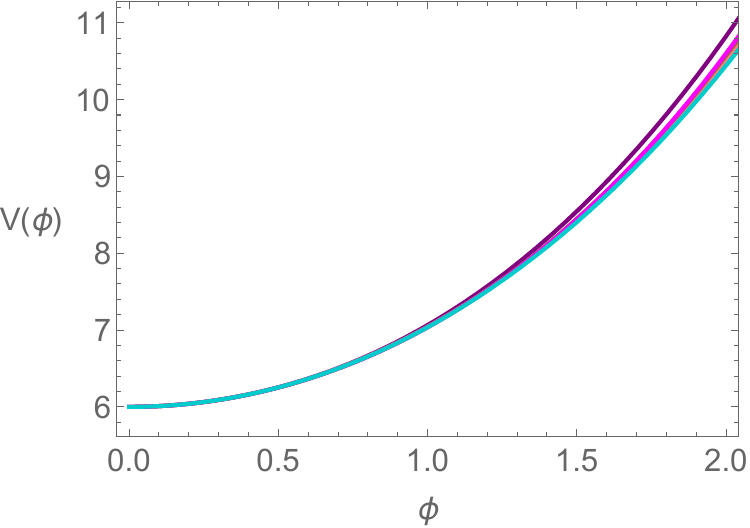} \label{}}
\quad
     {\includegraphics[width=6.8cm]{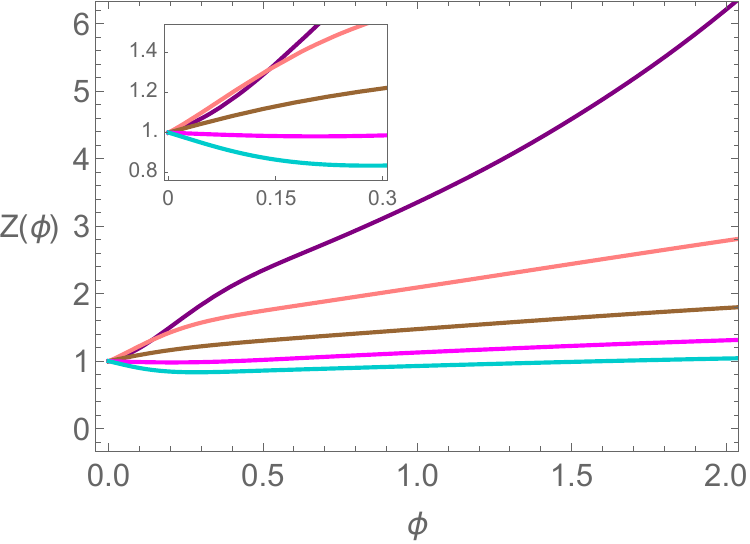} \label{}}
 \caption{Trained dilaton potential $\{V(\phi),\, Z(\phi)\}$ derived from $T$-linear resistivity data \eqref{RESISDATA} with $\rho_{0}=(0,\, 0.2,\, 0.4,\, 0.6,\, 0.8)$ (purple,\, pink,\, brown,\, magenta,\, cyan).}\label{MLFIG8}
\end{figure}

\begin{figure}[]
 \centering
     {\includegraphics[width=15cm]{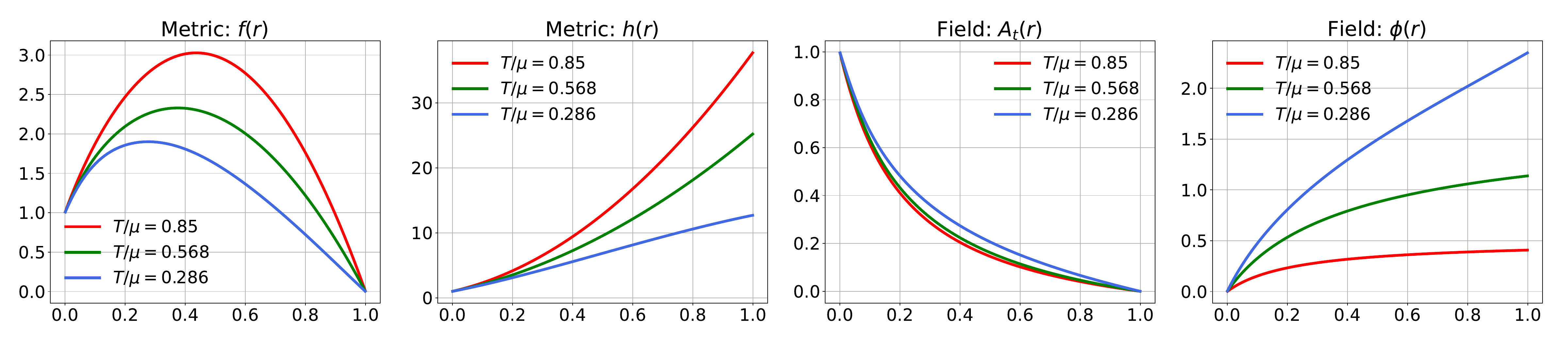} \label{}}
 \vspace{-0.5cm}
 \caption{Trained fields $\{f(r),\, h(r),\, A_t(r),\, \phi(r)\}$ obtained from PINNs at various temperatures with $\rho_0=0.8$.}\label{MLFIG9}
\end{figure}

The trained dilaton potentials obtained using PINNs are presented in Fig. \ref{MLFIG8}, with representative trained fields for $\rho_0 = 0.8$ shown in Fig. \ref{MLFIG9}. Table. \ref{TABLE2} summarizes the total loss and the trained value of $\beta/\mu$ for each $\rho_0$.
\begin{table}[h]
\centering
\begin{tabular}{cccccccc}
    $\rho_0$      & 0 & 0.2 & 0.4 & 0.6  & 0.8    \\
\hline
\hline
Loss $L$ & 0.0060  & 0.0060 & 0.0065 & 0.0050 & 0.0077 \\
$\beta/\mu$ & 10.17  & 9.94 & 9.92 & 10.00 & 9.90  
\end{tabular}
\caption{The total loss $L$ and $\beta/\mu$ after training when $\bar{A}=0.5$.}\label{TABLE2}
\end{table}

\begin{figure}[]
  \centering
     {\includegraphics[width=6.6cm]{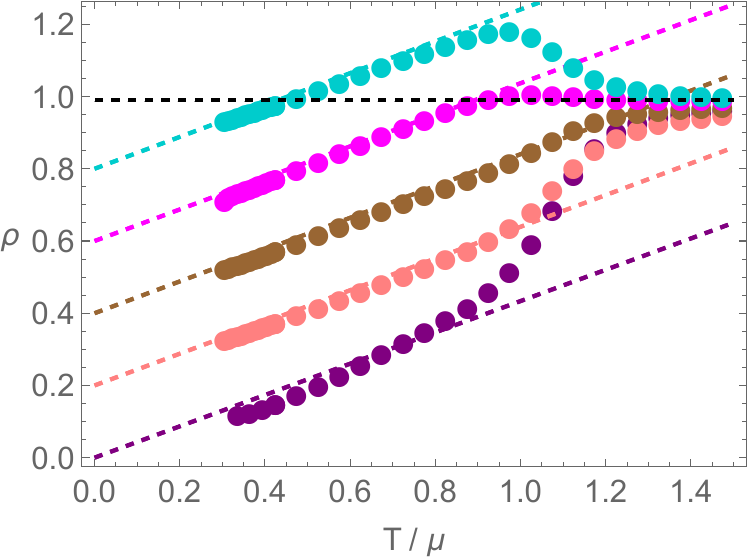} \label{}}
\quad
     {\includegraphics[width=7cm]{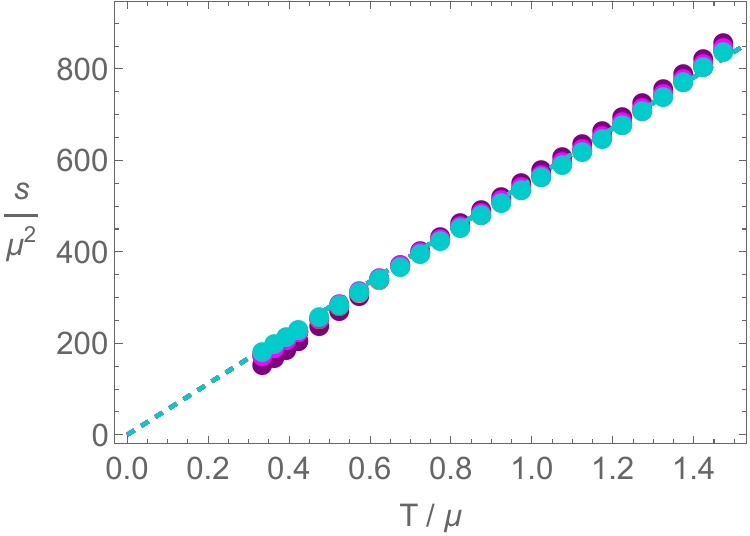} \label{}}
 \caption{Resistivity $\rho$ and thermal entropy $s$ with $\rho_{0}=(0,\, 0.2,\, 0.4,\, 0.6,\, 0.8)$ (purple,\, pink,\, brown,\, magenta,\, cyan). Colored dashed lines represent $T$-linear resistivity \eqref{RESISDATA}, while the black dashed line corresponds to \eqref{LAMRE} with $\beta/\mu$ from Table. \ref{TABLE2}. The slopes of thermal entropy $s/\mu^2$ are almost identical as required by machine learning process.}\label{MLFIG11}
\end{figure}

In essence, our findings indicate that the residual resistivity $\rho_0$ affects the dilaton potentials in a manner similar to $\bar{A}$, as follows: 
(I) the potential $V(\phi)$ remains nearly unchanged across different values of $\rho_0$;
(II) the slope of $Z(\phi)$ in the small $\phi$ limit shows dependence on $\rho_0$, which leads to the overshooting behavior in the resistivity, as illustrated in the left panel of Fig. \ref{MLFIG11}.
\begin{figure}[]
  \centering
     {\includegraphics[width=7.3cm]{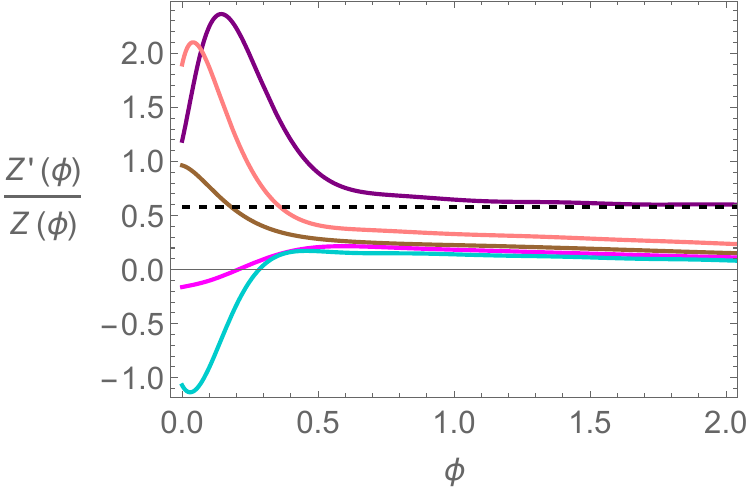} \label{}}
 \caption{$Z'(\phi)/Z(\phi)$ with $\rho_{0}=(0,\, 0.2,\, 0.4,\, 0.6,\, 0.8)$ (purple,\, pink,\, brown,\, magenta,\, cyan). The black dashed line corresponds to the Gubser-Rocha model value $\gamma = 1/\sqrt{3}\approx0.57735$.}\label{MLFIG10}
\end{figure}

However, there is a distinct feature that sets $\rho_0$ apart from $\bar{A}$. In the large $\phi$ limit, $Z(\phi)$ does not exhibit exponential growth as $\rho_0$ increases. Instead, the potential $Z(\phi)$ tends to ``flatten", as shown in the right panel of Fig. \ref{MLFIG8}. This flattening is further highlighted in Fig. \ref{MLFIG10}. We also find that at finite $\rho_0$, $Z(\phi)$ cannot be fitted with \eqref{EXPOFAC} at the large $\phi$ regime.

Before concluding this section, we leave two remarks regarding the residual resistivity. First, the flattening of $Z(\phi)$ at finite residual resistivity may be interpreted with the potential of the linear axion model. In this model, the resistivity is constant \eqref{LAMRE}, which can be viewed as a special case of residual resistivity where $Z(\phi)$ flattens to $Z(\phi)=1$.

Second, previous studies, such as \cite{Gouteraux:2014hca,Jeong:2018tua}, have investigated UV-completed holographic models that yield residual resistivity with analytic background solutions. However, the dilaton potentials in those models differ from ours, particularly because they exhibit a non-vanishing residual specific heat. In contrast, our model retains a linear specific heat, as shown in the right panel of Fig. \ref{MLFIG11}.\\

For all the data-driven dilaton potentials presented in this manuscript, including those shown in Fig. \ref{MLFIG4} and Fig. \ref{MLFIG8}, we find that they can be represented by the functional form:
\begin{align}\label{}
\begin{split}
\mathcal{F}(\phi) = \sum_{i=0}^{c_{\text{max}}} c_{i}  \, \phi^{i} \,,
\end{split}
\end{align}
where $\mathcal{F}$ represents $V$ or $Z$, and $c_{\text{max}}=12$. This polynomial expression may prove valuable for future analyses. However, to maintain clarity and brevity, we do not include the specific numerical coefficients in this manuscript. Readers seeking the complete expressions can access them through the GitHub repository provided at \href{https://github.com/sicobysico/ML_Linear-T-Resistivity}{this link}.

%
\subsubsection{Deep learning-assisted UV completion}\label{}
One central objective in holographic duality is to determine the bulk action that best describes a given set of boundary data. Machine learning provides a valuable tool for solving such inverse problems. Here, we explore how deep learning can assist in obtaining UV-completed potentials from a given $T$-linear resistivity.

Our deep learning analysis, discussed in previous sections, suggests that within the EMDA framework \eqref{action}, the minimal conditions for UV-completed potentials are
\begin{align}\label{UVCOMCON1}
V(\phi) = 6 \cosh \frac{\phi}{\sqrt{3}} \,, \qquad
Z(\phi) \approx  
\begin{cases}
\alpha(\bar{A}) \,\, e^{\frac{\phi}{\sqrt{3}}}  \quad\qquad\,\,\,\, (\text{low} \,T \,\,\,\text{or}\,\,\, \text{large} \,\phi) \,, \\
1 + Z^{(1)} \phi + \cdots  \quad (\text{high} \,T \,\,\,\text{or}\,\,\, \text{small} \,\phi) \,.
\end{cases}
\end{align}
Here, $V(\phi)$ is found by the fact that it remains largely unaffected by the slope of the $T$-linear resistivity and corresponds to the potential of the Gubser-Rocha model \eqref{GRaction}. Additionally, recall the relevant expressions in \eqref{Z1FIT} and \eqref{EXPOFAC}.

The key insight is that the boundary data --specifically, the slope $\bar{A}$ and the $T$ dependence of resistivity (linearity)-- determines the functional form of $Z(\phi)$ at low $T$. Meanwhile, the extent to which $T$-linear resistivity remains robust at higher $T$ dictates the approach to $Z(\phi=0) = 1$, including the coefficient $Z^{(1)}$ and higher-order corrections, in other words,
\begin{align}\label{}
Z(\phi) \approx  
\begin{cases}
\alpha(\bar{A}) \,\, e^{\frac{\phi}{\sqrt{3}}}  \quad\qquad\,\,\,\, (\text{low} \,T \,\,\,\,\text{: slope and linearity}) \,, \\
1 + Z^{(1)} \phi + \cdots  \quad (\text{high} \,T \,\,\text{: robustness of $T$-linear resistivity}) \,.
\end{cases}
\end{align}

\begin{figure}[]
  \centering
     {\includegraphics[width=7cm]{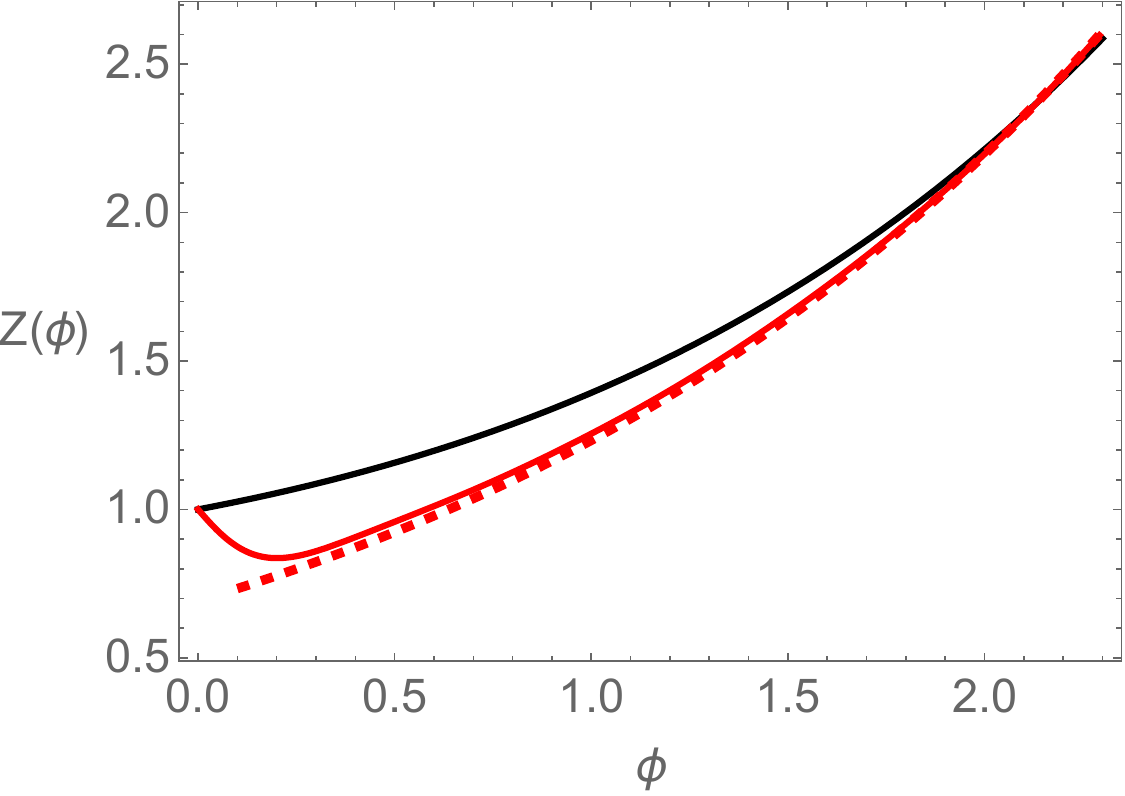} \label{}}
\quad
     {\includegraphics[width=6.7cm]{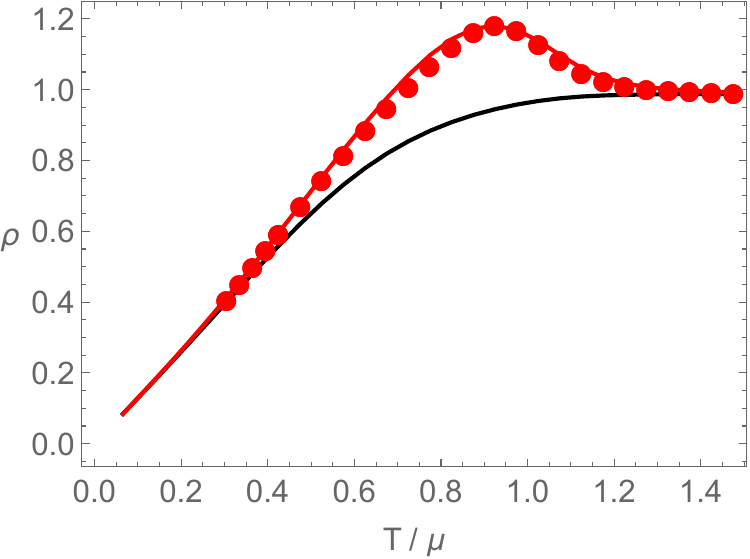} \label{}}
 \caption{$Z(\phi)$ potentials and resistivity for $\bar{A} = 1.5$. \textbf{Left panel:} the solid red line corresponds to \eqref{UVCOMCON2} with $\tilde{\alpha}=8$, while the black line represents $\tilde{\alpha}=0.7$. The red dashed line represents the fitting function \eqref{EXPOFAC}. \textbf{Right panel:} the solid lines correspond to the respective potentials shown in the left panel, while dots are taken from Fig. \ref{MLFIG7}. The red data correspond to a robustness of $T$-linear resistivity up to $T/\mu \approx 0.85 \, (\rho \approx 1)$, while the black data correspond to $T/\mu \approx 0.4 \, (\rho \approx 0.5)$.}\label{MLFIGUV}
\end{figure}

With this understanding, we can systematically construct UV-completed potentials for a given $\bar{A}$ and a specified temperature range where $T$-linear resistivity remains robust. As an example, let us consider $\bar{A} = 1.5$. If the $T$-linear resistivity persists up to $T/\mu = 0.85$, the resulting potential and resistivity are shown in red in Fig. \ref{MLFIG4} and Fig. \ref{MLFIG7}. For clarity, we present the same data in Fig. \ref{MLFIGUV}. 
The left panel shows that the low-$T$ fitting function (dashed line), $\alpha(\bar{A}) \, e^{\frac{\phi}{\sqrt{3}}}$, approximates well $Z(\phi)$ until $\phi \approx 0.8$, where $Z(\phi) \approx 1$. This suggests that $T$-linear resistivity remains robust up to $\rho \approx 1/Z(\phi \approx 0.8) \approx 1$, a result confirmed in the right panel.

To construct a potential with the same slope ($\bar{A}=1.5$) but a reduced temperature range of robustness for $T$-linear resistivity, we consider that $\alpha(\bar{A}) \, e^{\frac{\phi}{\sqrt{3}}}$ (\textit{i.e.,} $T$-linear resistivity) starts to deviate when $Z(\phi)$ is larger (\textit{i.e.,} smaller $\rho$). For instance, if the deviation appears at $Z(\phi \approx 1.9) = 2$ (implying that $\alpha(\bar{A}) \, e^{\frac{\phi}{\sqrt{3}}}$ is a good approximation when $\phi\geq1.9$), we expect that $T$-linear resistivity remains robust only up to $\rho \approx 0.5$.

As a proof of concept, we provide a minimal deep learning-assisted UV completion, given by
\begin{align}\label{UVCOMCON2}
V(\phi) = 6 \cosh \frac{\phi}{\sqrt{3}} \,, \qquad
Z(\phi) = \alpha \,\, e^{\frac{\phi}{\sqrt{3}}} \,+\, \left[1-\alpha\right] \,\, e^{-\tilde{\alpha} \, \phi} \,,
\end{align}
which follows the asymptotic behaviors in \eqref{UVCOMCON1} with $\tilde{\alpha}>-1/\sqrt{3}$. Here, the parameter $\tilde{\alpha}$ controls the near AdS behavior with $Z^{(1)} = {\alpha}/{\sqrt{3}} + (\alpha-1) \, \tilde{\alpha}$ for the given slope of resistivity $\alpha$.

For the case $\bar{A}=1.5$, where $\alpha \approx 0.7$ (see Fig. \ref{MLFIG6}), we find that our potential obtained via deep learning (solid red line in the left panel of Fig. \ref{MLFIGUV}) can be well approximated by \eqref{UVCOMCON2} when $\tilde{\alpha}=8$.\footnote{The minimal model \eqref{UVCOMCON2} provides a good approximation to the potentials obtained from deep learning (see Fig. \ref{MLFIG4}) such as $\bar{A}=1.5, 1.3, 1.1, 0.9$. However, when the slope is too low, as in the cases of $\bar{A}=0.7$ and $0.5$, it becomes less accurate in the small $\phi$ limit.} By varying $\tilde{\alpha}$, different potentials can be constructed. For instance, choosing $\tilde{\alpha}=0.7$ (solid black line in the left panel of Fig. \ref{MLFIGUV}) reproduces the scenario described earlier, where deviation appears near $Z(\phi \approx 1.9) = 2$.

By numerically solving the equations of motion \eqref{Ourmodeleq} with these UV-completed potentials, we compute the resistivity: the right panel of Fig. \ref{MLFIGUV}. The results confirm that 
(I) the red solid line, corresponding to \eqref{UVCOMCON2} with $\tilde{\alpha}=8$, reproduces the resistivity profile obtained via our deep learning method (red dots),
(II) the black solid line demonstrates that $T$-linear behavior persists up to $\rho \approx 0.5$. Additionally, we find that both cases yield a linear specific heat.

%
\section{Conclusion}\label{sec4}
We have conducted a deep learning-based investigation into the $T$-linear resistivity within the framework of holography. Employing physics-informed neural networks (PINNs)~\cite{RAISSI2019686}, which incorporate both data loss and physics loss (the latter derived from differential equations as a form of regularization during the training process), we successfully deduce the dilaton potentials $V(\phi)$ and $Z(\phi)$ in the holographic Einstein-Maxwell-Dilaton-Axion (EMDA) theories described by Eq.\eqref{action}.

For the data loss, inspired by experimental observations of strange metals (see the discussion near Eqs. \eqref{EXPRES} and \eqref{EXPRES2}), we utilized PINNs to explore the role of $T$-linear resistivity \eqref{RESISDATA} in holography, alongside the linear specific heat characteristic of strange metal phases in cuprates. In particular, we examined the influence of varying the slope $\bar{A}$ (of $T$-linear resistivity) and residual resistivity $\rho_0$ (at $T=0$) on the data-driven dilaton potentials. Through this, we evaluated the affirmative potential of machine learning in holographic condensed matter theory to address a realistic physical problem: better understanding on a universal transport properties observed in strange metals.\footnote{In Appendix \ref{App_neuralode}, we validate all our findings using another machine learning approach known as Neural Ordinary Differential Equations~\cite{Chen:2018wjc}.}

Our key findings are summarized as follows:
\begin{itemize}
\item{When $\bar{A}=1$ and $\rho_0=0$, the PINNs rediscovered one of the most prominent holographic models exhibiting $T$-linear resistivity: the Gubser-Rocha model~\cite{Gubser:2009qt,Davison:2013txa,Jeong:2018tua}.}
\item{The dilaton potential $V(\phi)$ remains nearly identical by changes in either $\bar{A}$ or $\rho_0$. This invariance is illustrated in the left panels of Figs.~\ref{MLFIG4} and \ref{MLFIG8}.}
\item{In the large $\phi$ limit (low $T$ regime), the dilaton coupling to the Maxwell term $Z(\phi)$ exhibits exponential growth when $\rho_0=0$, $Z(\phi) \approx  \alpha \, e^{\gamma \, \phi}$, with the exponential factor $\gamma = 1/\sqrt{3}$ observed across all $\bar{A}$ values studied, while $\alpha$ depends on $\bar{A}$. This behavior indicates that the infrared (IR) geometry is fixed to be the conformal to AdS$_2 \times \mathbb{R}^2$ given in \eqref{CONADS2R2}, characteristic of a semi-local quantum liquid~\cite{Iqbal2012}, and is associated with the marginal Fermi liquid phenomenology developed to describe experimental results in cuprates~\cite{Varma:1989aa}. Conversely, for $\rho_0 \neq 0$, this exponential growth ceases, and $Z(\phi)$ flattens, as shown in the right panel of Fig.~\ref{MLFIG8} and Fig. \ref{MLFIG10}.}
\item{
In the small $\phi$ limit (high $T$ regime), $Z(\phi)$ can be expanded as $Z(\phi) = 1 + Z^{(1)} \phi + \cdots$, where the leading constant term ensures the asymptotic AdS boundary (\textit{i.e.,} UV completion). This leads to a constant resistivity at high $T$, given by $\rho_{\text{const}} := 1/(1+\mu^2/\beta^2)$. The subleading terms in $Z(\phi)$ can be ``adjusted" to control how robust the $T$-linear resistivity observed at low $T$ remains at higher $T$. In particular, the ratio $\rho/ \rho_{\text{const}}$ can approach unity either from above, $\rho/ \rho_{\text{const}} \rightarrow 1^{+}$ (a behavior referred to as overshooting), or from below, $\rho/ \rho_{\text{const}} \rightarrow 1^{-}$: see the illustration of it in Fig. \ref{MLFIGUV}.
}
\end{itemize}

Last but not least, we further comment on the $T$-linear resistivity at $\rho_0=0$ with varying $\bar{A}$. Our results demonstrate that machine learning can uncover robust mechanisms for strong momentum relaxation, producing $T$-linear resistivity consistent with IR scaling analyses. Specifically, one can check that within the EMDA theories \eqref{action}~\cite{Charmousis:2010zz, Davison:2013txa, Gouteraux:2014hca, Kim:2015wba, Zhou:2015dha, Ge:2016lyn, Cremonini:2016avj, Chen:2017gsl, Blauvelt:2017koq, Ahn:2017kvc, Jeong:2018tua, Ahn:2019lrh}, where the axion-dilaton coupling $\frac{Y(\phi)}{2}\sum_{i=1}^{2} (\partial \chi_i)^2$ is simplified to $Y(\phi)=1$, the corresponding IR geometry takes the form of the called $\eta$ geometry:
\begin{align} \label{}
\begin{split}
\dd s^2 = \xi^{-\eta} \left(\frac{- \dd t^2 + \mathcal{L}^2 \dd \xi^2}{\xi^2} + \sum_{i=1}^{2}\dd x_{i}^{2}\right)  \,,\qquad 
\mathcal{L}^2 := \frac{2(1+\eta)^2}{2 - \beta^2} \,.
\end{split}
\end{align}
Here, the scaling parameters satisfy
\begin{align}
\begin{split}
z \rightarrow \infty \,, \qquad \theta \rightarrow -\infty \,, \qquad {\theta}/{z} \rightarrow -\eta \,.
\end{split}
\end{align}
In this geometry, the resistivity and entropy scale at low $T$ as
\begin{align} \label{}
\begin{split}
\rho \approx T^{\,\eta} \,, \qquad s \approx T^{\,\eta} \,.
\end{split}
\end{align}
As such, imposing $T$-linear resistivity and linear specific heat as we did in machine learning process corresponds to $\eta=1$, recovering the IR geometry conformal to AdS$_2 \times \mathbb{R}^2$ given in \eqref{CONADS2R2}. This consistency demonstrates that the IR geometry is unaffected by variations in $\bar{A}$, as the slope changes while the temperature scaling exponent remains fixed at $\eta=1$. It underscores the agreement between our machine learning framework and the IR geometric analysis.

It is worth noting that, in our analysis, we did not impose exponential dilaton potentials in the IR regime, nor did we assume a logarithmically running dilaton behavior. These conditions have been key assumptions in previous holographic studies of $T$-linear resistivity \cite{Charmousis:2010zz, Davison:2013txa, Gouteraux:2014hca, Kim:2015wba, Zhou:2015dha, Ge:2016lyn, Cremonini:2016avj, Chen:2017gsl, Blauvelt:2017koq, Ahn:2017kvc, Jeong:2018tua, Ahn:2019lrh}, which rely on IR geometry analysis. More generally, it remains possible that alternative holographic mechanisms could account for $T$-linear resistivity without relying on IR scaling arguments. However, our machine learning approach suggests that, at least within the EMDA theories considered in Eq.~\eqref{action}, IR scaling analysis emerges as the most natural -- or potentially the only -- viable explanation for $T$-linear resistivity. Establishing this result is a nontrivial task.\\

With the demonstrated feasibility of data-driven machine learning models for capturing cuprate phenomenology, the next significant challenge lies in identifying a unified holographic framework capable of reproducing all relevant physical observables, including all the universal transport anomalies of strange metals~\cite{Phillips:2022nxs}. Achieving this requires a systematic comparison of various inversely-solved holographic models including the driven potentials presented in this work.

A particularly important goal in this direction is to identify the dual gravity theory that simultaneously explains $T$-linear resistivity and the Hall angle~\cite{Kim:2010zq,Blake:2014yla,Zhou:2015dha,Kim:2015wba,Chen:2017gsl,Blauvelt:2017koq,Ahn:2023ciq} in the presence of a finite magnetic field, as well as Home’s law in superconductors~\cite{Homes:2004wv,Zaanen:2004aa,Erdmenger:2015qqa,Kim:2015dna,Kim:2016hzi,Kim:2016jjk,Jeong:2021wiu}. 

\begin{figure}[]
  \centering
     {\includegraphics[width=4.8cm]{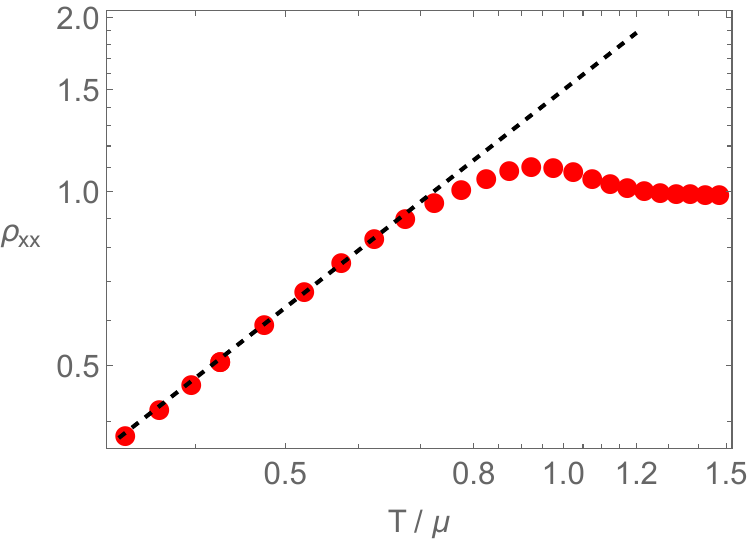} \label{}}
     {\includegraphics[width=4.8cm]{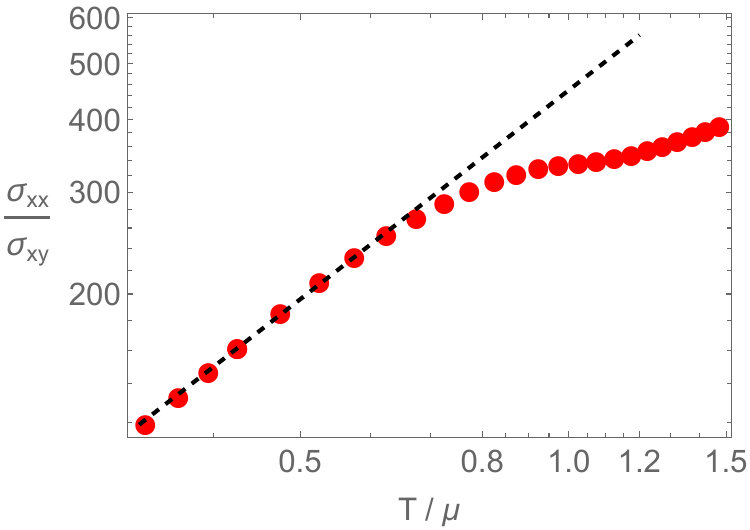} \label{}}   

     {\includegraphics[width=4.8cm]{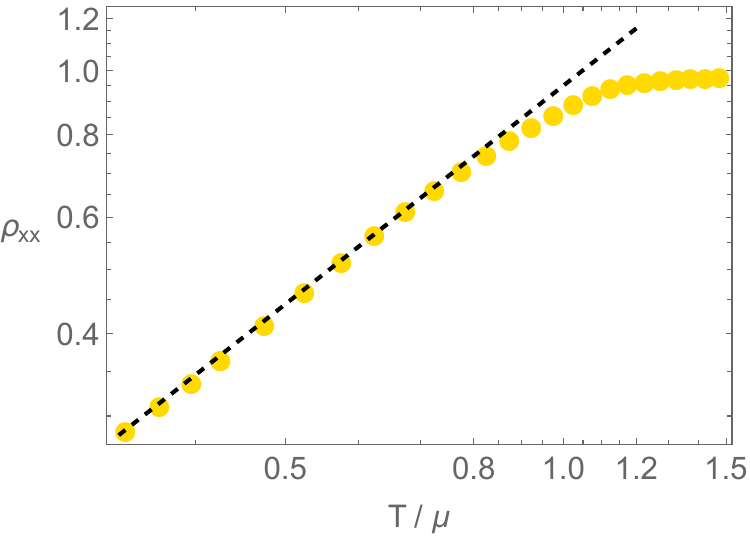} \label{}}
     {\includegraphics[width=4.8cm]{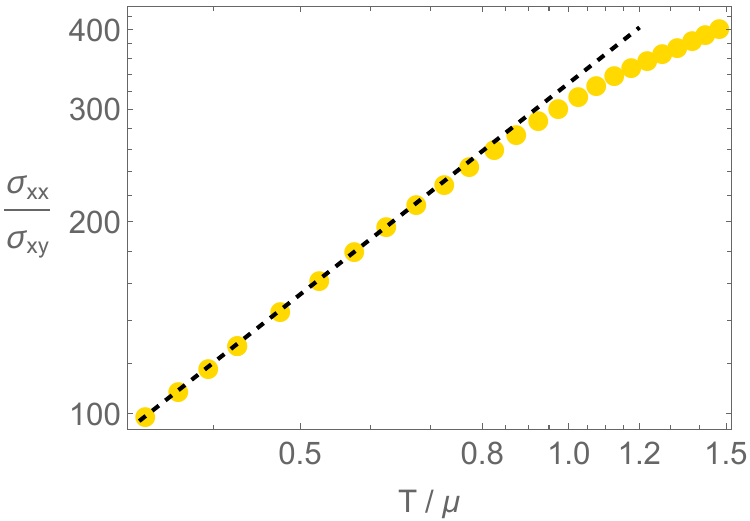} \label{}}   
     
     {\includegraphics[width=4.8cm]{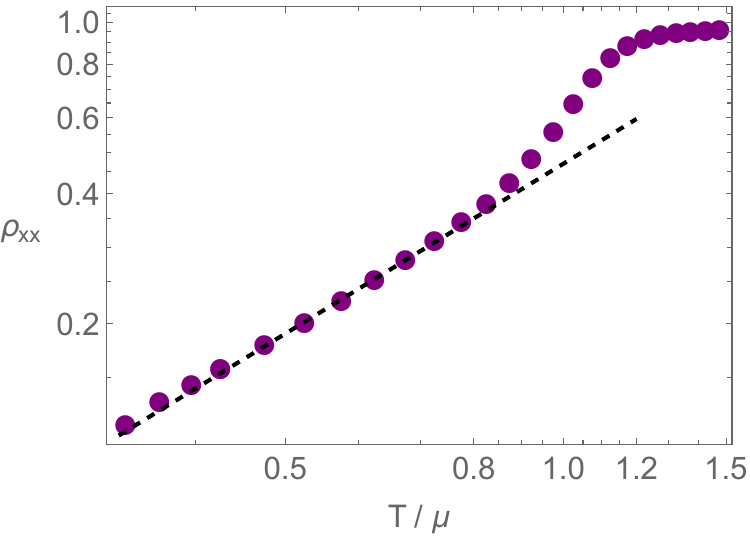} \label{}}
     {\includegraphics[width=4.8cm]{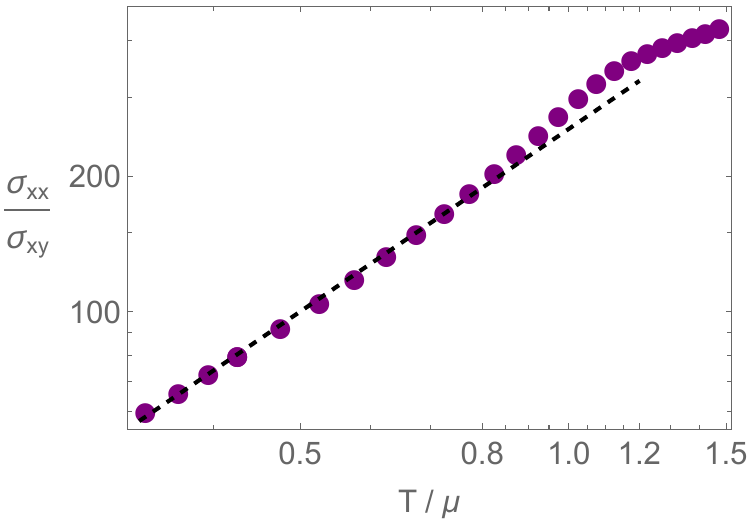} \label{}}
 \caption{Resistivity and Hall angle at finite magnetic field ($B/\mu^2=1$) with $\rho_0=0$ and $\bar{A}=(1.5\,,1.1\,,0.5)$ (red, yellow, purple). From the top to bottom panel, the fittings curves (dashed lines) are $(\rho_{xx},\,\sigma_{xx}/\sigma_{xy}) \approx\, $($T^{1.25},\, T^{1.2}$),\, ($T^{1.1},\, T^{1.1}$),\, ($T^{1.31},\, T^{1.35}$).}\label{MLFIG12}
\end{figure}

That said, it is tempting to explore whether the machine learning-generated dilaton potentials presented in this work can simultaneously exhibit $T$-linear resistivity and the Hall angle. To proceed, we adopt the results of~\cite{Blake:2014yla,Amoretti:2015gna,Blake:2015ina}, expressing the relevant DC conductivities in terms of horizon data. The corresponding expressions are
\begin{align}\label{DCFORHORI}
\begin{split}
\sigma_{xx} = \frac{h(1) \beta^2 \left( q^2 +B^2 Z_H^2 + h(1) \beta^2 Z_H \right)}{B^2 q^2 + \left( B^2 Z_H + h(1) \beta^2 \right)^2}  \,, \quad \,
\sigma_{xy} = \frac{B q \left( q^2 +B^2 Z_H^2 + 2 h(1) \beta^2 Z_H \right)}{B^2 q^2 + \left( B^2 Z_H + h(1) \beta^2 \right)^2}  \,.
\end{split} 
\end{align}
Here, the U(1) gauge field includes an external magnetic field, $B$, specified as $A=A_t(r) \dd t + B/2 (x \dd y - y \dd x)$. From these expressions, the longitudinal electric resistivity and the Hall angle are defined as
\begin{equation}
    \rho_{xx}=\frac{\sigma_{xx}}{\sigma_{xx}^2+\sigma_{xy}^2}, \qquad  \cot \Theta_H=\frac{\sigma_{xx}}{\sigma_{xy}} \,.
\end{equation}
By numerically solving the equations of motion for the EMDA theories with finite $B$, using our machine learning-generated dilaton potentials (trained to impose $T$-linear resistivity at $B=0$), we find that the resulting Hall angle $\cot \Theta_H$ does not scale as $T^2$. Instead, our analysis yields the fitting results shown in Fig. \ref{MLFIG12}, indicating deviations from the expected universal behavior ($\rho_{xx} \approx T$ and $\cot \Theta_H \approx T^2$).

This finding suggests that incorporating the Hall angle into the machine learning framework requires further refinement and detailed investigation. Future studies may benefit from proposed scenarios in~\cite{Ahn:2023ciq}, which discuss potential improvements for capturing the Hall angle within holographic models.
For additional context, refer to early studies on scaling behavior in systems with Lifshitz-hyperscaling critical points~\cite{Hartnoll:2015aa,Karch:2014mba,Karch:2015zqd,Amoretti:2016cad}.
Additionally, another well-known universality of high-$T_c$ superconductors, Home’s law~\cite{Homes:2004wv,Zaanen:2004aa,Erdmenger:2015qqa,Kim:2015dna,Kim:2016hzi,Kim:2016jjk,Jeong:2021wiu}, also warrants exploration within a machine learning-driven framework to uncover the underlying universal mechanisms. These directions represent promising avenues for future research, which we plan to address in forthcoming studies~\cite{wipMLteam}.

%
\acknowledgments
It is a pleasure to thank {Yongjun Ahn} for valuable discussions and correspondence.  
HSJ is supported by the Spanish MINECO ‘Centro de Excelencia Severo Ochoa' program under grant SEV-2012-0249, the Comunidad de Madrid ‘Atracci\'on de Talento’ program (ATCAM) grant 2020-T1/TIC-20495, the Spanish Research Agency via grants CEX2020-001007-S and PID2021-123017NB-I00, funded by MCIN/AEI/10.13039/501100011033, and ERDF A way of making Europe. 
B. Ahn was supported by Basic Science Research Program through the National Research Foundation of Korea funded by the Ministry of Education (NRF-2022R1I1A1A01064342).
This work was supported by the Basic Science Research Program through the National Research Foundation of Korea
(NRF) funded by the Ministry of Science, ICT \& Future Planning (NRF-2021R1A2C1006791) and the AI-based GIST Research Scientist Project grant funded by the GIST in 2025. This work was also supported by Creation of the
Quantum Information Science R\&D Ecosystem (Grant No. 2022M3H3A106307411) through the
National Research Foundation of Korea (NRF) funded by the Korean government (Ministry of
Science and ICT).
We would also like to thank the Institute for Theoretical Physics (IFT) Madrid for their hospitality during the program \textit{Quantum Gravity of Open Systems}, where parts of this work were undertaken.
All authors contributed equally to this paper and should be considered as co-first authors.

%
\appendix
\section{Neural ordinary differential equations}\label{App_neuralode}
Scientific Machine Learning (Scientific ML) is an emerging field with diverse applications across multiple disciplines. It combines the interpretability of traditional scientific structures, such as ordinary and partial differential equations (ODEs/PDEs), with the expressive power of neural networks. This integration enables the development of models that are both scientifically grounded and capable of addressing complex problems.  

The rapid growth of Scientific ML can be attributed to three widely adopted methodologies: Neural Ordinary Differential Equations (Neural ODEs)~\cite{Chen:2018wjc,Dupont:2019aa,Massaroli:2020aa,Yan:2019aa}, Universal Differential Equations~\cite{Rackauckas:2020aa,gmd-16-6671-2023,Teshima:2020aa,Bournez:2017aa}, and Physics-Informed Neural Networks (PINNs)~\cite{RAISSI2019686,Cai2021,Karniadakis_2021}. Each of these approaches highlights the synergy between data-driven techniques and fundamental scientific principles, paving the way for advancements in modeling and simulation.\\

In this section, we will verify the results presented in the main text (obtained by PINNs) using the Neural ODEs. It is the method of machine learning to find the unknown function $\Theta(r)$ of \eqref{PINNODEEQ} by engaging the deep neural network $\mathcal{D}(r)$ as the approximator for $\Theta(r)$.
The ODE can be rewritten as follows:
\begin{equation}
    \partial_r\mathcal{F} = \mathcal{H}(r,\mathcal{F};\mathcal{D}) \;.
\end{equation}
$\mathcal{F}$ is specified by the numerical methods for solving ODE. For example, Euler method yields the numerical solution by
\begin{equation}\label{ResNetEM}
    \mathcal{F}(r_N) = \mathcal{F}_{\text{ini}} + \sum_{n=0}^{N-1} \mathcal{H}(r_n,\mathcal{F}(r_n);\mathcal{D}(r_n))\cdot\Delta r \,.
\end{equation}
One can express \eqref{ResNetEM} in the general form given by 
\begin{equation}
    \mathcal{F}(r_{\text{fin}}) = \textrm{ODE\;Solver}\left[\mathcal{H};\, \mathcal{F}_{\text{ini}},\, (r_{\text{ini}},\,r_{\text{fin}});\, \mathcal{D}(r) \right] \,.
\end{equation}
After that, $\mathcal{D}(r)$ is optimized by minimizing the loss function consisting of the final condition and data:
\begin{equation}
    L = L_{\text{cond}} + L_{\text{data}} \,,
\end{equation}
where
\begin{equation}
    L_{\text{cond}} = |\mathcal{F}(r_{\text{fin}}) - \mathcal{F}_{\text{fin}}| \;.
\end{equation}

To prepare the equations for our ODE solver based on the \texttt{torchode} extension, we rewrite \eqref{Ourmodeleq} in terms of $f''(r)$, $h''(r)$, $A_t''(r)$, and $\phi''(r)$. The potentials $V(\phi)$ and $Z(\phi)$ are defined in the same form as \eqref{Ourmodelpotentials}. We have 6 shooting parameters: $f(1)$, $f'(1)$, $h(1)$, $A_t(1)$, $A_t'(1)$, and $\phi(1)$. Among these, some parameters are constrained by \eqref{Ourmodelcond2}, while the remaining are treated as trainable variables. Then, the loss function consists of 
\begin{equation}
    L_{\text{cond}} = |f(0) - 1| + |h(0) - 1| + |A_t(0) - \mu| + |\phi(0)| \;,
\end{equation}
which encodes the conditions from \eqref{Ourmodelcond}, along with the $L_{\text{data}}$ defined in \eqref{LOSS2}.

\begin{figure}[]
  \centering
     {\includegraphics[width=7cm]{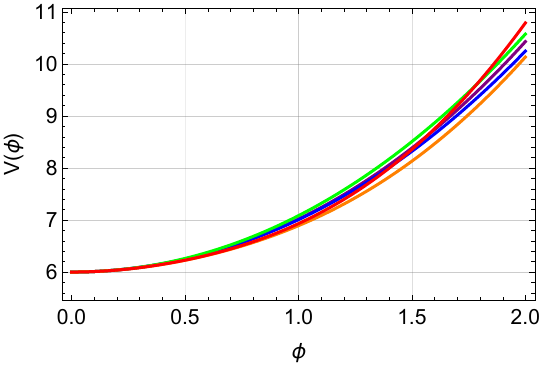} \label{}}
\quad
     {\includegraphics[width=6.8cm]{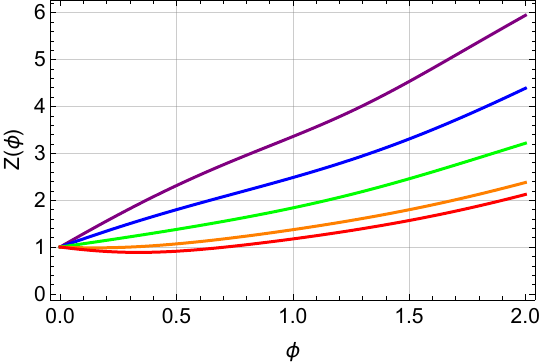} \label{}}
 \caption{Trained dilaton potentials $\{V(\phi),\, Z(\phi)\}$ derived from $T$-linear resistivity data \eqref{RESISDATA} by the neural ODEs with $\rho_0=0$ and $\bar{A}=(1.5,\, 1.3,\, 1.0,\, 0.7,\, 0.5)$ (red, orange, green, blue, purple).}\label{NeuralODEFIG1}
\end{figure}

\begin{figure}[]
  \centering
     {\includegraphics[width=7cm]{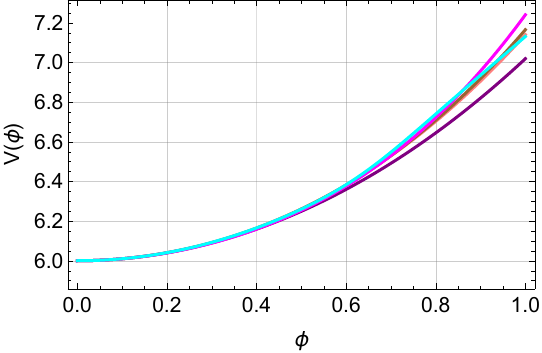} \label{}}
\quad
     {\includegraphics[width=6.8cm]{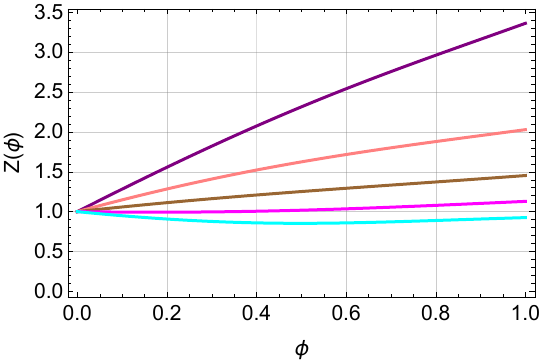} \label{}}
 \caption{Trained dilaton potential $\{V(\phi),\, Z(\phi)\}$ derived from $T$-linear resistivity data \eqref{RESISDATA} by the neural ODEs with $\rho_{0}=(0,\, 0.2,\, 0.4,\, 0.6,\, 0.8)$ (purple,\, pink,\, brown,\, magenta,\, cyan).}\label{NeuralODEFIG2}
\end{figure}

The potentials $V(\phi)$, $Z(\phi)$ trained by the neural ODEs are presented in Fig. \ref{NeuralODEFIG1} and \ref{NeuralODEFIG2}, showing consistent behaviors to the results of the PINNs which discussed in Section~\ref{sec32}. Likewise, as shown in Table. \ref{NeuralODETABLE1} and \ref{NeuralODETABLE2}, the final loss and $\beta/\mu$ exhibit no significant differences, further demonstrating the consistency between the two methods.\\

Both Neural ODEs and PINNs offer distinct advantages in scientific machine learning. PINNs, however, stand out as a particularly powerful and versatile method, making them especially suitable for modeling complex physical systems. Neural ODEs are advantageous for their memory efficiency and ability to produce smooth solutions. In contrast, PINNs excel at addressing a broader range of physical phenomena and demonstrate superior performance in solving complex, multi-dimensional problems, making them a preferred choice for many scientific and engineering applications. 

A key strength of PINNs lies in their ability to solve complex (ordinary and partial) differential equations, offering a robust and efficient alternative to traditional numerical solvers. This is particularly true when the differential equations themselves are highly intricate, where Neural ODEs may not be a viable option. By embedding physical laws and constraints directly into the neural network architecture, PINNs generate solutions that are not only data-driven but also consistent with fundamental physical principles. Furthermore, their capability for rapid inference after training enhances their utility in real-time applications, which are critical in various practical settings, where the need for accurate and physically consistent solutions is paramount.

\begin{table}[h]
\centering
\begin{tabular}{cccccccc}
    $\bar{A}$      & 1.5 & 1.3 & 1.0 & 0.7 & 0.5    \\
\hline
\hline
Loss $L$ & 0.0055 & 0.0043 & 0.0098 & 0.011 & 0.028  \\
$\beta/\mu$ & 9.97 & 9.92 & 9.95 & 10.05 & 10.14   
\end{tabular}
\caption{The total loss $L$ and $\beta/\mu$ after training by the neural ODEs with $\rho_0=0$.}\label{NeuralODETABLE1}
\end{table}
\begin{table}[h]
\centering
\begin{tabular}{cccccccc}
    $\rho_0$      & 0 & 0.2 & 0.4 & 0.6  & 0.8    \\
\hline
\hline
Loss $L$ & 0.028  & 0.0070 & 0.0023 & 0.0061 & 0.0063 \\
$\beta/\mu$ & 10.14  & 10.13 & 10.10 & 10.05 & 10.05  
\end{tabular}
\caption{The total loss $L$ and $\beta/\mu$ after training by the neural ODEs when $\bar{A}=0.5$.}\label{NeuralODETABLE2}
\end{table}

\bibliographystyle{JHEP}

\begin{thebibliography}{100}

\bibitem{Coleman_2005}
P.~Coleman and A.~J. Schofield, \emph{Quantum criticality},
  \href{http://dx.doi.org/10.1038/nature03279}{\emph{Nature} {\bf 433} (Jan.,
  2005) 226--229}.

\bibitem{Zaanen:2010yk}
J.~Zaanen, \emph{{A Modern, but way too short history of the theory of
  superconductivity at a high temperature}},
  \href{http://arxiv.org/abs/1012.5461}{{\tt 1012.5461}}.

\bibitem{Liu_2012}
H.~Liu, \emph{From black holes to strange metals},
  \href{http://dx.doi.org/10.1063/pt.3.1616}{\emph{Physics Today} {\bf 65}
  (June, 2012) 68--69}.

\bibitem{Faulkner:2010da}
T.~Faulkner, N.~Iqbal, H.~Liu, J.~McGreevy and D.~Vegh, \emph{{From Black Holes
  to Strange Metals}},  \href{http://arxiv.org/abs/1003.1728}{{\tt 1003.1728}}.

\bibitem{Phillips:2022nxs}
P.~W. Phillips, N.~E. Hussey and P.~Abbamonte, \emph{{Stranger than metals}},
  \href{http://dx.doi.org/10.1126/science.abh4273}{\emph{Science} {\bf 377}
  (2022) abh4273}, [\href{http://arxiv.org/abs/2205.12979}{{\tt 2205.12979}}].

\bibitem{Lohneysen:2007aa}
H.~v.~L{\"o}hneysen, \emph{Fermi-liquid instabilities at magnetic quantum phase
  transitions},
  \href{http://dx.doi.org/10.1103/RevModPhys.79.1015}{\emph{Reviews of Modern
  Physics} {\bf 79} (2007) 1015--1075}.

\bibitem{Sachdev:1993aa}
S.~Sachdev, \emph{Gapless spin-fluid ground state in a random quantum
  heisenberg magnet},
  \href{http://dx.doi.org/10.1103/PhysRevLett.70.3339}{\emph{Physical Review
  Letters} {\bf 70} (1993) 3339--3342}.

\bibitem{Guo:2020aa}
H.~Guo, Y.~Gu and S.~Sachdev, \emph{Linear in temperature resistivity in the
  limit of zero temperature from the time reparameterization soft mode},
  \href{http://dx.doi.org/https://doi.org/10.1016/j.aop.2020.168202}{\emph{Annals
  of Physics} {\bf 418} (2020) 168202},
  [\href{http://arxiv.org/abs/2004.05182}{{\tt 2004.05182}}].

\bibitem{Chowdhury:2022aa}
D.~Chowdhury, \emph{Sachdev-ye-kitaev models and beyond: Window into non-fermi
  liquids}, \href{http://dx.doi.org/10.1103/RevModPhys.94.035004}{\emph{Reviews
  of Modern Physics} {\bf 94} (2022) }.

\bibitem{Guo:2022aa}
H.~Guo, \emph{Large-n theory of critical fermi surfaces. ii. conductivity},
  \href{http://dx.doi.org/10.1103/PhysRevB.106.115151}{\emph{Physical Review B}
  {\bf 106} (2022) }.

\bibitem{Patel_2023}
A.~A. Patel, H.~Guo, I.~Esterlis and S.~Sachdev, \emph{Universal theory of
  strange metals from spatially random interactions},
  \href{http://dx.doi.org/10.1126/science.abq6011}{\emph{Science} {\bf 381}
  (Aug., 2023) 790--793}.

\bibitem{Li:2024aa}
C.~Li, D.~Valentinis, A.~A. Patel, H.~Guo, J.~Schmalian, S.~Sachdev et~al.,
  \emph{Strange metal and superconductor in the two-dimensional
  yukawa-sachdev-ye-kitaev model},
  \href{http://dx.doi.org/https://doi.org/10.1103/PhysRevLett.133.186502}{\emph{Physical
  Review Letters} {\bf 133} (2024) 186502},
  [\href{http://arxiv.org/abs/2406.07608}{{\tt 2406.07608}}].

\bibitem{Sachdev:2024aa}
S.~Sachdev, \emph{Lectures on the quantum phase transitions of metals},
  \href{http://arxiv.org/abs/2407.15919}{{\tt 2407.15919}}.

\bibitem{Wang:2024utm}
Y.-L. Wang, X.-H. Ge and S.-J. Sin, \emph{{Linear-T Resistivity from Spatially
  Random Vector Coupling}},  \href{http://arxiv.org/abs/2406.11170}{{\tt
  2406.11170}}.

\bibitem{Wang:2025oiz}
Y.-L. Wang, Y.-K. Han, X.-H. Ge and S.-J. Sin, \emph{{Hall Angle of a Spatially
  Random Vector Model}},  \href{http://arxiv.org/abs/2501.07792}{{\tt
  2501.07792}}.

\bibitem{Varma:1989aa}
C.~M. Varma, \emph{Phenomenology of the normal state of cu-o high-temperature
  superconductors},
  \href{http://dx.doi.org/10.1103/PhysRevLett.63.1996}{\emph{Physical Review
  Letters} {\bf 63} (1989) 1996--1999}.

\bibitem{Zaanen:2015oix}
J.~Zaanen, Y.-W. Sun, Y.~Liu and K.~Schalm, \emph{{Holographic Duality in
  Condensed Matter Physics}}.
\newblock Cambridge Univ. Press, 2015.

\bibitem{Hartnoll:2016apf}
S.~A. Hartnoll, A.~Lucas and S.~Sachdev, \emph{{Holographic quantum matter}},
  \href{http://arxiv.org/abs/1612.07324}{{\tt 1612.07324}}.

\bibitem{Baggioli:2019rrs}
M.~Baggioli, \emph{{Applied Holography}: {A Practical Mini-Course}},  other
  thesis, Madrid, IFT, 2019.
\newblock 10.1007/978-3-030-35184-7.

\bibitem{Zaanen:2021llz}
J.~Zaanen, \emph{{Lectures on quantum supreme matter}},
  \href{http://arxiv.org/abs/2110.00961}{{\tt 2110.00961}}.

\bibitem{Hussey__2004}
N.~E. Hussey~‖, K.~Takenaka and H.~Takagi, \emph{Universality of the
  mott--ioffe--regel limit in metals},
  \href{http://dx.doi.org/10.1080/14786430410001716944}{\emph{Philosophical
  Magazine} {\bf 84} (Sept., 2004) 2847--2864}.

\bibitem{Kim:2010zq}
B.~S. Kim, E.~Kiritsis and C.~Panagopoulos, \emph{{Holographic quantum
  criticality and strange metal transport}},
  \href{http://dx.doi.org/10.1088/1367-2630/14/4/043045}{\emph{New J. Phys.}
  {\bf 14} (2012) 043045}, [\href{http://arxiv.org/abs/1012.3464}{{\tt
  1012.3464}}].

\bibitem{Blake:2014yla}
M.~Blake and A.~Donos, \emph{{Quantum Critical Transport and the Hall Angle}},
  \href{http://arxiv.org/abs/1406.1659}{{\tt 1406.1659}}.

\bibitem{Zhou:2015dha}
Z.~Zhou, J.-P. Wu and Y.~Ling, \emph{{DC and Hall conductivity in holographic
  massive Einstein-Maxwell-Dilaton gravity}},
  \href{http://dx.doi.org/10.1007/JHEP08(2015)067}{\emph{JHEP} {\bf 08} (2015)
  067}, [\href{http://arxiv.org/abs/1504.00535}{{\tt 1504.00535}}].

\bibitem{Kim:2015wba}
K.-Y. Kim, K.~K. Kim, Y.~Seo and S.-J. Sin, \emph{{Thermoelectric
  Conductivities at Finite Magnetic Field and the Nernst Effect}},
  \href{http://dx.doi.org/10.1007/JHEP07(2015)027}{\emph{JHEP} {\bf 07} (2015)
  027}, [\href{http://arxiv.org/abs/1502.05386}{{\tt 1502.05386}}].

\bibitem{Chen:2017gsl}
Z.-N. Chen, X.-H. Ge, S.-Y. Wu, G.-H. Yang and H.-S. Zhang,
  \emph{{Magnetothermoelectric DC conductivities from holography models with
  hyperscaling factor in Lifshitz spacetime}},
  \href{http://dx.doi.org/10.1016/j.nuclphysb.2017.09.016}{\emph{Nucl. Phys.}
  {\bf B924} (2017) 387--405}, [\href{http://arxiv.org/abs/1709.08428}{{\tt
  1709.08428}}].

\bibitem{Blauvelt:2017koq}
E.~Blauvelt, S.~Cremonini, A.~Hoover, L.~Li and S.~Waskie, \emph{{Holographic
  model for the anomalous scalings of the cuprates}},
  \href{http://dx.doi.org/10.1103/PhysRevD.97.061901}{\emph{Phys. Rev.} {\bf
  D97} (2018) 061901}, [\href{http://arxiv.org/abs/1710.01326}{{\tt
  1710.01326}}].

\bibitem{Ahn:2023ciq}
Y.~Ahn, M.~Baggioli, H.-S. Jeong and K.-Y. Kim, \emph{{Inability of linear
  axion holographic Gubser-Rocha model to capture all the transport anomalies
  of strange metals}},
  \href{http://dx.doi.org/10.1103/PhysRevB.108.235104}{\emph{Phys. Rev. B} {\bf
  108} (2023) 235104}, [\href{http://arxiv.org/abs/2307.04433}{{\tt
  2307.04433}}].

\bibitem{Homes:2004wv}
C.~Homes, S.~Dordevic, M.~Strongin, D.~Bonn, R.~Liang et~al., \emph{{Universal
  scaling relation in high-temperature superconductors}},
  \href{http://dx.doi.org/10.1038/nature02673}{\emph{Nature} {\bf 430} (2004)
  539}, [\href{http://arxiv.org/abs/cond-mat/0404216}{{\tt cond-mat/0404216}}].

\bibitem{Zaanen:2004aa}
J.~Zaanen, \emph{Superconductivity: Why the temperature is high},
  {\emph{Nature} {\bf 430} (07, 2004) 512--513}.

\bibitem{Erdmenger:2015qqa}
J.~Erdmenger, B.~Herwerth, S.~Klug, R.~Meyer and K.~Schalm, \emph{{S-Wave
  Superconductivity in Anisotropic Holographic Insulators}},
  \href{http://dx.doi.org/10.1007/JHEP05(2015)094}{\emph{JHEP} {\bf 05} (2015)
  094}, [\href{http://arxiv.org/abs/1501.07615}{{\tt 1501.07615}}].

\bibitem{Kim:2015dna}
K.-Y. Kim, K.~K. Kim and M.~Park, \emph{{A Simple Holographic Superconductor
  with Momentum Relaxation}},
  \href{http://dx.doi.org/10.1007/JHEP04(2015)152}{\emph{JHEP} {\bf 04} (2015)
  152}, [\href{http://arxiv.org/abs/1501.00446}{{\tt 1501.00446}}].

\bibitem{Kim:2016hzi}
K.~K. Kim, M.~Park and K.-Y. Kim, \emph{{Ward identity and Homes' law in a
  holographic superconductor with momentum relaxation}},
  \href{http://dx.doi.org/10.1007/JHEP10(2016)041}{\emph{JHEP} {\bf 10} (2016)
  041}, [\href{http://arxiv.org/abs/1604.06205}{{\tt 1604.06205}}].

\bibitem{Kim:2016jjk}
K.-Y. Kim and C.~Niu, \emph{{Homes' law in Holographic Superconductor with
  Q-lattices}}, \href{http://dx.doi.org/10.1007/JHEP10(2016)144}{\emph{JHEP}
  {\bf 10} (2016) 144}, [\href{http://arxiv.org/abs/1608.04653}{{\tt
  1608.04653}}].

\bibitem{Jeong:2021wiu}
H.-S. Jeong and K.-Y. Kim, \emph{{Homes\textquoteright{} law in holographic
  superconductor with linear-T resistivity}},
  \href{http://dx.doi.org/10.1007/JHEP03(2022)060}{\emph{JHEP} {\bf 03} (2022)
  060}, [\href{http://arxiv.org/abs/2112.01153}{{\tt 2112.01153}}].

\bibitem{Charmousis:2010zz}
C.~Charmousis, B.~Gouteraux, B.~Kim, E.~Kiritsis and R.~Meyer, \emph{{Effective
  Holographic Theories for low-temperature condensed matter systems}},
  \href{http://dx.doi.org/10.1007/JHEP11(2010)151}{\emph{JHEP} {\bf 1011}
  (2010) 151}, [\href{http://arxiv.org/abs/1005.4690}{{\tt 1005.4690}}].

\bibitem{Davison:2013txa}
R.~A. Davison, K.~Schalm and J.~Zaanen, \emph{{Holographic duality and the
  resistivity of strange metals}},
  \href{http://dx.doi.org/10.1103/PhysRevB.89.245116}{\emph{Phys. Rev.} {\bf
  B89} (2014) 245116}, [\href{http://arxiv.org/abs/1311.2451}{{\tt
  1311.2451}}].

\bibitem{Gouteraux:2014hca}
B.~Gout{\'e}raux, \emph{{Charge transport in holography with momentum
  dissipation}}, \href{http://dx.doi.org/10.1007/JHEP04(2014)181}{\emph{JHEP}
  {\bf 1404} (2014) 181}, [\href{http://arxiv.org/abs/1401.5436}{{\tt
  1401.5436}}].

\bibitem{Ge:2016lyn}
X.-H. Ge, Y.~Tian, S.-Y. Wu, S.-F. Wu and S.-F. Wu, \emph{{Linear and quadratic
  in temperature resistivity from holography}},
  \href{http://dx.doi.org/10.1007/JHEP11(2016)128}{\emph{JHEP} {\bf 11} (2016)
  128}, [\href{http://arxiv.org/abs/1606.07905}{{\tt 1606.07905}}].

\bibitem{Cremonini:2016avj}
S.~Cremonini, H.-S. Liu, H.~Lu and C.~N. Pope, \emph{{DC Conductivities from
  Non-Relativistic Scaling Geometries with Momentum Dissipation}},
  \href{http://dx.doi.org/10.1007/JHEP04(2017)009}{\emph{JHEP} {\bf 04} (2017)
  009}, [\href{http://arxiv.org/abs/1608.04394}{{\tt 1608.04394}}].

\bibitem{Ahn:2017kvc}
H.-S. Jeong, Y.~Ahn, D.~Ahn, C.~Niu, W.-J. Li and K.-Y. Kim, \emph{{Thermal
  diffusivity and butterfly velocity in anisotropic Q-Lattice models}},
  \href{http://dx.doi.org/10.1007/JHEP01(2018)140}{\emph{JHEP} {\bf 01} (2018)
  140}, [\href{http://arxiv.org/abs/1708.08822}{{\tt 1708.08822}}].

\bibitem{Jeong:2018tua}
H.-S. Jeong, K.-Y. Kim and C.~Niu, \emph{{Linear-$T$ resistivity at high
  temperature}}, \href{http://dx.doi.org/10.1007/JHEP10(2018)191}{\emph{JHEP}
  {\bf 10} (2018) 191}, [\href{http://arxiv.org/abs/1806.07739}{{\tt
  1806.07739}}].

\bibitem{Ahn:2019lrh}
Y.~Ahn, H.-S. Jeong, D.~Ahn and K.-Y. Kim, \emph{{Linear-$T$ resistivity from
  low to high temperature: axion-dilaton theories}},
  \href{http://dx.doi.org/10.1007/JHEP04(2020)153}{\emph{JHEP} {\bf 04} (2020)
  153}, [\href{http://arxiv.org/abs/1907.12168}{{\tt 1907.12168}}].

\bibitem{Andrade:2013gsa}
T.~Andrade and B.~Withers, \emph{{A simple holographic model of momentum
  relaxation}}, \href{http://dx.doi.org/10.1007/JHEP05(2014)101}{\emph{JHEP}
  {\bf 1405} (2014) 101}, [\href{http://arxiv.org/abs/1311.5157}{{\tt
  1311.5157}}].

\bibitem{Vegh:2013sk}
D.~Vegh, \emph{{Holography without translational symmetry}},
  \href{http://arxiv.org/abs/1301.0537}{{\tt 1301.0537}}.

\bibitem{Baggioli:2021xuv}
M.~Baggioli, K.-Y. Kim, L.~Li and W.-J. Li, \emph{{Holographic Axion Model: a
  simple gravitational tool for quantum matter}},
  \href{http://dx.doi.org/10.1007/s11433-021-1681-8}{\emph{Sci. China Phys.
  Mech. Astron.} {\bf 64} (2021) 270001},
  [\href{http://arxiv.org/abs/2101.01892}{{\tt 2101.01892}}].

\bibitem{Natsuume:2014sfa}
M.~Natsuume, \emph{{AdS/CFT Duality User Guide}}, vol.~903.
\newblock Lect.Notes Phys, 2015,
  \href{http://dx.doi.org/10.1007/978-4-431-55441-7}{10.1007/978-4-431-55441-7}.

\bibitem{PhysRevLett.120.171602}
L.~Alberte, M.~Ammon, A.~Jim\'enez-Alba, M.~Baggioli and O.~Pujol\`as,
  \emph{Holographic phonons},
  \href{http://dx.doi.org/10.1103/PhysRevLett.120.171602}{\emph{Phys. Rev.
  Lett.} {\bf 120} (Apr, 2018) 171602}.

\bibitem{Ammon:2019wci}
M.~Ammon, M.~Baggioli and A.~Jim\'enez-Alba, \emph{{A Unified Description of
  Translational Symmetry Breaking in Holography}},
  \href{http://dx.doi.org/10.1007/JHEP09(2019)124}{\emph{JHEP} {\bf 09} (2019)
  124}, [\href{http://arxiv.org/abs/1904.05785}{{\tt 1904.05785}}].

\bibitem{Baggioli:2022pyb}
M.~Baggioli and B.~Gout\'eraux, \emph{{Colloquium: Hydrodynamics and holography
  of charge density wave phases}},
  \href{http://dx.doi.org/10.1103/RevModPhys.95.011001}{\emph{Rev. Mod. Phys.}
  {\bf 95} (2023) 011001}, [\href{http://arxiv.org/abs/2203.03298}{{\tt
  2203.03298}}].

\bibitem{Balm:2022bju}
F.~Balm et~al., \emph{{T-linear resistivity, optical conductivity, and
  Planckian transport for a holographic local quantum critical metal in a
  periodic potential}},
  \href{http://dx.doi.org/10.1103/PhysRevB.108.125145}{\emph{Phys. Rev. B} {\bf
  108} (2023) 125145}, [\href{http://arxiv.org/abs/2211.05492}{{\tt
  2211.05492}}].

\bibitem{Davison:2014lua}
R.~A. Davison and B.~Gout{\'e}raux, \emph{{Momentum dissipation and effective
  theories of coherent and incoherent transport}},
  \href{http://arxiv.org/abs/1411.1062}{{\tt 1411.1062}}.

\bibitem{Blake:2016jnn}
M.~Blake and A.~Donos, \emph{{Diffusion and Chaos from near AdS$_2$ horizons}},
  \href{http://dx.doi.org/10.1007/JHEP02(2017)013}{\emph{JHEP} {\bf 02} (2017)
  013}, [\href{http://arxiv.org/abs/1611.09380}{{\tt 1611.09380}}].

\bibitem{Amoretti:2016cad}
A.~Amoretti, M.~Baggioli, N.~Magnoli and D.~Musso, \emph{{Chasing the cuprates
  with dilatonic dyons}},
  \href{http://dx.doi.org/10.1007/JHEP06(2016)113}{\emph{JHEP} {\bf 06} (2016)
  113}, [\href{http://arxiv.org/abs/1603.03029}{{\tt 1603.03029}}].

\bibitem{Blake:2017qgd}
M.~Blake, R.~A. Davison and S.~Sachdev, \emph{{Thermal diffusivity and chaos in
  metals without quasiparticles}},  \href{http://arxiv.org/abs/1705.07896}{{\tt
  1705.07896}}.

\bibitem{Baggioli:2017ojd}
M.~Baggioli and W.-J. Li, \emph{{Diffusivities bounds and chaos in holographic
  Horndeski theories}},
  \href{http://dx.doi.org/10.1007/JHEP07(2017)055}{\emph{JHEP} {\bf 07} (2017)
  055}, [\href{http://arxiv.org/abs/1705.01766}{{\tt 1705.01766}}].

\bibitem{Giataganas:2017koz}
D.~Giataganas, U.~G\"ursoy and J.~F. Pedraza, \emph{{Strongly-coupled
  anisotropic gauge theories and holography}},
  \href{http://dx.doi.org/10.1103/PhysRevLett.121.121601}{\emph{Phys. Rev.
  Lett.} {\bf 121} (2018) 121601}, [\href{http://arxiv.org/abs/1708.05691}{{\tt
  1708.05691}}].

\bibitem{Davison:2018ofp}
R.~A. Davison, S.~A. Gentle and B.~Gout\'eraux, \emph{{Slow relaxation and
  diffusion in holographic quantum critical phases}},
  \href{http://dx.doi.org/10.1103/PhysRevLett.123.141601}{\emph{Phys. Rev.
  Lett.} {\bf 123} (2019) 141601}, [\href{http://arxiv.org/abs/1808.05659}{{\tt
  1808.05659}}].

\bibitem{Blake:2018leo}
M.~Blake, R.~A. Davison, S.~Grozdanov and H.~Liu, \emph{{Many-body chaos and
  energy dynamics in holography}},
  \href{http://dx.doi.org/10.1007/JHEP10(2018)035}{\emph{JHEP} {\bf 10} (2018)
  035}, [\href{http://arxiv.org/abs/1809.01169}{{\tt 1809.01169}}].

\bibitem{Jeong:2019zab}
H.-S. Jeong, K.-Y. Kim, Y.~Seo, S.-J. Sin and S.-Y. Wu, \emph{{Holographic
  Spectral Functions with Momentum Relaxation}},
  \href{http://dx.doi.org/10.1103/PhysRevD.102.026017}{\emph{Phys. Rev. D} {\bf
  102} (2020) 026017}, [\href{http://arxiv.org/abs/1910.11034}{{\tt
  1910.11034}}].

\bibitem{Arean:2020eus}
D.~Arean, R.~A. Davison, B.~Gout\'eraux and K.~Suzuki, \emph{{Hydrodynamic
  Diffusion and Its Breakdown near AdS2 Quantum Critical Points}},
  \href{http://dx.doi.org/10.1103/PhysRevX.11.031024}{\emph{Phys. Rev. X} {\bf
  11} (2021) 031024}, [\href{http://arxiv.org/abs/2011.12301}{{\tt
  2011.12301}}].

\bibitem{Liu:2021qmt}
Y.~Liu and X.-M. Wu, \emph{{Breakdown of hydrodynamics from holographic pole
  collision}}, \href{http://dx.doi.org/10.1007/JHEP01(2022)155}{\emph{JHEP}
  {\bf 01} (2022) 155}, [\href{http://arxiv.org/abs/2111.07770}{{\tt
  2111.07770}}].

\bibitem{Jeong:2021zhz}
H.-S. Jeong, K.-Y. Kim and Y.-W. Sun, \emph{{Bound of diffusion constants from
  pole-skipping points: spontaneous symmetry breaking and magnetic field}},
  \href{http://dx.doi.org/10.1007/JHEP07(2021)105}{\emph{JHEP} {\bf 07} (2021)
  105}, [\href{http://arxiv.org/abs/2104.13084}{{\tt 2104.13084}}].

\bibitem{Wu:2021mkk}
N.~Wu, M.~Baggioli and W.-J. Li, \emph{{On the universality of AdS$_{2}$
  diffusion bounds and the breakdown of linearized hydrodynamics}},
  \href{http://dx.doi.org/10.1007/JHEP05(2021)014}{\emph{JHEP} {\bf 05} (2021)
  014}, [\href{http://arxiv.org/abs/2102.05810}{{\tt 2102.05810}}].

\bibitem{Jeong:2021zsv}
H.-S. Jeong, K.-Y. Kim and Y.-W. Sun, \emph{{The breakdown of
  magneto-hydrodynamics near AdS$_{2}$ fixed point and energy diffusion
  bound}}, \href{http://dx.doi.org/10.1007/JHEP02(2022)006}{\emph{JHEP} {\bf
  02} (2022) 006}, [\href{http://arxiv.org/abs/2105.03882}{{\tt 2105.03882}}].

\bibitem{Huh:2021ppg}
K.-B. Huh, H.-S. Jeong, K.-Y. Kim and Y.-W. Sun, \emph{{Upper bound of the
  charge diffusion constant in holography}},
  \href{http://dx.doi.org/10.1007/JHEP07(2022)013}{\emph{JHEP} {\bf 07} (2022)
  013}, [\href{http://arxiv.org/abs/2111.07515}{{\tt 2111.07515}}].

\bibitem{Jeong:2022luo}
H.-S. Jeong, K.-Y. Kim and Y.-W. Sun, \emph{{Quasi-normal modes of dyonic black
  holes and magneto-hydrodynamics}},
  \href{http://arxiv.org/abs/2203.02642}{{\tt 2203.02642}}.

\bibitem{Baggioli:2022uqb}
M.~Baggioli, S.~Grieninger, S.~Grozdanov and Z.~Lu, \emph{{Aspects of
  univalence in holographic axion models}},
  \href{http://dx.doi.org/10.1007/JHEP11(2022)032}{\emph{JHEP} {\bf 11} (2022)
  032}, [\href{http://arxiv.org/abs/2205.06076}{{\tt 2205.06076}}].

\bibitem{Jeong:2023ynk}
H.-S. Jeong, \emph{{Quantum chaos and pole-skipping in a semilocally critical
  IR fixed point}},  \href{http://arxiv.org/abs/2309.13412}{{\tt 2309.13412}}.

\bibitem{Ahn:2024aiw}
Y.~Ahn, V.~Jahnke, H.-S. Jeong, C.-W. Ji, K.-Y. Kim and M.~Nishida, \emph{{On
  pole-skipping with gauge-invariant variables in holographic axion theories}},
   \href{http://arxiv.org/abs/2402.12951}{{\tt 2402.12951}}.

\bibitem{Zhao:2023qms}
Z.~Zhao, W.~Cai and S.~Ishigaki, \emph{{Doped Holographic Superconductors in
  Gubser-Rocha model}},  \href{http://arxiv.org/abs/2309.14851}{{\tt
  2309.14851}}.

\bibitem{Baggioli:2014roa}
M.~Baggioli and O.~Pujolas, \emph{{Electron-Phonon Interactions,
  Metal-Insulator Transitions, and Holographic Massive Gravity}},
  \href{http://dx.doi.org/10.1103/PhysRevLett.114.251602}{\emph{Phys. Rev.
  Lett.} {\bf 114} (2015) 251602}, [\href{http://arxiv.org/abs/1411.1003}{{\tt
  1411.1003}}].

\bibitem{Alberte:2015isw}
L.~Alberte, M.~Baggioli, A.~Khmelnitsky and O.~Pujolas, \emph{{Solid Holography
  and Massive Gravity}},
  \href{http://dx.doi.org/10.1007/JHEP02(2016)114}{\emph{JHEP} {\bf 02} (2016)
  114}, [\href{http://arxiv.org/abs/1510.09089}{{\tt 1510.09089}}].

\bibitem{Amoretti:2016bxs}
A.~Amoretti, D.~Are\'an, R.~Argurio, D.~Musso and L.~A. Pando~Zayas, \emph{{A
  holographic perspective on phonons and pseudo-phonons}},
  \href{http://dx.doi.org/10.1007/JHEP05(2017)051}{\emph{JHEP} {\bf 05} (2017)
  051}, [\href{http://arxiv.org/abs/1611.09344}{{\tt 1611.09344}}].

\bibitem{Alberte:2017oqx}
L.~Alberte, M.~Ammon, A.~Jim\'enez-Alba, M.~Baggioli and O.~Pujol\`as,
  \emph{{Holographic Phonons}},
  \href{http://dx.doi.org/10.1103/PhysRevLett.120.171602}{\emph{Phys. Rev.
  Lett.} {\bf 120} (2018) 171602}, [\href{http://arxiv.org/abs/1711.03100}{{\tt
  1711.03100}}].

\bibitem{Amoretti:2017frz}
A.~Amoretti, D.~Are\'an, B.~Gout\'eraux and D.~Musso, \emph{{Effective
  holographic theory of charge density waves}},
  \href{http://dx.doi.org/10.1103/PhysRevD.97.086017}{\emph{Phys. Rev. D} {\bf
  97} (2018) 086017}, [\href{http://arxiv.org/abs/1711.06610}{{\tt
  1711.06610}}].

\bibitem{Amoretti:2017axe}
A.~Amoretti, D.~Aren, B.~Goutraux and D.~Musso, \emph{{DC resistivity of
  quantum critical, charge density wave states from gauge-gravity duality}},
  \href{http://dx.doi.org/10.1103/PhysRevLett.120.171603}{\emph{Phys. Rev.
  Lett.} {\bf 120} (2018) 171603}, [\href{http://arxiv.org/abs/1712.07994}{{\tt
  1712.07994}}].

\bibitem{Alberte:2017cch}
L.~Alberte, M.~Ammon, M.~Baggioli, A.~Jim\'enez and O.~Pujol\`as, \emph{{Black
  hole elasticity and gapped transverse phonons in holography}},
  \href{http://dx.doi.org/10.1007/JHEP01(2018)129}{\emph{JHEP} {\bf 01} (2018)
  129}, [\href{http://arxiv.org/abs/1708.08477}{{\tt 1708.08477}}].

\bibitem{Andrade:2017cnc}
T.~Andrade, M.~Baggioli, A.~Krikun and N.~Poovuttikul, \emph{{Pinning of
  longitudinal phonons in holographic spontaneous helices}},
  \href{http://dx.doi.org/10.1007/JHEP02(2018)085}{\emph{JHEP} {\bf 02} (2018)
  085}, [\href{http://arxiv.org/abs/1708.08306}{{\tt 1708.08306}}].

\bibitem{Amoretti:2018tzw}
A.~Amoretti, D.~Are\'an, B.~Gout\'eraux and D.~Musso, \emph{{Universal
  relaxation in a holographic metallic density wave phase}},
  \href{http://dx.doi.org/10.1103/PhysRevLett.123.211602}{\emph{Phys. Rev.
  Lett.} {\bf 123} (2019) 211602}, [\href{http://arxiv.org/abs/1812.08118}{{\tt
  1812.08118}}].

\bibitem{Amoretti:2019cef}
A.~Amoretti, D.~Are\'an, B.~Gout\'eraux and D.~Musso, \emph{{Diffusion and
  universal relaxation of holographic phonons}},
  \href{http://dx.doi.org/10.1007/JHEP10(2019)068}{\emph{JHEP} {\bf 10} (2019)
  068}, [\href{http://arxiv.org/abs/1904.11445}{{\tt 1904.11445}}].

\bibitem{Baggioli:2019abx}
M.~Baggioli and S.~Grieninger, \emph{{Zoology of solid \textbackslash{}\& fluid
  holography \textemdash{} Goldstone modes and phase relaxation}},
  \href{http://dx.doi.org/10.1007/JHEP10(2019)235}{\emph{JHEP} {\bf 10} (2019)
  235}, [\href{http://arxiv.org/abs/1905.09488}{{\tt 1905.09488}}].

\bibitem{Amoretti:2019kuf}
A.~Amoretti, D.~Are\'an, B.~Gout\'eraux and D.~Musso, \emph{{Gapless and gapped
  holographic phonons}},
  \href{http://dx.doi.org/10.1007/JHEP01(2020)058}{\emph{JHEP} {\bf 01} (2020)
  058}, [\href{http://arxiv.org/abs/1910.11330}{{\tt 1910.11330}}].

\bibitem{Ammon:2019apj}
M.~Ammon, M.~Baggioli, S.~Gray and S.~Grieninger, \emph{{Longitudinal Sound and
  Diffusion in Holographic Massive Gravity}},
  \href{http://dx.doi.org/10.1007/JHEP10(2019)064}{\emph{JHEP} {\bf 10} (2019)
  064}, [\href{http://arxiv.org/abs/1905.09164}{{\tt 1905.09164}}].

\bibitem{Baggioli:2020edn}
M.~Baggioli, S.~Grieninger and L.~Li, \emph{{Magnetophonons \& type-B
  Goldstones from Hydrodynamics to Holography}},
  \href{http://dx.doi.org/10.1007/JHEP09(2020)037}{\emph{JHEP} {\bf 09} (2020)
  037}, [\href{http://arxiv.org/abs/2005.01725}{{\tt 2005.01725}}].

\bibitem{Amoretti:2021fch}
A.~Amoretti, D.~Arean, D.~K. Brattan and N.~Magnoli, \emph{{Hydrodynamic
  magneto-transport in charge density wave states}},
  \href{http://dx.doi.org/10.1007/JHEP05(2021)027}{\emph{JHEP} {\bf 05} (2021)
  027}, [\href{http://arxiv.org/abs/2101.05343}{{\tt 2101.05343}}].

\bibitem{Amoretti:2021lll}
A.~Amoretti, D.~Arean, D.~K. Brattan and L.~Martinoia, \emph{{Hydrodynamic
  magneto-transport in holographic charge density wave states}},
  \href{http://dx.doi.org/10.1007/JHEP11(2021)011}{\emph{JHEP} {\bf 11} (2021)
  011}, [\href{http://arxiv.org/abs/2107.00519}{{\tt 2107.00519}}].

\bibitem{Wang:2021jfu}
X.-J. Wang and W.-J. Li, \emph{{Holographic phonons by gauge-axion coupling}},
  \href{http://dx.doi.org/10.1007/JHEP07(2021)131}{\emph{JHEP} {\bf 07} (2021)
  131}, [\href{http://arxiv.org/abs/2105.07225}{{\tt 2105.07225}}].

\bibitem{Zhong:2022mok}
Y.-Y. Zhong and W.-J. Li, \emph{{Transverse Goldstone mode in holographic
  fluids with broken translations}},
  \href{http://dx.doi.org/10.1140/epjc/s10052-022-10430-w}{\emph{Eur. Phys. J.
  C} {\bf 82} (2022) 511}, [\href{http://arxiv.org/abs/2202.05437}{{\tt
  2202.05437}}].

\bibitem{Bajec:2024jez}
M.~Bajec, S.~Grozdanov and A.~Soloviev, \emph{{Spectra of correlators in the
  relaxation time approximation of kinetic theory}},
  \href{http://arxiv.org/abs/2403.17769}{{\tt 2403.17769}}.

\bibitem{RezaMohammadiMozaffar:2016lbo}
M.~Reza Mohammadi~Mozaffar, A.~Mollabashi and F.~Omidi, \emph{{Non-local Probes
  in Holographic Theories with Momentum Relaxation}},
  \href{http://dx.doi.org/10.1007/JHEP10(2016)135}{\emph{JHEP} {\bf 10} (2016)
  135}, [\href{http://arxiv.org/abs/1608.08781}{{\tt 1608.08781}}].

\bibitem{Yekta:2020wup}
D.~M. Yekta, H.~Babaei-Aghbolagh, K.~Babaei~Velni and H.~Mohammadzadeh,
  \emph{{Holographic complexity for black branes with momentum relaxation}},
  \href{http://dx.doi.org/10.1103/PhysRevD.104.086025}{\emph{Phys. Rev. D} {\bf
  104} (2021) 086025}, [\href{http://arxiv.org/abs/2009.01340}{{\tt
  2009.01340}}].

\bibitem{Li:2019rpp}
Y.-Z. Li and X.-M. Kuang, \emph{{Probes of holographic thermalization in a
  simple model with momentum relaxation}},
  \href{http://dx.doi.org/10.1016/j.nuclphysb.2020.115043}{\emph{Nucl. Phys. B}
  {\bf 956} (2020) 115043}, [\href{http://arxiv.org/abs/1911.11980}{{\tt
  1911.11980}}].

\bibitem{Zhou:2019xzc}
Y.-T. Zhou, X.-M. Kuang, Y.-Z. Li and J.-P. Wu, \emph{{Holographic subregion
  complexity under a thermal quench in an Einstein-Maxwell-axion theory with
  momentum relaxation}},
  \href{http://dx.doi.org/10.1103/PhysRevD.101.106024}{\emph{Phys. Rev. D} {\bf
  101} (2020) 106024}, [\href{http://arxiv.org/abs/1912.03479}{{\tt
  1912.03479}}].

\bibitem{Huang:2019zph}
Y.-f. Huang, Z.-j. Shi, C.~Niu, C.-y. Zhang and P.~Liu, \emph{{Mixed State
  Entanglement for Holographic Axion Model}},
  \href{http://dx.doi.org/10.1140/epjc/s10052-020-7921-y}{\emph{Eur. Phys. J.
  C} {\bf 80} (2020) 426}, [\href{http://arxiv.org/abs/1911.10977}{{\tt
  1911.10977}}].

\bibitem{Jeong:2022zea}
H.-S. Jeong, K.-Y. Kim and Y.-W. Sun, \emph{{Holographic entanglement density
  for spontaneous symmetry breaking}},
  \href{http://dx.doi.org/10.1007/JHEP06(2022)078}{\emph{JHEP} {\bf 06} (2022)
  078}, [\href{http://arxiv.org/abs/2203.07612}{{\tt 2203.07612}}].

\bibitem{HosseiniMansoori:2022hok}
S.~A. Hosseini~Mansoori, O.~Luongo, S.~Mancini, M.~Mirjalali, M.~Rafiee and
  A.~Tavanfar, \emph{{Planar black holes in holographic axion gravity: Islands,
  Page times, and scrambling times}},
  \href{http://dx.doi.org/10.1103/PhysRevD.106.126018}{\emph{Phys. Rev. D} {\bf
  106} (2022) 126018}, [\href{http://arxiv.org/abs/2209.00253}{{\tt
  2209.00253}}].

\bibitem{Ahn:2024gjf}
B.~Ahn, H.-S. Jeong, K.-Y. Kim and K.~Yun, \emph{{Deep learning bulk spacetime
  from boundary optical conductivity}},
  \href{http://dx.doi.org/10.1007/JHEP03(2024)141}{\emph{JHEP} {\bf 03} (2024)
  141}, [\href{http://arxiv.org/abs/2401.00939}{{\tt 2401.00939}}].

\bibitem{Ahn:2024jkk}
B.~Ahn, H.-S. Jeong, K.-Y. Kim and K.~Yun, \emph{{Holographic reconstruction of
  black hole spacetime: machine learning and entanglement entropy}},
  \href{http://dx.doi.org/10.1007/JHEP01(2025)025}{\emph{JHEP} {\bf 01} (2025)
  025}, [\href{http://arxiv.org/abs/2406.07395}{{\tt 2406.07395}}].

\bibitem{Donos:2014cya}
A.~Donos and J.~P. Gauntlett, \emph{{Thermoelectric DC conductivities from
  black hole horizons}},
  \href{http://dx.doi.org/10.1007/JHEP11(2014)081}{\emph{JHEP} {\bf 11} (2014)
  081}, [\href{http://arxiv.org/abs/1406.4742}{{\tt 1406.4742}}].

\bibitem{Kim:2014bza}
K.-Y. Kim, K.~K. Kim, Y.~Seo and S.-J. Sin, \emph{{Coherent/incoherent metal
  transition in a holographic model}},
  \href{http://dx.doi.org/10.1007/JHEP12(2014)170}{\emph{JHEP} {\bf 12} (2014)
  170}, [\href{http://arxiv.org/abs/1409.8346}{{\tt 1409.8346}}].

\bibitem{Kim:2015sma}
K.-Y. Kim, K.~K. Kim, Y.~Seo and S.-J. Sin, \emph{{Gauge Invariance and
  Holographic Renormalization}},
  \href{http://dx.doi.org/10.1016/j.physletb.2015.07.058}{\emph{Phys. Lett.}
  {\bf B749} (2015) 108--114}, [\href{http://arxiv.org/abs/1502.02100}{{\tt
  1502.02100}}].

\bibitem{Gubser:2009qt}
S.~S. Gubser and F.~D. Rocha, \emph{{Peculiar properties of a charged dilatonic
  black hole in $AdS_5$}},
  \href{http://dx.doi.org/10.1103/PhysRevD.81.046001}{\emph{Phys.Rev.} {\bf
  D81} (2010) 046001}, [\href{http://arxiv.org/abs/0911.2898}{{\tt
  0911.2898}}].

\bibitem{Iqbal2012}
N.~Iqbal, H.~Liu and M.~Mezei, \emph{Semi-local quantum liquids},
  \href{http://dx.doi.org/10.1007/JHEP04(2012)086}{\emph{Journal of High Energy
  Physics} {\bf 2012} (Apr, 2012) 86}.

\bibitem{Goldstein2010}
K.~Goldstein, S.~Kachru, S.~Prakash and S.~P. Trivedi, \emph{Holography of
  charged dilaton black holes},
  \href{http://dx.doi.org/10.1007/JHEP08(2010)078}{\emph{Journal of High Energy
  Physics} {\bf 2010} (Aug, 2010) 78}.

\bibitem{Charmousis2010}
C.~Charmousis, B.~Gout{\'e}raux, B.~Soo~Kim, E.~Kiritsis and R.~Meyer,
  \emph{Effective holographic theories for low-temperature condensed matter
  systems}, \href{http://dx.doi.org/10.1007/JHEP11(2010)151}{\emph{Journal of
  High Energy Physics} {\bf 2010} (Nov, 2010) 151}.

\bibitem{Hartnoll2012}
S.~A. Hartnoll and E.~Shaghoulian, \emph{Spectral weight in holographic scaling
  geometries}, \href{http://dx.doi.org/10.1007/JHEP07(2012)078}{\emph{Journal
  of High Energy Physics} {\bf 2012} (Jul, 2012) 78}.

\bibitem{Liu:2024gxr}
X.-L. Liu, J.~Nian and L.~A. Pando~Zayas, \emph{{Quantum Corrections to
  Holographic Strange Metal at Low Temperature}},
  \href{http://arxiv.org/abs/2410.11487}{{\tt 2410.11487}}.

\bibitem{Kiritsis:2015oxa}
E.~Kiritsis and J.~Ren, \emph{{On Holographic Insulators and Supersolids}},
  \href{http://dx.doi.org/10.1007/JHEP09(2015)168}{\emph{JHEP} {\bf 09} (2015)
  168}, [\href{http://arxiv.org/abs/1503.03481}{{\tt 1503.03481}}].

\bibitem{Ling:2016yxy}
Y.~Ling, Z.~Xian and Z.~Zhou, \emph{{Power Law of Shear Viscosity in
  Einstein-Maxwell-Dilaton-Axion model}},
  \href{http://dx.doi.org/10.1088/1674-1137/41/2/023104}{\emph{Chin. Phys.}
  {\bf C41} (2017) 023104}, [\href{http://arxiv.org/abs/1610.08823}{{\tt
  1610.08823}}].

\bibitem{Bhattacharya:2014dea}
J.~Bhattacharya, S.~Cremonini and B.~Gout{\'e}raux, \emph{{Intermediate
  scalings in holographic RG flows and conductivities}},
  \href{http://arxiv.org/abs/1409.4797}{{\tt 1409.4797}}.

\bibitem{Cvetic:1999xp}
M.~Cvetic, M.~J. Duff, P.~Hoxha, J.~T. Liu, H.~Lu, J.~X. Lu et~al.,
  \emph{{Embedding AdS black holes in ten-dimensions and eleven-dimensions}},
  \href{http://dx.doi.org/10.1016/S0550-3213(99)00419-8}{\emph{Nucl. Phys. B}
  {\bf 558} (1999) 96--126}, [\href{http://arxiv.org/abs/hep-th/9903214}{{\tt
  hep-th/9903214}}].

\bibitem{Faulkner:2009wj}
T.~Faulkner, H.~Liu, J.~McGreevy and D.~Vegh, \emph{{Emergent quantum
  criticality, Fermi surfaces, and AdS(2)}},
  \href{http://dx.doi.org/10.1103/PhysRevD.83.125002}{\emph{Phys.Rev.} {\bf
  D83} (2011) 125002}, [\href{http://arxiv.org/abs/0907.2694}{{\tt
  0907.2694}}].

\bibitem{Chowdhury:2021qpy}
D.~Chowdhury, A.~Georges, O.~Parcollet and S.~Sachdev, \emph{{Sachdev-Ye-Kitaev
  models and beyond: Window into non-Fermi liquids}},
  \href{http://dx.doi.org/10.1103/RevModPhys.94.035004}{\emph{Rev. Mod. Phys.}
  {\bf 94} (2022) 035004}, [\href{http://arxiv.org/abs/2109.05037}{{\tt
  2109.05037}}].

\bibitem{Hartnoll:2014lpa}
S.~A. Hartnoll, \emph{{Theory of universal incoherent metallic transport}},
  \href{http://dx.doi.org/10.1038/nphys3174}{\emph{Nature Phys.} {\bf 11}
  (2015) 54}, [\href{http://arxiv.org/abs/1405.3651}{{\tt 1405.3651}}].

\bibitem{Wang:2023rca}
Z.~Wang, X.-H. Ge and S.~Ishigaki, \emph{{Dependence of the critical
  temperature and disorder in holographic superconductors on superfluid
  density}},  \href{http://arxiv.org/abs/2312.16029}{{\tt 2312.16029}}.

\bibitem{Lu:2024qxj}
C.-Y. Lu, X.-H. Ge and S.-J. Sin, \emph{{Holographic fermions in the Dyonic
  Gubser-Rocha black hole}},  \href{http://arxiv.org/abs/2412.20160}{{\tt
  2412.20160}}.

\bibitem{Ling:2013nxa}
Y.~Ling, C.~Niu, J.-P. Wu and Z.-Y. Xian, \emph{{Holographic Lattice in
  Einstein-Maxwell-Dilaton Gravity}},
  \href{http://dx.doi.org/10.1007/JHEP11(2013)006}{\emph{JHEP} {\bf 1311}
  (2013) 006}, [\href{http://arxiv.org/abs/1309.4580}{{\tt 1309.4580}}].

\bibitem{Eede:2023rrv}
S.~T. V.~d. Eede, T.~J.~N. van Stralen, C.~F.~J. Flipse and H.~T.~C. Stoof,
  \emph{{Plasmons in a layered strange metal using the gauge-gravity duality}},
  \href{http://dx.doi.org/10.1103/PhysRevB.109.085119}{\emph{Phys. Rev. B} {\bf
  109} (2024) 085119}, [\href{http://arxiv.org/abs/2311.03142}{{\tt
  2311.03142}}].

\bibitem{Kim:2017dgz}
K.-Y. Kim and C.~Niu, \emph{{Diffusion and Butterfly Velocity at Finite
  Density}}, \href{http://dx.doi.org/10.1007/JHEP06(2017)030}{\emph{JHEP} {\bf
  06} (2017) 030}, [\href{http://arxiv.org/abs/1704.00947}{{\tt 1704.00947}}].

\bibitem{Liu:2024tqe}
Y.~Liu, Y.-W. Sun and X.-M. Wu, \emph{{Holographic Schwinger-Keldysh effective
  field theories including a non-hydrodynamic mode}},
  \href{http://arxiv.org/abs/2411.16306}{{\tt 2411.16306}}.

\bibitem{Arean:2024pzo}
D.~Are\'an, H.-S. Jeong, J.~F. Pedraza and L.-C. Qu, \emph{{Kasner interiors
  from analytic hairy black holes}},
  \href{http://dx.doi.org/10.1007/JHEP11(2024)138}{\emph{JHEP} {\bf 11} (2024)
  138}, [\href{http://arxiv.org/abs/2407.18430}{{\tt 2407.18430}}].

\bibitem{Hinton_2006}
G.~E. Hinton and R.~R. Salakhutdinov, \emph{Reducing the dimensionality of data
  with neural networks},
  \href{http://dx.doi.org/10.1126/science.1127647}{\emph{Science} {\bf 313}
  (July, 2006) 504--507}.

\bibitem{Bengio2007ScalingLA}
Y.~Bengio and Y.~LeCun, \emph{Scaling learning algorithms towards ai},  2007.

\bibitem{LeCun_2015}
Y.~LeCun, Y.~Bengio and G.~Hinton, \emph{Deep learning},
  \href{http://dx.doi.org/10.1038/nature14539}{\emph{Nature} {\bf 521} (May,
  2015) 436--444}.

\bibitem{Carleo:2019ptp}
G.~Carleo, I.~Cirac, K.~Cranmer, L.~Daudet, M.~Schuld, N.~Tishby et~al.,
  \emph{{Machine learning and the physical sciences}},
  \href{http://dx.doi.org/10.1103/RevModPhys.91.045002}{\emph{Rev. Mod. Phys.}
  {\bf 91} (2019) 045002}, [\href{http://arxiv.org/abs/1903.10563}{{\tt
  1903.10563}}].

\bibitem{RUEHLE20201}
F.~Ruehle, \emph{Data science applications to string theory},
  \href{http://dx.doi.org/https://doi.org/10.1016/j.physrep.2019.09.005}{\emph{Physics
  Reports} {\bf 839} (2020) 1--117}.

\bibitem{Carleo:2017aa}
G.~Carleo and M.~Troyer, \emph{Solving the quantum many-body problem with
  artificial neural networks},
  \href{http://dx.doi.org/https://doi.org/10.1126/science.aag2302}{\emph{Science}
  {\bf 355} (2017) 602}, [\href{http://arxiv.org/abs/1606.02318}{{\tt
  1606.02318}}].

\bibitem{Bedolla_2020}
E.~Bedolla, L.~C. Padierna and R.~Casta{\~n}eda-Priego, \emph{Machine learning
  for condensed matter physics},
  \href{http://dx.doi.org/10.1088/1361-648x/abb895}{\emph{Journal of Physics:
  Condensed Matter} {\bf 33} (Nov., 2020) 053001}.

\bibitem{Boehnlein:2022aa}
A.~Boehnlein, \emph{Machine learning in nuclear physics},
  \href{http://dx.doi.org/10.1103/RevModPhys.94.031003}{\emph{Reviews of Modern
  Physics} {\bf 94} (2022) }.

\bibitem{Hashimoto:2018ftp}
K.~Hashimoto, S.~Sugishita, A.~Tanaka and A.~Tomiya, \emph{{Deep learning and
  the AdS/CFT correspondence}},
  \href{http://dx.doi.org/10.1103/PhysRevD.98.046019}{\emph{Phys. Rev. D} {\bf
  98} (2018) 046019}, [\href{http://arxiv.org/abs/1802.08313}{{\tt
  1802.08313}}].

\bibitem{Tanaka_2021}
A.~Tanaka, A.~Tomiya and K.~Hashimoto, \emph{Deep Learning and Physics}.
\newblock Springer Singapore, 2021,
  \href{http://dx.doi.org/10.1007/978-981-33-6108-9}{10.1007/978-981-33-6108-9}.

\bibitem{Song:2020agw}
M.~Song, M.~S.~H. Oh, Y.~Ahn and K.-Y. Kima, \emph{{AdS/Deep-Learning made
  easy: simple examples}},
  \href{http://dx.doi.org/10.1088/1674-1137/abfc36}{\emph{Chin. Phys. C} {\bf
  45} (2021) 073111}, [\href{http://arxiv.org/abs/2011.13726}{{\tt
  2011.13726}}].

\bibitem{You:2017guh}
Y.-Z. You, Z.~Yang and X.-L. Qi, \emph{{Machine Learning Spatial Geometry from
  Entanglement Features}},
  \href{http://dx.doi.org/10.1103/PhysRevB.97.045153}{\emph{Phys. Rev. B} {\bf
  97} (2018) 045153}, [\href{http://arxiv.org/abs/1709.01223}{{\tt
  1709.01223}}].

\bibitem{Gan:2017nyt}
W.-C. Gan and F.-W. Shu, \emph{{Holography as deep learning}},
  \href{http://dx.doi.org/10.1142/S0218271817430209}{\emph{Int. J. Mod. Phys.
  D} {\bf 26} (2017) 1743020}, [\href{http://arxiv.org/abs/1705.05750}{{\tt
  1705.05750}}].

\bibitem{Lee:2017skk}
J.-W. Lee, \emph{{Quantum fields as deep learning}},
  \href{http://dx.doi.org/10.3938/jkps.76.684}{\emph{J. Korean Phys. Soc.} {\bf
  76} (2020) 684--687}, [\href{http://arxiv.org/abs/1708.07408}{{\tt
  1708.07408}}].

\bibitem{Swingle:2009bg}
B.~Swingle, \emph{{Entanglement Renormalization and Holography}},
  \href{http://dx.doi.org/10.1103/PhysRevD.86.065007}{\emph{Phys. Rev. D} {\bf
  86} (2012) 065007}, [\href{http://arxiv.org/abs/0905.1317}{{\tt 0905.1317}}].

\bibitem{Hashimoto:2018bnb}
K.~Hashimoto, S.~Sugishita, A.~Tanaka and A.~Tomiya, \emph{{Deep Learning and
  Holographic QCD}},
  \href{http://dx.doi.org/10.1103/PhysRevD.98.106014}{\emph{Phys. Rev. D} {\bf
  98} (2018) 106014}, [\href{http://arxiv.org/abs/1809.10536}{{\tt
  1809.10536}}].

\bibitem{Hashimoto:2020jug}
K.~Hashimoto, H.-Y. Hu and Y.-Z. You, \emph{{Neural ordinary differential
  equation and holographic quantum chromodynamics}},
  \href{http://dx.doi.org/10.1088/2632-2153/abe527}{\emph{Mach. Learn. Sci.
  Tech.} {\bf 2} (2021) 035011}, [\href{http://arxiv.org/abs/2006.00712}{{\tt
  2006.00712}}].

\bibitem{Akutagawa:2020yeo}
T.~Akutagawa, K.~Hashimoto and T.~Sumimoto, \emph{{Deep Learning and AdS/QCD}},
  \href{http://dx.doi.org/10.1103/PhysRevD.102.026020}{\emph{Phys. Rev. D} {\bf
  102} (2020) 026020}, [\href{http://arxiv.org/abs/2005.02636}{{\tt
  2005.02636}}].

\bibitem{Hashimoto:2021ihd}
K.~Hashimoto, K.~Ohashi and T.~Sumimoto, \emph{{Deriving the dilaton potential
  in improved holographic QCD from the meson spectrum}},
  \href{http://dx.doi.org/10.1103/PhysRevD.105.106008}{\emph{Phys. Rev. D} {\bf
  105} (2022) 106008}, [\href{http://arxiv.org/abs/2108.08091}{{\tt
  2108.08091}}].

\bibitem{Hashimoto:2022eij}
K.~Hashimoto, K.~Ohashi and T.~Sumimoto, \emph{{Deriving the dilaton potential
  in improved holographic QCD from the chiral condensate}},
  \href{http://dx.doi.org/10.1093/ptep/ptad026}{\emph{PTEP} {\bf 2023} (2023)
  033B01}, [\href{http://arxiv.org/abs/2209.04638}{{\tt 2209.04638}}].

\bibitem{Mansouri:2024uwc}
M.~Mansouri, K.~Bitaghsir~Fadafan and X.~Chen, \emph{{Holographic complex
  potential of a quarkonium from deep learning}},
  \href{http://arxiv.org/abs/2406.06285}{{\tt 2406.06285}}.

\bibitem{Luo:2024iwf}
O.-Y. Luo, X.~Chen, F.-P. Li, X.-H. Li and K.~Zhou, \emph{{Neural Network
  Modeling of Heavy-Quark Potential from Holography}},
  \href{http://arxiv.org/abs/2408.03784}{{\tt 2408.03784}}.

\bibitem{Yan:2020wcd}
Y.-K. Yan, S.-F. Wu, X.-H. Ge and Y.~Tian, \emph{{Deep learning black hole
  metrics from shear viscosity}},
  \href{http://dx.doi.org/10.1103/PhysRevD.102.101902}{\emph{Phys. Rev. D} {\bf
  102} (4, 2020) 101902}, [\href{http://arxiv.org/abs/2004.12112}{{\tt
  2004.12112}}].

\bibitem{Gu:2024lrz}
Z.-F. Gu, Y.-K. Yan and S.-F. Wu, \emph{{Neural ODEs for holographic transport
  models without translation symmetry}},
  \href{http://arxiv.org/abs/2401.09946}{{\tt 2401.09946}}.

\bibitem{Chen:2024ckb}
X.~Chen and M.~Huang, \emph{{Machine learning holographic black hole from
  lattice QCD equation of state}},
  \href{http://dx.doi.org/10.1103/PhysRevD.109.L051902}{\emph{Phys. Rev. D}
  {\bf 109} (2024) L051902}, [\href{http://arxiv.org/abs/2401.06417}{{\tt
  2401.06417}}].

\bibitem{Bea:2024xgv}
Y.~Bea, R.~Jimenez, D.~Mateos, S.~Liu, P.~Protopapas, P.~Taranc\'on-\'Alvarez
  et~al., \emph{{Gravitational duals from equations of state}},
  \href{http://dx.doi.org/10.1007/JHEP07(2024)087}{\emph{JHEP} {\bf 07} (2024)
  087}, [\href{http://arxiv.org/abs/2403.14763}{{\tt 2403.14763}}].

\bibitem{Cai:2024eqa}
R.-G. Cai, S.~He, L.~Li and H.-A. Zeng, \emph{{Neural Ordinary Differential
  Equations for Mapping the Magnetic QCD Phase Diagram via Holography}},
  \href{http://arxiv.org/abs/2406.12772}{{\tt 2406.12772}}.

\bibitem{Li:2022zjc}
K.~Li, Y.~Ling, P.~Liu and M.-H. Wu, \emph{{Learning the black hole metric from
  holographic conductivity}},
  \href{http://dx.doi.org/10.1103/PhysRevD.107.066021}{\emph{Phys. Rev. D} {\bf
  107} (2023) 066021}, [\href{http://arxiv.org/abs/2209.05203}{{\tt
  2209.05203}}].

\bibitem{Kim:2024car}
S.~Kim, K.~K. Kim and Y.~Seo, \emph{{Phase Diagram from Nonlinear Interaction
  between Superconducting Order and Density: Toward Data-Based Holographic
  Superconductor}},  \href{http://arxiv.org/abs/2410.06523}{{\tt 2410.06523}}.

\bibitem{Hashimoto:2024yev}
K.~Hashimoto, K.~Matsuo, M.~Murata, G.~Ogiwara and D.~Takeda,
  \emph{{Machine-learning emergent spacetime from linear response in future
  tabletop quantum gravity experiments}},
  \href{http://arxiv.org/abs/2411.16052}{{\tt 2411.16052}}.

\bibitem{Park:2022fqy}
C.~Park, C.-O. Hwang, K.~Cho and S.-J. Kim, \emph{{Dual geometry of
  entanglement entropy via deep learning}},
  \href{http://dx.doi.org/10.1103/PhysRevD.106.106017}{\emph{Phys. Rev. D} {\bf
  106} (2022) 106017}, [\href{http://arxiv.org/abs/2205.04445}{{\tt
  2205.04445}}].

\bibitem{Park:2023slm}
C.~Park, S.~Kim and J.~H. Lee, \emph{{Holography Transformer}},
  \href{http://arxiv.org/abs/2311.01724}{{\tt 2311.01724}}.

\bibitem{Legros_2018}
A.~Legros, S.~Benhabib, W.~Tabis, F.~Lalibert{\'e}, M.~Dion, M.~Lizaire et~al.,
  \emph{Universal t-linear resistivity and planckian dissipation in overdoped
  cuprates}, \href{http://dx.doi.org/10.1038/s41567-018-0334-2}{\emph{Nature
  Physics} {\bf 15} (Nov., 2018) 142--147}.

\bibitem{Fournier:1998aa}
P.~Fournier, \emph{Insulator-metal crossover near optimal doping in
  pr$_{2-x}$ce$_{x}$cuo$_{4\pm\delta}$: Anomalous normal-state low temperature
  resistivity},
  \href{http://dx.doi.org/10.1103/PhysRevLett.81.4720}{\emph{Physical Review
  Letters} {\bf 81} (1998) 4720--4723}.

\bibitem{Dagan:2004aa}
Y.~Dagan, \emph{Evidence for a quantum phase transition in
  pr$_{2-x}$ce$_{x}$cuo$_{4-\delta}$ from transport measurements},
  \href{http://dx.doi.org/10.1103/PhysRevLett.92.167001}{\emph{Physical Review
  Letters} {\bf 92} (2004) }.

\bibitem{Tafti:2014aa}
F.~F. Tafti, \emph{Nernst effect in the electron-doped cuprate superconductor
  pr$_{2-x}$ce$_{x}$cuo$_{4}$: Superconducting fluctuations, upper critical
  field h$_{c2}$, and the origin of the t$_c$ dome},
  \href{http://dx.doi.org/10.1103/PhysRevB.90.024519}{\emph{Physical Review B}
  {\bf 90} (2014) }.

\bibitem{Jin_2011}
K.~Jin, N.~P. Butch, K.~Kirshenbaum, J.~Paglione and R.~L. Greene, \emph{Link
  between spin fluctuations and electron pairing in copper oxide
  superconductors}, \href{http://dx.doi.org/10.1038/nature10308}{\emph{Nature}
  {\bf 476} (Aug., 2011) 73--75}.

\bibitem{Sarkar:2017aa}
T.~Sarkar, \emph{Fermi surface reconstruction and anomalous low-temperature
  resistivity in electron-doped la$_{2-x}$ce$_{x}$cuo$_{4}$},
  \href{http://dx.doi.org/10.1103/PhysRevB.96.155449}{\emph{Physical Review B}
  {\bf 96} (2017) }.

\bibitem{Martin:1990aa}
S.~Martin, \emph{Normal-state transport properties of
  bi$_{2+x}$sr$_{2-y}$cuo$_{6+\delta}$ crystals},
  \href{http://dx.doi.org/10.1103/PhysRevB.41.846}{\emph{Physical Review B}
  {\bf 41} (1990) 846--849}.

\bibitem{Cooper_2009}
R.~A. Cooper, Y.~Wang, B.~Vignolle, O.~J. Lipscombe, S.~M. Hayden, Y.~Tanabe
  et~al., \emph{Anomalous criticality in the electrical resistivity of
  la$_{2-x}$sr$_{x}$cuo$_{4}$},
  \href{http://dx.doi.org/10.1126/science.1165015}{\emph{Science} {\bf 323}
  (Jan., 2009) 603--607}.

\bibitem{Daou_2008}
R.~Daou, N.~Doiron-Leyraud, D.~LeBoeuf, S.~Y. Li, F.~Lalibert{\'e},
  O.~Cyr-Choini{\`e}re et~al., \emph{Linear temperature dependence of
  resistivity and change in the fermi surface at the pseudogap critical point
  of a high-tc superconductor},
  \href{http://dx.doi.org/10.1038/nphys1109}{\emph{Nature Physics} {\bf 5}
  (Nov., 2008) 31--34}.

\bibitem{Collignon:2017aa}
C.~Collignon, \emph{Fermi-surface transformation across the pseudogap critical
  point of the cuprate superconductor la$_{1.6-x}$nd$_{0.4}$sr$_{x}$cuo$_{4}$},
  \href{http://dx.doi.org/10.1103/PhysRevB.95.224517}{\emph{Physical Review B}
  {\bf 95} (2017) }.

\bibitem{Doiron_Leyraud_2017}
N.~Doiron-Leyraud, O.~Cyr-Choini{\`e}re, S.~Badoux, A.~Ataei, C.~Collignon,
  A.~Gourgout et~al., \emph{Pseudogap phase of cuprate superconductors confined
  by fermi surface topology},
  \href{http://dx.doi.org/10.1038/s41467-017-02122-x}{\emph{Nature
  Communications} {\bf 8} (Dec., 2017) }.

\bibitem{Hartnoll:2021ydi}
S.~A. Hartnoll and A.~P. Mackenzie, \emph{{Colloquium: Planckian dissipation in
  metals}}, \href{http://dx.doi.org/10.1103/RevModPhys.94.041002}{\emph{Rev.
  Mod. Phys.} {\bf 94} (2022) 041002},
  [\href{http://arxiv.org/abs/2107.07802}{{\tt 2107.07802}}].

\bibitem{sachdev2011}
S.~Sachdev, \emph{Quantum phase transitions}.
\newblock Cambridge University Press, Cambridge, second ed.~ed., 2011.

\bibitem{Bruin_2013}
J.~A.~N. Bruin, H.~Sakai, R.~S. Perry and A.~P. Mackenzie, \emph{Similarity of
  scattering rates in metals showing t -linear resistivity},
  \href{http://dx.doi.org/10.1126/science.1227612}{\emph{Science} {\bf 339}
  (Feb., 2013) 804--807}.

\bibitem{RAISSI2019686}
M.~Raissi, P.~Perdikaris and G.~Karniadakis, \emph{Physics-informed neural
  networks: A deep learning framework for solving forward and inverse problems
  involving nonlinear partial differential equations},
  \href{http://dx.doi.org/https://doi.org/10.1016/j.jcp.2018.10.045}{\emph{Journal
  of Computational Physics} {\bf 378} (2019) 686--707}.

\bibitem{Raissi:2017aa}
M.~Raissi, P.~Perdikaris and G.~E. Karniadakis, \emph{Physics informed deep
  learning (part i): Data-driven solutions of nonlinear partial differential
  equations},  \href{http://arxiv.org/abs/1711.10561}{{\tt 1711.10561}}.

\bibitem{Raissi:2017ab}
M.~Raissi, P.~Perdikaris and G.~E. Karniadakis, \emph{Physics informed deep
  learning (part ii): Data-driven discovery of nonlinear partial differential
  equations},  \href{http://arxiv.org/abs/1711.10566}{{\tt 1711.10566}}.

\bibitem{Karniadakis_2021}
G.~E. Karniadakis, I.~G. Kevrekidis, L.~Lu, P.~Perdikaris, S.~Wang and L.~Yang,
  \emph{Physics-informed machine learning},
  \href{http://dx.doi.org/10.1038/s42254-021-00314-5}{\emph{Nature Reviews
  Physics} {\bf 3} (May, 2021) 422--440}.

\bibitem{Toscano:2024aa}
J.~D. Toscano, V.~Oommen, A.~J. Varghese, Z.~Zou, N.~A. Daryakenari, C.~Wu
  et~al., \emph{From pinns to pikans: Recent advances in physics-informed
  machine learning},  \href{http://arxiv.org/abs/2410.13228}{{\tt 2410.13228}}.

\bibitem{Cuomo_2022}
S.~Cuomo, V.~S. Di~Cola, F.~Giampaolo, G.~Rozza, M.~Raissi and F.~Piccialli,
  \emph{Scientific machine learning through physics--informed neural networks:
  Where we are and what's next},
  \href{http://dx.doi.org/10.1007/s10915-022-01939-z}{\emph{Journal of
  Scientific Computing} {\bf 92} (July, 2022) }.

\bibitem{Farea_2024}
A.~Farea, O.~Yli-Harja and F.~Emmert-Streib, \emph{Understanding
  physics-informed neural networks: Techniques, applications, trends, and
  challenges}, \href{http://dx.doi.org/10.3390/ai5030074}{\emph{AI} {\bf 5}
  (Aug., 2024) 1534--1557}.

\bibitem{Ganga:2024aa}
S.~Ganga and Z.~Uddin, \emph{Exploring physics-informed neural networks: From
  fundamentals to applications in complex systems},
  \href{http://arxiv.org/abs/2410.00422}{{\tt 2410.00422}}.

\bibitem{Raissi:2024aa}
M.~Raissi, P.~Perdikaris, N.~Ahmadi and G.~E. Karniadakis,
  \emph{Physics-informed neural networks and extensions},
  \href{http://arxiv.org/abs/2408.16806}{{\tt 2408.16806}}.

\bibitem{Zhao_2024}
C.~Zhao, F.~Zhang, W.~Lou, X.~Wang and J.~Yang, \emph{A comprehensive review of
  advances in physics-informed neural networks and their applications in
  complex fluid dynamics},
  \href{http://dx.doi.org/10.1063/5.0226562}{\emph{Physics of Fluids} {\bf 36}
  (Oct., 2024) }.

\bibitem{Cai2021}
S.~Cai, Z.~Mao, Z.~Wang, M.~Yin and G.~E. Karniadakis, \emph{Physics-informed
  neural networks (pinns) for fluid mechanics: a review},
  \href{http://dx.doi.org/10.1007/s10409-021-01148-1}{\emph{Acta Mechanica
  Sinica} {\bf 37} (Dec, 2021) 1727--1738}.

\bibitem{HBWJ2023}
B.~Huang and J.~Wang, \emph{Applications of physics-informed neural networks in
  power systems - a review},
  \href{http://dx.doi.org/10.1109/TPWRS.2022.3162473}{\emph{IEEE Transactions
  on Power Systems} {\bf 38} (2023) 572--588}.

\bibitem{Lawal_2022}
Z.~K. Lawal, H.~Yassin, D.~T.~C. Lai and A.~Che~Idris, \emph{Physics-informed
  neural network (pinn) evolution and beyond: A systematic literature review
  and bibliometric analysis},
  \href{http://dx.doi.org/10.3390/bdcc6040140}{\emph{Big Data and Cognitive
  Computing} {\bf 6} (Nov., 2022) 140}.

\bibitem{Raissi_2020}
M.~Raissi, A.~Yazdani and G.~E. Karniadakis, \emph{Hidden fluid mechanics:
  Learning velocity and pressure fields from flow visualizations},
  \href{http://dx.doi.org/10.1126/science.aaw4741}{\emph{Science} {\bf 367}
  (Feb., 2020) 1026--1030}.

\bibitem{HORNIK1989359}
K.~Hornik, M.~Stinchcombe and H.~White, \emph{Multilayer feedforward networks
  are universal approximators},
  \href{http://dx.doi.org/https://doi.org/10.1016/0893-6080(89)90020-8}{\emph{Neural
  Networks} {\bf 2} (1989) 359--366}.

\bibitem{Kingma:2014vow}
D.~P. Kingma and J.~Ba, \emph{{Adam: A Method for Stochastic Optimization}},
  12, 2014.
\newblock \href{http://arxiv.org/abs/1412.6980}{{\tt 1412.6980}}.

\bibitem{towardsdatasciencePhysicsInformed}
I.~Henderson, ``Physics informed neural networks (pinns): An intuitive guide.''
  \url{https://towardsdatascience.com/physics-informed-neural-networks-pinns-an-intuitive-guide-fff138069563}.

\bibitem{Smith:2015aa}
L.~N. Smith, \emph{Cyclical learning rates for training neural networks},
  \href{http://arxiv.org/abs/1506.01186}{{\tt 1506.01186}}.

\bibitem{Smith:2017aa}
L.~N. Smith and N.~Topin, \emph{Super-convergence: Very fast training of neural
  networks using large learning rates},
  \href{http://arxiv.org/abs/1708.07120}{{\tt 1708.07120}}.

\bibitem{Chen:2018wjc}
R.~T.~Q. Chen, Y.~Rubanova, J.~Bettencourt and D.~Duvenaud, \emph{{Neural
  Ordinary Differential Equations}}, {\emph{Advances in Neural Information
  Processing Systems 31} (2018) }, [\href{http://arxiv.org/abs/1806.07366}{{\tt
  1806.07366}}].

\bibitem{Amoretti:2015gna}
A.~Amoretti and D.~Musso, \emph{{Magneto-transport from momentum dissipating
  holography}}, \href{http://dx.doi.org/10.1007/JHEP09(2015)094}{\emph{JHEP}
  {\bf 09} (2015) 094}, [\href{http://arxiv.org/abs/1502.02631}{{\tt
  1502.02631}}].

\bibitem{Blake:2015ina}
M.~Blake, A.~Donos and N.~Lohitsiri, \emph{{Magnetothermoelectric Response from
  Holography}}, \href{http://dx.doi.org/10.1007/JHEP08(2015)124}{\emph{JHEP}
  {\bf 08} (2015) 124}, [\href{http://arxiv.org/abs/1502.03789}{{\tt
  1502.03789}}].

\bibitem{Hartnoll:2015aa}
S.~A. Hartnoll, \emph{Scaling theory of the cuprate strange metals},
  \href{http://dx.doi.org/10.1103/PhysRevB.91.155126}{\emph{Physical Review B}
  {\bf 91} (2015) }.

\bibitem{Karch:2014mba}
A.~Karch, \emph{{Conductivities for Hyperscaling Violating Geometries}},
  \href{http://dx.doi.org/10.1007/JHEP06(2014)140}{\emph{JHEP} {\bf 06} (2014)
  140}, [\href{http://arxiv.org/abs/1405.2926}{{\tt 1405.2926}}].

\bibitem{Karch:2015zqd}
A.~Karch, K.~Limtragool and P.~W. Phillips, \emph{{Unparticles and Anomalous
  Dimensions in the Cuprates}},
  \href{http://dx.doi.org/10.1007/JHEP03(2016)175}{\emph{JHEP} {\bf 03} (2016)
  175}, [\href{http://arxiv.org/abs/1511.02868}{{\tt 1511.02868}}].

\bibitem{wipMLteam}
B.~Ahn, H.-S. Jeong, C.-W. Ji, K.-Y. Kim and K.~Yun, \emph{working in
  progress}, .

\bibitem{Dupont:2019aa}
E.~Dupont, A.~Doucet and Y.~W. Teh, \emph{Augmented neural odes},
  {\emph{Advances in Neural Information Processing Systems 32} (2019) },
  [\href{http://arxiv.org/abs/1904.01681}{{\tt 1904.01681}}].

\bibitem{Massaroli:2020aa}
S.~Massaroli, M.~Poli, J.~Park, A.~Yamashita and H.~Asama, \emph{Dissecting
  neural odes}, {\emph{Advances in Neural Information Processing Systems 33}
  (2020) }, [\href{http://arxiv.org/abs/2002.08071}{{\tt 2002.08071}}].

\bibitem{Yan:2019aa}
H.~Yan, J.~Du, V.~Y.~F. Tan and J.~Feng, \emph{On robustness of neural ordinary
  differential equations},  \href{http://arxiv.org/abs/1910.05513}{{\tt
  1910.05513}}.

\bibitem{Rackauckas:2020aa}
C.~Rackauckas, Y.~Ma, J.~Martensen, C.~Warner, K.~Zubov, R.~Supekar et~al.,
  \emph{Universal differential equations for scientific machine learning},
  \href{http://arxiv.org/abs/2001.04385}{{\tt 2001.04385}}.

\bibitem{gmd-16-6671-2023}
J.~Bolibar, F.~Sapienza, F.~Maussion, R.~Lguensat, B.~Wouters and F.~P\'erez,
  \emph{Universal differential equations for glacier ice flow modelling},
  \href{http://dx.doi.org/10.5194/gmd-16-6671-2023}{\emph{Geoscientific Model
  Development} {\bf 16} (2023) 6671--6687}.

\bibitem{Teshima:2020aa}
T.~Teshima, K.~Tojo, M.~Ikeda, I.~Ishikawa and K.~Oono, \emph{Universal
  approximation property of neural ordinary differential equations},
  \href{http://arxiv.org/abs/2012.02414}{{\tt 2012.02414}}.

\bibitem{Bournez:2017aa}
O.~Bournez and A.~Pouly, \emph{A universal ordinary differential equation},
  \href{http://dx.doi.org/https://doi.org/10.23638/LMCS-16%281%3A28%292020}{\emph{Logical
  Methods in Computer Science, Volume 16, Issue 1 (February 28, 2020)
  lmcs:4437} (02, 2017) }, [\href{http://arxiv.org/abs/1702.08328}{{\tt
  1702.08328}}].

\end{thebibliography}

\providecommand{\href}[2]{#2}\begingroup\raggedright\endgroup

\end{document}